\documentclass[aps, pra, 10pt, twocolumn, groupedaddress, superscriptaddress, nofootinbib]{revtex4-2}

\usepackage[british]{babel} 
\usepackage{times}[tx]
\usepackage{amsmath}
\usepackage{amsfonts}
\usepackage{amsthm}
\usepackage{amssymb}

\DeclareMathOperator{\arctanh}{arctanh}

\usepackage{amssymb} 
\usepackage{mathtools}
\usepackage{physics}  

\usepackage{graphicx} 
\graphicspath{{figures/}} 
\usepackage{multirow}

\usepackage[dvipsnames]{xcolor} 
\definecolor{myblue}{RGB}{34,31,150}

\usepackage[breaklinks=true,colorlinks=true,linkcolor=myblue,urlcolor=myblue,citecolor=myblue]{hyperref} 

\begin{document}

\title{On the role of symmetry and geometry in global quantum sensing}

\author{Julia Boeyens}
\affiliation{Naturwissenschaftlich-Technische Fakultät, Universität Siegen, Siegen 57068, Germany}

\author{Jonas Glatthard}
\affiliation{Department of Physics and Astronomy, University of Exeter, Exeter EX4 4QL, United Kingdom}
\affiliation{School of Physics and Astronomy,
University of Nottingham, Nottingham NG7 2RD, United Kingdom}

\author{Edward Gandar}
\affiliation{Department of Physics and Astronomy, University of Exeter, Exeter EX4 4QL, United Kingdom}

\author{Stefan Nimmrichter}
\affiliation{Naturwissenschaftlich-Technische Fakultät, Universität Siegen, Siegen 57068, Germany}

\author{Luis A. Correa}
\affiliation{Instituto Universitario de Estudios Avanzados (IUdEA), Universidad de La Laguna, La Laguna 38203, Spain}
\affiliation{Sección de Física, Facultad de Ciencias, Universidad de La Laguna, La Laguna 38203, Spain}
\affiliation{Department of Physics and Astronomy, University of Exeter, Exeter EX4 4QL, United Kingdom}

\author{Jes\'{u}s Rubio}
\email{j.rubiojimenez@surrey.ac.uk}
\affiliation{School of Mathematics and Physics, University of Surrey, Guildford GU2 7XH, United Kingdom}

\date{\today}
   
\begin{abstract}

Global quantum sensing enables parameter estimation across arbitrary ranges with a finite number of measurements. 
Among the various existing formulations, the Bayesian paradigm stands as a flexible approach for optimal protocol design under minimal assumptions.
Within this paradigm, however, there are two fundamentally different ways to capture prior ignorance and uninformed estimation; namely, requiring invariance of the prior distribution under specific parameter transformations, or adhering to the geometry of a state space. 
In this paper we carefully examine the practical consequences of both the invariance-based and the geometry-based approaches, and show how to apply them in relevant examples of rate and coherence estimation in noisy settings.
We find that, while the invariance-based approach often leads to simpler priors and estimators and is more broadly applicable in adaptive scenarios, the geometry-based one can lead to faster posterior convergence in a well-defined measurement setting.
Crucially, by employing 
the notion of location-isomorphic parameters, we are able to unify the two formulations into a single practical and versatile framework for optimal global quantum sensing, detailing when and how each set of assumptions should be employed to tackle any given estimation task. We thus provide a blueprint for the design of novel high-precision quantum sensors.

\end{abstract}

\maketitle


\section{Introduction}

Precise knowledge of a system's parameters is crucial for the control of quantum devices \cite{degen2017, doherty2000} and their application to the study of nature \cite{schnabel_quantum_2010, belenchia2021quantum, carney2021mechanical, kimball2023, ye2024essayquantumsensing, bass2024quantum, demille2024quantum}. 
Strategies to estimate such parameters are typically designed to optimise sensing performance over many measurements, both in conventional phase estimation \cite{xiao1987, Giovanetti2006, Dorner2009, demkowicz2015quantum} and in emerging applications of open-system metrology \cite{Correa2015, Wu2020, saleem2023, Mirkhalaf2024}. These strategies generally rely on local estimation, which assumes that the true value of the parameter is approximately known within a small window \cite{paris2009, demkowicz2015quantum}. 

However, in many practical applications, the available data for estimating these parameters is inherently limited \cite{Teklu_2009, lumino2018experimental, rubio2018quantum, Bavaresco2024, meyer2023quantum, Cimini_2024}. For instance, quantum optical and optomechanical platforms are increasingly put forward as sensors to probe fundamental physics and, possibly, discover or rule out yet unobserved phenomena, such as quantum signatures of gravity, dark matter particles, and spontaneous collapse \cite{buchmueller2023largescale, bose2024massivequantumsystemsinterfaces, braun2025metrologygravitationaleffectsmechanical, ye2024essayquantumsensing, sherrill2023, amaral2024vector, amaral2024first}. 
Signals of such type are expected to be weak and poorly characterised, while the corresponding experiments would be difficult to realise and the data, sparse. This not only calls for suitable quantum-enhanced sensing schemes, but also for a prescription of optimal estimation strategies from finite statistics. 

The Bayesian paradigm is well suited for this task, as noticed, e.g, in the context of quantum--classical model selection \cite{ralph2018dynamical} and tests of collapse models \cite{schrinski2020macroscopicity, schrinski2020quantumclassical, schrinski2023testing, schrinski2023macroscopic, laing2024bayesian}.
Within this paradigm, any information available prior to performing a measurement is quantified through a prior distribution, which is subsequently updated as new data are acquired \cite{jaynes2003probability, toussaint2011bayesian}. 
This procedure is general and can be applied to any sample size.
Consequently, Bayesian techniques naturally enable \emph{global} estimation and sensing, allowing the unknown parameter to initially lie within a window of arbitrary size.

Bayesian techniques readily provide the optimisation equations needed to construct optimal estimation strategies \cite{jaynes2003probability, helstrom1976quantum}.
This arises from the explicit incorporation of prior information. 
As we shall see, prior information, or rather its absence, can also be used to derive loss functions---quantifying the errors we aim to minimise---from first principles. 
In complex scenarios, however, solving these optimisation equations can be challenging or even infeasible, motivating the widespread use of lower error bounds, either Bayesian \cite{tsang2012zivzakai, tsang2016quantum, mehboudi2021fundamental} or frequentist \cite{yan2018frequentist, gessner2023hierarchies}, as a practical compromise. 
Yet, the estimation strategies suggested by error bounds often lack generality, being optimal only under specific circumstances \cite{demkowicz2015quantum, rubio2018non, salmon2023only, dennis2024}.
This motivates a redirection of efforts towards the search for new methods to optimise Bayesian quantum estimation strategies \emph{exactly} \cite{bavaresco2024designing, kurdzialek2024quantum}.

One such method uses the notion of \emph{location-isomorphic} parameter \cite{rubio2024first, overton2024five}, i.e., a parameter that does not represent a location---such as temperature, decay rate, or probability of success---but can be mapped to a location parameter through an appropriate transformation.  
For instance, scale parameters are mapped to locations via a logarithmic transformation.  
Since, for location parameters, the optimal estimation strategy can be written in closed form in both classical \cite{linden2014bayesian, jaynes2003probability} and quantum settings \cite{personick1971application, helstrom1976quantum}, such a mapping extends these optimal strategies to any location-isomorphic parameter. 

Furthermore, the class of estimation problems that are location-isomorphic is remarkably broad. 
For instance, global quantum thermometry \cite{rubio2021global, mehboudi2021fundamental, boeyens2021uninformed, chang2024global, mok2021optimal, alves2022, mukhopadhyay2025current}, based on either quantum scale estimation \cite{rubio2021global, rubio2023quantum} or geometric considerations via the notion of thermodynamic length \cite{jorgensen2021bayesian}, was an early realisation of location-isomorphic estimation. 
In addition, scale-invariant errors have been applied to length-scale estimation for massive oscillators \cite{volkoff2025}.
More importantly, location-isomorphic strategies have been experimentally validated in the estimation of temperature \cite{Glatthard2022, hewitt2024controlling} and atom number \cite{overton2024five} in cold-atom platforms. 
In general, based on available examples, the location-isomorphic formalism appears to be applicable whenever the parameter of interest lies within an interval lacking a cyclic topology. 
The only notable counterexample we are aware of is phase estimation.

Realising the full potential of location-isomorphic estimation requires a clear and practical methodology to determine how a given parameter can be transformed into a location parameter.
The present work is, to the best of our knowledge, the first to provide such a methodology.
Namely, we derive an explicit expression for the function that transforms any given parameter into a location parameter. This is found to be proportional to the indefinite integral of a prior distribution representing maximum ignorance.
Since such a function fully determines optimal estimators, POVMs, and errors \cite{rubio2024first, overton2024five}, the problem effectively reduces to the well-known task of identifying suitable ignorance priors \cite{kass1996the, jaynes2003probability, toussaint2011bayesian}.

The search for ignorance priors is typically done following two classic approaches.
One relies on the underlying symmetries of the estimation problem \cite{jaynes1968prior,harney_bayesian_2003, eatonsudderth2004,tanaka2012, demkowicz2020multi}, prominently advocated by \citet{jaynes2003probability} and later given a modern, rigorous formulation by \citet{linden2014bayesian}.
The aforementioned quantum scale estimation framework \cite{rubio2023quantum} is an example of this idea. 
The second approach relies on information geometry \cite{brody1996geometry, amari2016, goldberg2021}, which employs tools from differential geometry and a notion of distance derived from asymptotic estimation theory to study inference problems.
Bayesian information geometry \cite{kass1996the, jermyn2005invariant, snoussi2007, linden2014bayesian} utilises this distance to select the desired ignorance prior, which corresponds to the well-known Jeffreys's general rule \cite{jeffreys1946invariant}. This has been applied, e.g., to global quantum thermometry \cite{jorgensen2021bayesian} as mentioned above and in quantum state estimation \cite{slater1995,kwek1999quantum,tanaka2012,li2018}.

Here we show that both approaches can be unified within a single formalism, revealing that invariance principles and geometric arguments can, in fact, be seen as two manifestations of a location-isomorphic framework.
When focusing on specific symmetries, we further demonstrate that full POVM and probe optimisation becomes possible---something generally not achievable when relying solely on geometric approaches due to the reliance on a fixed probe state.
Ultimately, our methodology yields both practical algorithms for adaptive optimisation under finite sample sizes and arbitrary prior information, as well as fundamental precision limits.

In addition, by carefully tracking the assumptions underlying the construction of ignorance priors, we argue that focusing on specific symmetries is more suitable when the state or likelihood function---which relates measurands and parameters---is partially or entirely unknown. 
Conversely, when details of the measurement process are well known, allowing the corresponding statistical models to be fixed, information geometry can facilitate faster convergence of estimates to the true value and thus accelerate information acquisition.  
We illustrate these conclusions through representative case studies, including the estimation of rate parameters in Poisson processes, quantum coherence in a depolarising channel, and the lifetime of an atomic state undergoing spontaneous photon emission.

\begin{figure}[t]
    \includegraphics[trim={17cm 5cm 17cm 2cm},clip,width=\linewidth]{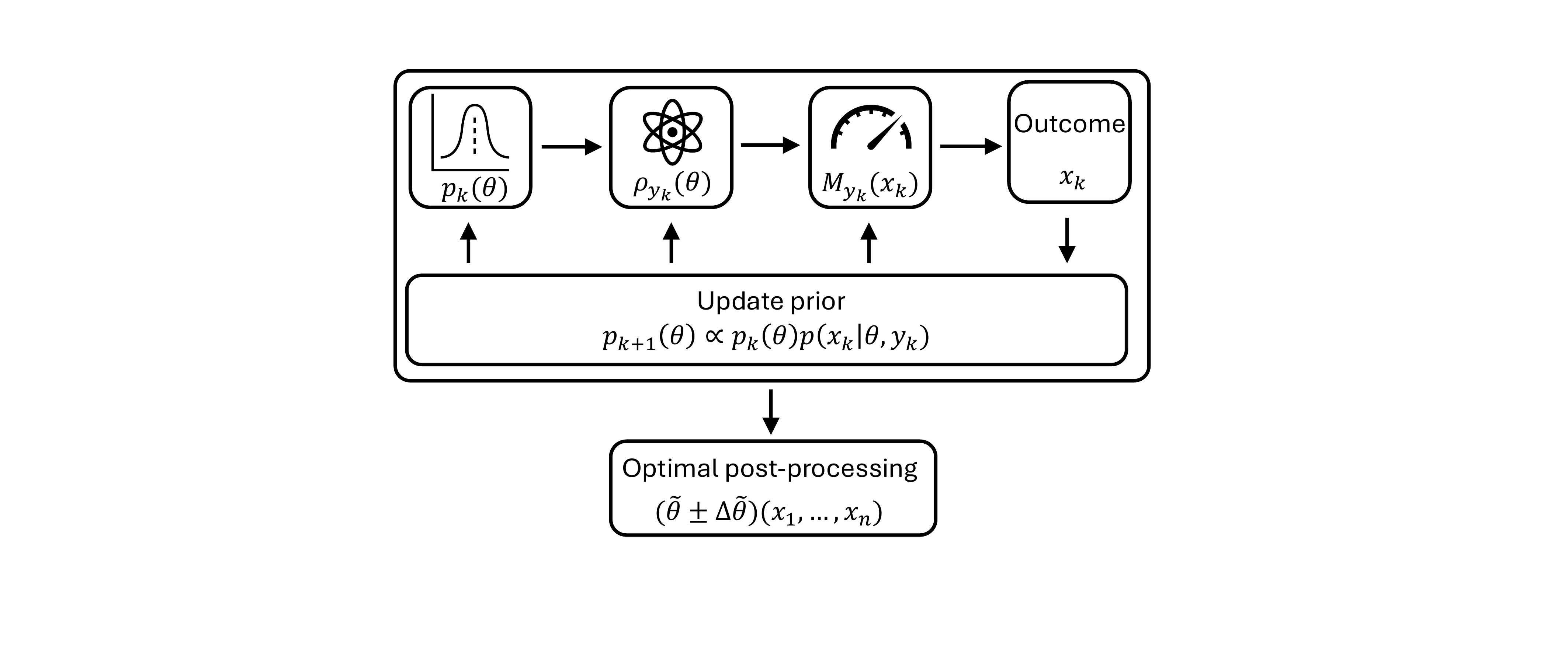}
    \caption{Diagram illustrating the reasoning behind adaptive Bayesian parameter estimation. 
    One begins with a prior distribution $p_0(\theta)$ and a probe state $\rho_{y_0}(\theta)$ that encodes information about the parameter of interest $\theta$. 
    Here, $y_k$ is the control parameter used in shot $k$. 
    A POVM $M_{y_k}(x_k)$ is applied, yielding outcome $x_k$ with likelihood $p(x_k | \theta, y_k) = \Tr\left[ M_{y_k}(x_k) \rho_{y_k}(\theta) \right]$. 
    The prior is updated via Bayes's rule and can be used to optimise the probe state and measurement settings for subsequent shots. 
    When desired, the full set of outcomes $x_1, \ldots, x_n$ can be post-processed to produce an estimate for the unknown parameter. 
    Further details are provided in Sec.~\ref{sec:simmetry-informed_estimation}.}
\label{fig:figure-1}
\end{figure}

\begin{figure*}[t]
    \includegraphics[trim={0cm 0.5cm 0cm 0cm},clip,width=\linewidth]{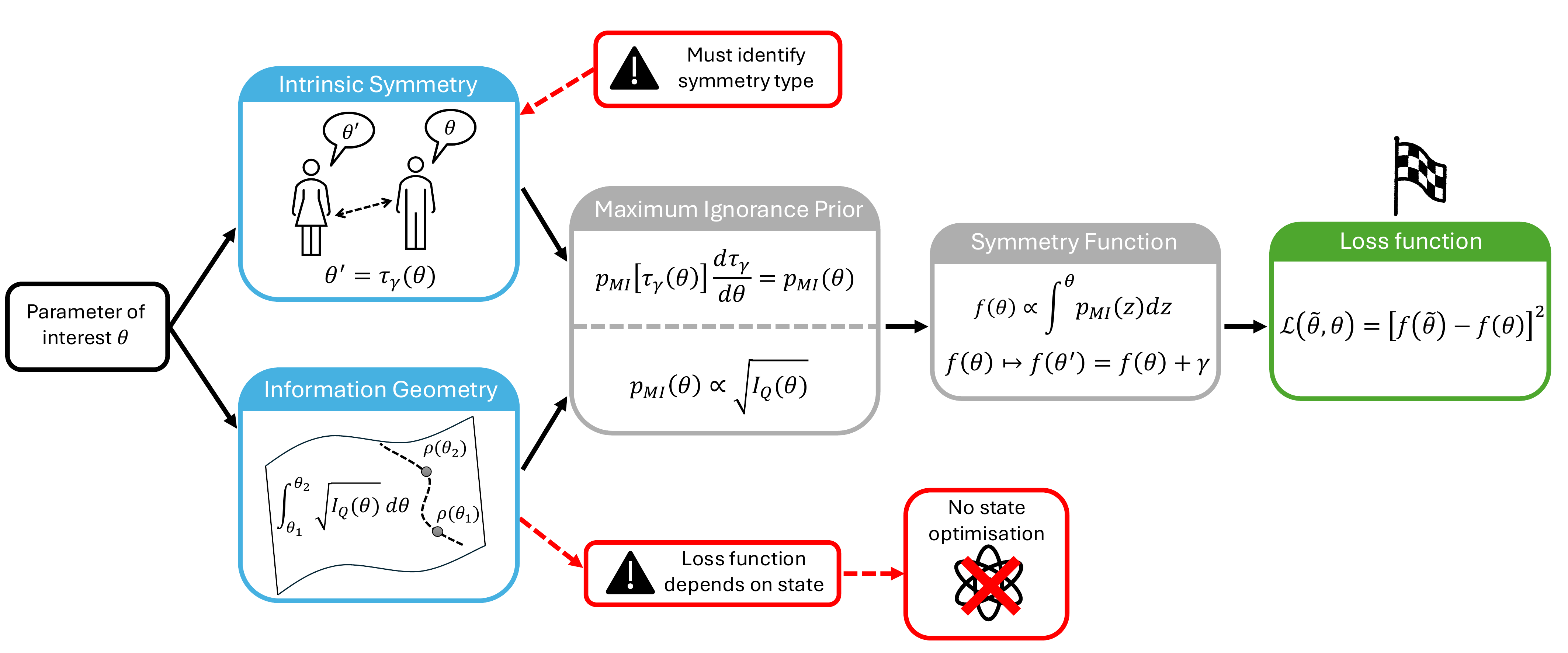}
    \caption{
    Diagram illustrating the construction of loss functions based on specific symmetries or the geometry of the state space. 
    Given the nature of the unknown parameter $\theta$, in the first approach, the ignorance prior satisfies the functional equation \eqref{eq:prior-transformation}, which requires identifying a symmetry $\tau_\gamma (\theta)$ of the problem. 
    In the information geometry approach, the geodesic length is given by Jeffreys's rule and is directly proportional to the cumulative probability under the ignorance prior. 
    In both frameworks, the symmetry function $f(\theta)$, calculated via Eq.~\eqref{eq:symmetry-equation}, can be computed to map the parameter to a location variable with an associated quadratic loss function (see Sec.~\ref{sec:simmetry-informed_estimation}).    
    Further details about the scope and limitations of these approaches are given in Sec.~\ref{sec:symmetry-section}}
\label{fig:figure-2}
\end{figure*}

In summary, this work introduces a unified and practical framework for Bayesian quantum estimation, including the first general prescription for location-isomorphic transformations based on ignorance priors. 
It clarifies when symmetry-based or geometric approaches are best suited for protocol design under finite data, offering a systematic alternative to increasingly complex error hierarchies or \emph{ad hoc} uncertainty quantifiers. 
A visual summary of our methodology is presented in Figs.~\ref{fig:figure-1} and \ref{fig:figure-2}. 
Fig.~\ref{fig:figure-1} illustrates the workings of adaptive quantum metrology protocols, while Fig.~\ref{fig:figure-2} provides the steps to construct a suitable loss function using either specific symmetries or information geometry. 
These results lay the foundation for advances in optimal measurement design and precision sensing across diverse physical platforms.

The paper is structured as follows. 
Sec.~\ref{sec:simmetry-informed_estimation} generalises previous formulations of location-isomorphic estimation to accommodate both specific symmetries and the geometric structure of states and likelihood functions. Additionally, it compiles the key formulas necessary for practical implementation.  
Sec.~\ref{sec:symmetry-section} introduces our proposed methodology for mapping unknown parameters to location parameters.  
Sec.~\ref{sec:examples} presents the case studies in a tutorial style.   Sec.~\ref{sec:remarks} summarises our conclusions on designing global quantum sensing protocols.  

\section{Location-isomorphic quantum estimation theory}\label{sec:simmetry-informed_estimation}

The ultimate goal of quantum metrology is to estimate an unknown parameter $\Theta$ from a set of measurement outcomes $\boldsymbol{x} = (x_1, \dots, x_{\mu})$ made on a quantum probe \cite{dunningham2006using, demkowicz2015quantum, degen2017quantum}.
In the Bayesian paradigm, the starting point is a prior probability, $p(\theta)$, for a hypothesis, $\theta \in [\theta_{\mathrm{min}}, \theta_{\mathrm{max}}]$, about the true value of $\Theta$.
This probability accounts for any information available prior to performing measurements \cite{jaynes1968prior, kass1996the}.
Assuming that  $\boldsymbol{x}$ represents independent runs of an estimation protocol, as is often the case in sensing experiments \cite{Glatthard2022, overton2024five}, the prior may be updated via Bayes's theorem into the posterior probability 
\begin{equation}
    p(\theta|\boldsymbol{x},\boldsymbol{y}) \propto p(\theta) \prod_{i=1}^{\mu} p(x_i|\theta, y_i)
    \label{eq:bayes}
\end{equation}
of likely parameter values compatible with the data.
Here, $\boldsymbol{y} = (y_1, \dots, y_{\mu})$ are the values taken by some control parameter and $p(x|\theta, y)$ is a likelihood function relating the outcome $x$ with the hypothesis $\theta$ and $y$.\footnote{
Control parameters often do not inform the value of the unknown parameter, i.e., $p(\theta|\boldsymbol{y}) \mapsto p(\theta)$, as is assumed in Eq.~\eqref{eq:bayes}.
Note that control parameters describe relevant experimental settings that can be either fixed or varied from shot to shot. 
}
For quantum systems the likelihood is given as 
\begin{equation}
    p(x|\theta,y) = \mathrm{Tr}[M_{y}(x)\rho_y (\theta)],
    \label{eq:born}
\end{equation}
with positive operator-valued measure (POVM) elements $M_y(x)$ and state $\rho_y(\theta)$. 

Thus, given some measurement outcomes $\boldsymbol{x}$ and a quantum model that determines the likelihood by virtue of Born's rule in Eq.~\eqref{eq:born}, the posterior distribution in Eq.~\eqref{eq:bayes}  encodes all the available information about $\Theta$.
However, it may be more expedient to assign a single parameter value $\tilde\theta_{\boldsymbol{y}} (x)$ to the data, the \emph{estimate}, as well as an error quantifier $\Delta\tilde\theta_{\boldsymbol{y}}(x)$.
The estimator $\tilde\theta_{\boldsymbol{y}}$ is a functional of the posterior that, together with the POVM and the probe state, could be chosen optimally to reduce the error, possibly attaining fundamental precision limits \cite{giovannetti2004quantum, demkowicz2015quantum}.

In the following, we combine the location-isomorphic results of Refs.~\cite{rubio2024first,overton2024five}---which address, respectively, a single-shot setting with POVM optimisation and a multi-shot setting for estimators only---into a single framework for optimal adaptive quantum metrology. 
This unified perspective serves as the foundation for subsequently considering not only group symmetries, but also the geometry of the state space, which is the central focus of this work  and allows for a comparison of these approaches on an even footing (cf. Sec.~\ref{sec:symmetry-section}).

\subsection{Prior ignorance and information loss}\label{sec:prior-loss}

A useful approach to optimal estimation is minimising a quantifier of information loss, averaged over the prior \cite{jaynes2003probability, linden2014bayesian}. In the specific case of location or shift parameters, this yields closed-form analytical expressions for the estimator and the error \cite{helstrom1976quantum}, although, in general, the optimisation is not analytically tractable. 
Location-isomorphic estimation mitigates this issue by mapping a broad class of problems onto the estimation of locations.
In this section, we revisit the mapping, focusing on the transformation of priors and loss functions, before addressing full optimisation in subsequent sections.

In the absence of any \emph{a priori} information about the true value of a location parameter $\Theta$, any hypothesis $\theta \in (-\infty, \infty)$ should be equally informative as a translation of it, i.e.,
\begin{equation}
    \theta' = \theta + \gamma,
    \label{eq:translation}
\end{equation}
with arbitrary $\gamma$.  
This is equivalent to imposing the condition \cite{jaynes2003probability, linden2014bayesian}
\begin{equation}
    p(\theta')d\theta' = p(\theta)d\theta,
    \label{eq:mi-constraint}
\end{equation}
on the prior, which leads to the functional equation
\begin{equation}
    p(\theta + \gamma) = p(\theta).
    \label{eq:functional-location}    
\end{equation}
This indicates that, without additional information, the prior which holds for a location parameter must be invariant under translations.
Eq.~\eqref{eq:functional-location} is solved by the familiar flat prior
\begin{equation}
    p_{\mathrm{MI}}(\theta) \propto 1,
    \label{eq:flat-prior}
\end{equation}
where the subscript stands for \emph{maximum ignorance} \cite{kass1996the, jaynes2003probability}. 
This prior is improper, as its integral over $(-\infty, \infty)$ diverges \cite{toussaint2011bayesian}.
However, one typically normalises Eq.~\eqref{eq:flat-prior} over a finite range, $[\theta_{\mathrm{min}}, \theta_{\mathrm{max}}]$, of practical interest. 
This gives 
 \begin{equation}
    p_{\mathrm{MI}}(\theta) = \frac{1}{\theta_{\mathrm{max}}-\theta_{\mathrm{min}}},
    \label{eq:flat-prior-normalised}
\end{equation}
which is also invariant under translations if $\theta_{\mathrm{max}}$ and $\theta_{\mathrm{min}}$ transform as in Eq.~\eqref{eq:translation}. 

Upon making an estimate $\tilde{\theta}$ of the true value $\Theta$, we need to further quantify the information loss incurred by this process should the hypothesis $\theta$ be correct.
We do so in an translation-invariant manner by calculating the degree of separation between $\tilde{\theta}$ and $\theta$ in the hypothesis space via the squared loss function
\begin{equation}
    \mathcal{L}(\tilde{\theta},\theta) = (\tilde{\theta} - \theta)^2.
    \label{eq:location-losses}
\end{equation}

Flat priors and squared errors are the default choice in most estimation and metrology studies. 
However, while these are appropriate to work with location parameters, they are not necessarily applicable in cases with no translation invariance \cite{kass1996the, holevo2011probabilistic, rubio2023quantum}---a fact often overlooked. 
To solve this problem, Ref.~\cite{rubio2024first} introduced a map  
\begin{equation}
    f\hspace{-0.2em}: \Theta \mapsto f(\Theta)
    \label{eq:map-isomorphic}
\end{equation}
such that $f(\Theta)$ is a location parameter, but $\Theta$ need not be.
We call such parameters \emph{location-isomorphic}. 

Since $\theta$ is a hypothesis for $\Theta$, $f(\theta) \in (-\infty, \infty)$ is the hypothesis for $f(\Theta)$ and absence of \emph{a priori} information is represented in this hypothesis space by $p_{\mathrm{MI}}[f(\theta)] \propto 1$. 
Given that $p_{\mathrm{MI}}(\theta)\,d\theta = p_{\mathrm{MI}}[f(\theta)]\,d f(\theta)$, we find the ignorance prior
\begin{equation}
    p_{\mathrm{MI}}(\theta) \propto \frac{df(\theta)}{d\theta}
    \label{eq:prior-isomorphic}
\end{equation}
in the original hypothesis space.
By normalising over the hypothesis range $[\theta_{\mathrm{min}},\theta_{\mathrm{max}}]$, we arrive at
\begin{equation}
    p_{\mathrm{MI}}(\theta) = \frac{|df(\theta)/d\theta|}{f(\theta_{\mathrm{max}})-f(\theta_{\mathrm{min}})}.
    \label{eq:prior-isomorphic-normalised}
\end{equation}
Likewise, the degree of separation between estimate $\tilde{\theta}$ and hypothesis $\theta$ is given by upgrading the squared loss in Eq.~\eqref{eq:location-losses} to the \emph{quadratic} loss function
\begin{equation}
    \mathcal{L}(\tilde{\theta}, \theta) = [f(\tilde{\theta})-f(\theta)]^2.
    \label{eq:quadratic}
\end{equation}
Naturally, these expressions recover the formalism for location parameters when $f(z) = z$ and are themselves invariant under translations
\begin{equation}
    f(\theta') = f(\theta) +  \gamma.
\end{equation}

As we shall see in Secs.~\ref{sec:symmetry-section} and \ref{sec:examples}, the mapping $f$ for location-isomorphic parameters can be determined by symmetries and constraints inherent to the underlying physics.
For this reason, we refer to $f$ as the \emph{symmetry function} and the resulting estimation framework as \emph{symmetry-informed}.
In addition, quadratic errors unify a wide range of scenarios under a single framework.  
For example, $f(z) = \log(z/z_u)$, with arbitrary positive constant $z_u$, leads to the logarithmic loss 
\begin{equation}
    \mathcal{L}(\tilde{\theta}, \theta) = \log^2\left( \frac{\tilde{\theta}}{\theta} \right)
    \label{eq:log-loss}
\end{equation}
used in global thermometry and scale estimation \cite{rubio2021global, mehboudi2021fundamental,boeyens2021uninformed, rubio2023quantum, chang2024global}.
Furthermore, $f(z) = 2\,\mathrm{arctanh(2z-1)}$ leads to the hyperbolic loss 
\begin{equation}
    \mathcal{L}(\tilde{\theta}, \theta) = 4\,\arctanh^2{\left( \frac{\tilde{\theta}-\theta}{\tilde{\theta}+\theta-2\tilde{\theta}\theta}\right)}
    \label{eq:hyperbolic-loss}
\end{equation}
employed in the estimation of weight parameters \cite{rubio2024first}, such as interferometric transmittance \cite{crowley2014tradeoff}, probability of success \cite{jaynes1968prior}, or superposition coefficients, where the aim is to estimate the weight distribution between two objects as $\Theta$ and $1-\Theta$, respectively (cf. Sec.~\ref{sec:coherence}).
Adopting the quadratic loss criterion in Eq.~\eqref{eq:quadratic}, the following two sections collect and expand on the expressions in Refs.~\cite{rubio2024first, overton2024five} needed to calculate optimal estimation strategies.

Note that the use of a loss function $\mathcal{L}(\tilde{\theta}, \theta)$ may be seen as akin to \emph{mathematically} measuring in the hypothesis space.
Given that this must be done prior to performing any experiment or optimisation, we may proceed as  
\begin{equation}
    \mathcal{L}(\tilde{\theta},\theta) = \bigg| \mathcal{A} \int_\theta^{\tilde{\theta}} dz\,p_{\mathrm{MI}}(z) \bigg|^k,
\end{equation}
where $\mathcal{A}$ is a constant setting the units of loss, while $k$ can be used to gauge the relative weight of outliers in the final error.
For $k =2$, by inserting the flat prior in Eq.~\eqref{eq:flat-prior-normalised} and choosing $\mathcal{A} = \theta_{\mathrm{max}}-\theta_{\mathrm{min}}$, we recover the squared loss function in Eq.~\eqref{eq:location-losses} and, by inserting the symmetry-informed prior in Eq.~\eqref{eq:prior-isomorphic-normalised} and choosing $\mathcal{A} = f(\theta_{\mathrm{max}})-f(\theta_{\mathrm{min}})$, we get the quadratic error in Eq.~\eqref{eq:quadratic}.

We note that the prior used in the construction of a loss function need not coincide with the prior used in a specific experimental inference task, because the two serve different roles: the former defines the class of loss functions appropriate for a given parameter type, which is fixed, while the latter encodes the specific prior knowledge available in a particular experimental scenario. Indeed, here we use maximum-ignorance priors to ensure that information losses are measured in the most conservative manner within a given class of estimation problems---locations, scales, etc. 
This holds irrespective of the prior $p(\theta)$ used for the actual estimation process, which may or may not coincide with $p_{\mathrm{MI}}(\theta)$, and provides an appealing rationale for systematically constructing loss functions from prior information, even beyond the quadratic case considered in this work.

\subsection{Optimal estimator and error for multiple shots}\label{sec:experiments-formalism}

We next proceed to derive optimal strategies based on quadratic loss functions.
In Bayesian theory, an estimator is considered \emph{optimal} with respect to a given loss function when it minimises the average information loss 
\begin{equation}
    \langle \bar{\mathcal{L}}_{\boldsymbol{y}} \rangle = \int d\boldsymbol{x}\,p(\boldsymbol{x}|\boldsymbol{y}) \bar{\mathcal{L}}_{\boldsymbol{y}}(\boldsymbol{x}),
    \label{eq:loss-multi}
\end{equation}
where the evidence is given by
\begin{equation}
    p(\boldsymbol{x}|\boldsymbol{y}) = \int d\theta\,p(\theta)\, p(\boldsymbol{x}|\theta , \,\boldsymbol{y})
    \label{eq:evidence}
\end{equation}
and the empirical (i.e., outcome-dependent) loss, by
\begin{equation}
    \bar{\mathcal{L}}_{\boldsymbol{y}}(\boldsymbol{x}) = \int\,d\theta\,p(\theta|\boldsymbol{x}, \boldsymbol{y})\,\mathcal{L}[\tilde{\theta}_{\boldsymbol{y}}(\boldsymbol{x}), \theta].
    \label{eq:classical-exp-loss}
\end{equation}
For the quadratic loss in Eq.~\eqref{eq:quadratic}, the optimal estimator $\tilde{\vartheta}_{\boldsymbol{y}}$ can be obtained analytically using calculus of variations and is given by
\begin{equation}
\tilde{\vartheta}_{\boldsymbol{y}}(\boldsymbol{x}) = f^{-1}\left[ \int d\theta p(\theta|\boldsymbol{x}, \boldsymbol{y})\,f(\theta) \right].
\label{eq:opt-est-f}
\end{equation}
Furthermore, since $f(\theta)$ represents a location parameter, the upper and lower bounds of the empirical error bars associated with Eq.~\eqref{eq:opt-est-f} can be cast as 
\begin{equation}
    f^{-1}\left\lbrace f[\tilde{\vartheta}_{\boldsymbol{y}}(\boldsymbol{x})] \pm \sqrt{\bar{\mathcal{L}}_{\boldsymbol{y}, \mathrm{opt}}(\boldsymbol{x})} \right\rbrace \simeq \tilde{\vartheta}_{\boldsymbol{y}}(\boldsymbol{x}) \pm \Delta \tilde{\vartheta}_{\boldsymbol{y}}(\boldsymbol{x}),
\end{equation}
where
\begin{equation}
\Delta \tilde{\vartheta}_{\boldsymbol{y}}(\boldsymbol{x}) = \frac{\sqrt{\bar{\mathcal{L}}_{\boldsymbol{y}, \mathrm{opt}}(\boldsymbol{x})}}{|f'[\tilde{\vartheta}_{\boldsymbol{y}}(\boldsymbol{x})]|}
\label{eq:opt-err-f}
\end{equation}
is a Bayesian analogue of the standard deviation.
Here, the Taylor expansion is justified whenever $\sqrt{\bar{\mathcal{L}}_{\boldsymbol{y}, \mathrm{opt}}(\boldsymbol{x})}$ is small, while the optimal empirical loss $\bar{\mathcal{L}}_{\boldsymbol{y}, \mathrm{opt}}(\boldsymbol{x})$ is calculated by inserting Eq.~\eqref{eq:opt-est-f} into Eq.~\eqref{eq:classical-exp-loss}.

Eqs.~\eqref{eq:opt-est-f} and \eqref{eq:opt-err-f} provide an optimal method for post-processing data in any experimental context---classical or quantum---where a quadratic loss function is a suitable quantifier of estimation error.
The use of any other estimator will necessarily result in a larger quadratic information loss.
Note as well that Eqs.~\eqref{eq:opt-est-f} and \eqref{eq:opt-err-f} are valid for any sequence of measurement data $\boldsymbol{x}$ and control parameters $\boldsymbol{y}$, without the need of assuming independent trials in Eq.~\eqref{eq:bayes} or even that the vectors $\boldsymbol{x}$ and $\boldsymbol{y}$ are of the same length \cite{overton2024five}.
	
\subsection{Optimal quantum strategy for individual shots}\label{sec:1shot-formulas}

Besides optimising the post-processing of the measured data, one may also optimise the measurement strategy itself.
To do so, we adopt a practical perspective and treat each experimental run as the fundamental unit for optimisation.

Let us first restrict to a single shot (i.e., we replace $\boldsymbol{x} \mapsto x$ and $\boldsymbol{y} \mapsto y$) and use the Born rule in Eq.~\eqref{eq:born} to write the average infromation loss in Eq.~\eqref{eq:loss-multi} as \cite{helstrom1976quantum}
\begin{equation}
    \langle \bar{\mathcal{L}}_{y} \rangle = \mathrm{Tr} \left\lbrace \int dx\, M_{y}(x) \,W[\tilde{\theta}_{y}(x)] \right\rbrace,
    \label{eq:mean-loss}
\end{equation}
where
\begin{equation}
    W[\tilde{\theta}_{y}(x)] = \int d\theta\,p(\theta)\,\rho_{y}(\theta) \,\mathcal{L}[\tilde{\theta}_{y}(x), \theta].
\end{equation}
Choosing the quadratic loss in Eq.~\eqref{eq:quadratic}, and given a state $\rho_{y}(\theta)$, the optimal estimation strategy is completely determined by an operator $\mathcal{S}_{y}$ that solves the Lyapunov equation
\begin{equation}
    \mathcal{S}_{y} \, \rho_{y, 0} + \rho_{y, 0} \, \mathcal{S}_{y} = 2 \rho_{y,1},
    \label{eq:lyapunov}
\end{equation}
where 
\begin{equation}
    \rho_{y,k} = \int d\theta\,p(\theta) \,\rho_{y}(\theta)\,f(\theta)^k.
    \label{eq:state-moments}
\end{equation}
Specifically, given the eigendecomposition 
\begin{equation}
    \mathcal{S}_{y} = \int ds\,\mathcal{P}_{y}(s)\,s,
\end{equation}
where $\mathcal{P}_{y}(s)\,\mathcal{P}_{y}(s') = \delta(s-s')\,\mathcal{P}_{y}(s')$, the optimal estimator and POVM minimising Eq.~\eqref{eq:mean-loss} are
\begin{subequations}
\label{eq:optimal_strategy}
\begin{align}
    \tilde{\vartheta}_{y}(s) &= f^{-1}(s),
    \label{eq:opt-est-1shot} \\
    \mathcal{M}_{y,f}(s) &= \mathcal{P}_{y}(s),
    \label{eq:opt-povm}
\end{align}
\end{subequations}
respectively. 
That is, the optimal strategy consists in projecting onto the eigenstates of an operator $\mathcal{S}_{y}$ and transforming its spectrum with the $f$ inverse map to produce estimates for $\Theta$. 

We note that, by using the identity \cite{rubio2023quantum}
\begin{equation}
    s = \frac{\mathrm{Tr}[\mathcal{P}_{y}(s)\,\rho_{y,1}]}{\mathrm{Tr}[\mathcal{P}_{y}(s) \,\rho_{y,0}]},
    \label{eq:special_identity}
\end{equation}
we may recover Eq.~\eqref{eq:opt-est-f} from Eq.~\eqref{eq:opt-est-1shot}, thus demonstrating the consistency between these two estimator-optimisation procedures.
Furthermore, selecting $f(z) = z$, one can establish a linear relationship between $\mathcal{S}_{y}$ and the symmetric logarithmic derivative used in local estimation \cite{demkowicz2020multi}. 
This can be achieved, for example, by performing a linear expansion of $\rho_y(\theta)$ around the mean of the prior distribution \cite{branford2021average}.

By inserting Eqs.~\eqref{eq:optimal_strategy} into Eq.~\eqref{eq:mean-loss}, we may write the averaged minimum information loss as
\begin{equation}
    \langle \bar{\mathcal{L}}_{y} \rangle_{\mathrm{min}} = \int d\theta\,p(\theta) f(\theta)^2 - \mathcal{G}_{y},
    \label{eq:minimum-loss}
\end{equation}
where
\begin{equation}
    \mathcal{G}_{y} = \mathrm{Tr}(\rho_{y, 0} \mathcal{S}_{y}^2) = \int ds\,p(s|y) f[\tilde{\vartheta}_y(s)]^2
    \label{eq:precision_gain}
\end{equation}
and $p(s\vert y)$ is defined analogously to Eq.~\eqref{eq:evidence}.
Noting that the first contribution to Eq.~\eqref{eq:minimum-loss} is fully determined by the prior information, we may interpret $\mathcal{G}_{y}$ as a quantifier of average precision gain due to the single measurement outcome $ x $. One should thus maximise $\mathcal{G}_{y}$ over the control parameter $ y $ to make the most of that first shot. Alternatively, if the prior $p(\theta)$ is replaced by
\begin{equation}
   p_{n-1}(\theta) \coloneqq
\begin{cases}
    p(\theta) & \text{if } n = 1 \\
    p(\theta) \prod_{i = 1}^{n-1} p(x_i|\theta, y_i) & \text{for } n > 1,
\end{cases}
\label{eq:adaptive-general}
\end{equation}
which gives the state of knowledge after the first $n-1$ measurements, the maximisation of Eq.~\eqref{eq:precision_gain} for the $n$-th shot provides a practical means to design a fully adaptive protocol \cite{mehboudi2021fundamental, Glatthard2022, smith2024adaptive, overton2024five}. 

All the above enables a multi-shot protocol where the POVM is adaptively optimised via Eq.~\eqref{eq:lyapunov}, the control parameter is optimised using Eq.~\eqref{eq:precision_gain}, and the measurement data is optimally processed by Eqs.~\eqref{eq:opt-est-f} and \eqref{eq:opt-err-f}. 
This set of equations provides the most advanced operational formulation of symmetry-informed estimation to date, and they are fully determined except for the choice of symmetry function, $f$. It is important to note that Eq.~\eqref{eq:precision_gain} may be used regardless of the particular measurement giving rise to the evidence, i.e., it is not limited to the optimal measurement $\mathcal{P}_y(s)$, but may be applied to any given setup. Next, we examine different approaches to choose the function $ f $, before presenting some illustrative examples in Sec.~\ref{sec:examples}.

\section{Calculation of symmetry functions}\label{sec:symmetry-section}

We have seen that the symmetry function $f$ is directly related to the notion of ignorance prior for location-isomorphic parameters, as per Eq.~\eqref{eq:prior-isomorphic}. 
By integrating this equation, we can write down the explicit formula
\begin{equation}
    f(\theta) \propto \int^\theta dz\,p_{\mathrm{MI}}(z).
    \label{eq:symmetry-equation}
\end{equation}
Therefore, calculating a symmetry function amounts to calculating the prior that best represents maximum ignorance about the unknown parameter, $p_{\mathrm{MI}}(z)$, for the class of estimation problems under consideration.  

Several approaches for constructing ignorance priors exist \cite{kass1996the, jaynes2003probability, linden2014bayesian}.  
Here, we focus on two broad categories.
The first one consists in imposing invariance of the prior under a specific parameter transformation that is either physically or statistically meaningful.
The second one exploits the geometric properties of the space of quantum states or likelihood functions.
The remainder of this section expands on each of these approaches and identifies their key assumptions and implications. 

\subsection{Invariance under specific transformations}\label{sec:specific-transformations}

Suppose two hypotheses, $\theta$ and $\theta'$, are related by a family of transformations
\begin{equation}
    \theta' = \tau_{\gamma}(\theta),
    \label{eq:fund_transformation}
\end{equation}
with free constant $\gamma$.
If $\tau_{\gamma}$ is a symmetry of the estimation problem, this induces an equivalence class between hypotheses, so that the ignorance prior we seek, $p_{\mathrm{MI}}(\theta)$, must satisfy Eq.~\eqref{eq:mi-constraint}. 
This leads to the functional equation
\begin{equation}
    p[\tau_{\gamma}(\theta)]\bigg\vert\frac{d \tau_{\gamma}(\theta)}{d\theta}\bigg\vert = p(\theta).
    \label{eq:prior-transformation}
\end{equation}
Consequently, if we can solve Eq.~\eqref{eq:prior-transformation} for given $\tau_\gamma$, we may use the resulting ignorance prior to compute a symmetry function $f$ via Eq.~\eqref{eq:symmetry-equation}.
The question then is how to identify a suitable transformation as per Eq.~\eqref{eq:fund_transformation}.
This requires a case-by-case approach, but typically two main scenarios arise: transformations arising from relationships with external variables such as control parameters or measurands, and transformations dictated by the intrinsic nature of the parameter.

As an example of a transformation dictated by the metrological context, suppose the magnitude of the control parameter $y$ is defined to be `large' or `small' relative to an unknown scale with hypothesis $\theta$, i.e., when $y/\theta \gg 1$ or $y/\theta \ll 1$, respectively \cite{rubio2023quantum}. 
Assuming that the scale of $y$ can be varied, we can rescale these quantities as
\begin{equation}
\begin{cases}
\theta' = \gamma \theta \eqqcolon \tau_{\gamma}(\theta) \\
y' = \gamma y 
\end{cases}
\label{eq:scale-symmetry1}
\end{equation}
without changing the ratio $y/\theta$.
Inserting this $\tau_{\gamma}$ into Eq.~\eqref{eq:prior-transformation} leads to 
\begin{equation}
    \gamma p(\gamma\theta) = p(\theta),
    \label{eq:scale-functional}
\end{equation}
and taking the derivative with respect to $\gamma$ and solving the resulting differential equation returns the well-known Jeffreys's prior\footnote{
Jeffreys derived both this particular prior, which is valid for scale parameters \cite{jaynes1968prior}, and a more general rule based on information geometry \cite{jeffreys1946invariant} (cf. Sec.~\ref{sec:geometry_invariance}).
Since we discuss both in this work, we refer to the latter as \emph{Jeffreys’s general rule} and reserve \emph{Jeffreys’s prior} for the specific case of scale parameters.
}
\begin{equation}
    p_{\mathrm{MI}}(\theta) \propto \frac{1}{\theta}.
\end{equation}
This leads, as per Eq.~\eqref{eq:symmetry-equation}, to the logarithmic symmetry function
\begin{equation}
    f(\theta) = c_1 \log\left(\frac{\theta}{\theta_u}\right),
    \label{eq:scale-symmetry-function}
\end{equation}
with free constants $c_1$ and $\theta_u$, from where the logarithmic loss in Eq.~\eqref{eq:log-loss} emerges upon selecting $c_1 = 1$. 

A second example of contextual transformation is letting $x$ be the measured position of an object and $\theta$ a hypothesis for its expected position.
If the origin of the coordinate system does not affect the measurement, we can shift it as
\begin{equation}
\begin{cases}
\theta' = \theta + \gamma \eqqcolon \tau_{\gamma}(\theta) \\
x' = x + \gamma
\end{cases}
\label{eq:translation-measurand}
\end{equation}
without affecting the value of the difference $x - \theta$.  
Since the hypothesis is translated, the formalism for location parameters in Sec.~\ref{sec:prior-loss} follows.

These examples show that, while the calculation of ignorance priors requires only a transformation of the hypothesis space, some symmetries inherently extend to the sampling space (through $x$) or the space of control parameters, thereby restricting not only the prior but also the states and likelihood functions.\footnote{ 
In view of this, priors are arguably \emph{not} inherently less objective than statistical models \cite{jaynes2003probability, linden2014bayesian}.
}
For instance, imposing Eq.~\eqref{eq:scale-symmetry1} on the matrix elements of the state as
\begin{equation}
    \langle n| \rho_{y'}(\theta')|m\rangle = \langle n|\rho_y(\theta)|m\rangle
\end{equation}
leads to the set of functional equations 
\begin{equation}
    \langle n|\rho_{\gamma y}(\gamma \theta)|m\rangle = \langle n|\rho_y(\theta)|m\rangle,
\end{equation}
which are solved by
\begin{equation}
    \langle n|\rho_y(\theta)|m\rangle = h_{nm}\left(\frac{y}{\theta}\right),
    \label{eq:state-ratio}
\end{equation}
where $h_{nm}$ is a free function for the $(n, m)$-th element.
Gibbs states, e.g., satisfy Eq.~\eqref{eq:state-ratio}, with $y/\theta$ representing an energy-to-temperature ratio.
Similarly, imposing Eq.~\eqref{eq:translation-measurand} on a likelihood model as 
\begin{equation}
    p(x'|\theta')dx' = p(x|\theta)dx
\end{equation}
leads to the functional equation
\begin{equation}
    p(x + \gamma| \theta + \gamma) = p(x|\theta),
\end{equation}
which is solved by the well-known family of shift-invariant models
\begin{equation}
    p(x|\theta) = h(x - \theta),
\end{equation}
with free function $h$.
In light of this observation, we may conversely infer relevant transformation-invariance properties of a parameter from a given state or likelihood model.
Such an approach has been widely explored in the literature \cite{toussaint2011bayesian,linden2014bayesian,kass1996the,eatonsudderth2004,harney_bayesian_2003} when only the measurand $x$ is involved and is best described from a group-theoretical perspective. 
This is however beyond the scope of this work. 

Contextual transformations present some limitations. 
Symmetries relative to the control parameter $y$ allow for all measurement types but require states that depend on the same type of $y$. 
If the measurand $x$ is also involved, only measurements compatible with the same type of $x$ can be performed. 
Nevertheless, there are scenarios where the intrinsic properties of the unknown parameter suffice to identify a relevant transformation independently of any metrological context, thus bypassing these limitations. 
We conclude this section by providing an example illustrating this second scenario.

Let $\{\Theta, 1 - \Theta\}$ denote the distribution of weights over the two-element set $\{e_0, e_1\}$. 
These could represent, e.g., the outcomes of a coin flip, which constitute a paradigmatic example in estimation theory \cite{jaynes2003probability, linden2014bayesian}, or quantum states.
Let $\theta \in (0, 1)$ be a hypothesis for $\Theta$ \cite{jaynes1968prior, rubio2024first}.
A relevant $\tau_\gamma$ can be identified by observing that the weight parameter $\Theta$ quantifies how likely, say, $e_0$ is. This corresponds to a probability assignment $p(e_0) = \theta$.
But we could equally consider a second hypothesis value $\theta'$ leading to a new assignment $p(e_0|I) = \theta'$.
Here, we introduce a proposition $I$ as a formal means to distinguish between the two probabilities, so that it provides no additional information about the true $\Theta$.
These probabilities are related via Bayes's theorem as 
\begin{equation}
    p(e_0|I) = \frac{p(e_0) p(I|e_0)}{p(e_0) p(I|e_0) + [1 - p(e_0)] p(I|e_1)},
\end{equation}
and this yields the M\"{o}bius transformations
\begin{equation}
    \theta' = \frac{\gamma \theta}{1 - \theta + \gamma \theta} \eqqcolon \tau_{\gamma}(\theta),
    \label{eq:mobius}
\end{equation}
with $\gamma = p(I|e_0)/p(I|e_1)$. 
These symmetry transformations can be naturally interpreted as a rescaling of the `odds' when written as 
\begin{equation}
    \frac{\theta'}{1-\theta'} = \gamma \frac{\theta}{1-\theta}.
\end{equation}

We obtain the corresponding ignorance prior by inserting Eq.~\eqref{eq:mobius} into Eq.~\eqref{eq:prior-transformation}, which renders the functional equation
\begin{equation}
    (1 - \theta + \gamma \theta)^2 p(\theta) = \gamma p\left( \frac{\gamma \theta}{1 - \theta + \gamma \theta} \right).
\end{equation}
By taking the derivative with respect to $\gamma$ and solving the resulting differential equation, we find
\begin{equation}\label{eq:weight-prior}
    p_{\mathrm{MI}}(\theta) \propto \frac{1}{\theta (1-\theta)}.
\end{equation}
In turn, this leads to the hyperbolic symmetry function
\begin{equation}
    f(\theta) = c_1 \arctanh{(2\theta - 1)} + c_2
    \label{eq:hyperbolic_symmetry_function}
\end{equation}
upon using Eq.~\eqref{eq:symmetry-equation},
with free constants $c_1$ and $c_2$.
The hyperbolic loss in Eq.~\eqref{eq:hyperbolic-loss} emerges when selecting $c_1 = 2$ and arbitrary $c_2$. 

The symmetry in Eq.~\eqref{eq:mobius} was derived \citet{jaynes1968prior} for the estimation of the probability of success in a binomial distribution, with the more general argument above provided in Ref.~\cite{rubio2024first}. 
It will be applied to the estimation of quantum coherence in Sec.~\ref{sec:coherence}.
This metrological framework directly follows from the nature of the parameter \( \Theta \)---a weight---and provides a practical illustration of a rationale for calculating symmetry functions making as few assumptions as possible.

\subsection{Information geometry}\label{sec:geometry_invariance}

A transformation-informed ignorance prior is a reasonable starting point in situations where the measurement system and strategy, mathematically represented by a likelihood function, are not yet known or may be updated adaptively as more data are taken. This also includes hypothesis tests and estimation tasks combining the data of many different past and future experiments. In many other practical applications, details about the measurement are well known and under control. In quantum estimation tasks, for example, a state $\rho_y(\theta)$ or even a likelihood function $p(x|\theta, y)$ are typically fixed. 
Hence, significantly more information about the experimental platform is incorporated  \emph{a priori} than in the transformation-informed approaches discussed in Sec.~\ref{sec:specific-transformations}, which rely solely on the structural properties of metrological variables.  
We now discuss how prior ignorance may be characterised in this more informed scenario and how this ultimately impacts the resulting symmetry functions.  

\subsubsection{Maximum ignorance from the space of likelihood functions}\label{sec:geometry_likelihood}

Consider the space $S_\theta$ of likelihood functions.  
This space can be treated as a differential statistical manifold, where each point corresponds to a likelihood $p(x|\theta)$, viewed as a function of $\theta$.  
For now, we omit the explicit dependence on the control parameter $y$. The Riemannian metric tensor $ g $, which is symmetric and positive definite, can be used to define distance (and angles) on the manifold \cite{calin2014geometric}.
In particular, the (squared) line element between the likelihoods $p(x|\theta)$ and $p(x|\theta + d\theta)$ can be expressed as
\begin{equation}
    dl^2 = g(\theta)\,d\theta^2.
\end{equation}
The exploitation of the geometrical properties of such manifold for inference is referred-to as \emph{information geometry} \cite{jermyn2005invariant, snoussi2007, jorgensen2021bayesian, Nielsen2022}.

A key property of a Riemannian metric is that any change of parametrisation leaves the distance between two points unchanged. Following \citet{rao_information_1945} and  \citet{jeffreys1946invariant}, a natural choice of Riemannian metric is the Fisher information, given by\footnote{
One way of deriving the Fisher information metric is employing the Kullback–Leibler divergence to calculate the statistical distance between two infinitesimally close likelihood functions \cite{calin2014geometric,jeffreys1946invariant}.
By Taylor expanding, it can be shown that $\mathcal{D}_{\mathrm{KL}}(p_\theta||p_{\theta+d\theta}) \approx \frac{1}{2} I(\theta) d \theta^2$, which may be intuitively interpreted as an `informational difference'.
Note that the same result can be found without the extra factor of $\frac{1}{2}$ by Taylor expanding the symmetric difference $\mathcal{D}_{\mathrm{KL}} (p_{\theta} || p_{\theta + d\theta}) + \mathcal{D}_{\mathrm{KL}}(p_{\theta + d\theta} || p_{\theta})$, as was originally done by \citet{jeffreys1946invariant}.
}
\begin{equation}
    \label{eq:fisher_info_metric}
    g(\theta) = \int \frac{dx}{p(x|\theta)} \left[\frac{\partial p(x|\theta)}{\partial \theta}\right]^2 \eqqcolon I(\theta).
\end{equation}
It can be shown that Eq.~\eqref{eq:fisher_info_metric} is indeed invariant under reparametrisations of the sampling space \cite{calin2014geometric,amari2007}. 
More importantly, the Fisher information metric transforms covariantly, i.e., given an invertible function between coordinates $\eta=\eta(\theta)$, it holds that $g(\theta) d \theta=g(\eta) d \eta$. 
This means that the geometry defined by the metric does not depend on the specific way in which the parameter of the likelihood function is expressed.

The metric induces a distance on the space $S_\theta$ via the geodesic length. The information-geometric approach equates the distance on the parameter space to the distance between the corresponding parametrised hypotheses. One may thus demand that the cumulative probability from an ignorance prior on an interval be proportional to the geodesic length of the corresponding interval in $S_\theta$. That is,
\begin{equation}
\int_{\theta_1}^{\theta_2} p_\mathrm{MI}(\theta) d\theta \propto
\int_{\theta_1}^{\theta_2} \sqrt{I(\theta)}d\theta,
\end{equation}
for all $\theta_1,\theta_2$. From this, it immediately follows that
\begin{equation}
    p_\mathrm{MI}(\theta) \propto \sqrt{I(\theta)},
    \label{eq:jeffreys_prior}
\end{equation}
which is the well-known \emph{Jeffreys's general rule}\footnote{
An alternative way of deducing Eq.~\eqref{eq:jeffreys_prior} is as follows. 
Suppose we follow a path $p[x|\theta(l)]$ within the space $S_\theta$, where $l$ parametrises the curve. 
The \emph{a priori} information about $\Theta$ must be consistent with how the values of $l$ are chosen, i.e., $p(\theta)d\theta = p(l) dl$.
But $l$ may be viewed as a location parameter---it quantifies displacement along the path---for which maximum ignorance is represented as $p(l) \propto 1$. 
Consequently, $p(\theta)d\theta \propto dl = \sqrt{I(\theta)}d\theta$, which leads to Jeffreys's general rule. 
} \cite{jeffreys1946invariant}.

While this is a purely geometric construction, a closer inspection reveals a meaningful metrological interpretation.  
First, given the invariance of the metric under reparametrisations of the hypothesis space---the property that typically motivates the use of Jeffreys's general rule \cite{jeffreys1946invariant, jermyn2005invariant, snoussi2007, eatonsudderth2010, jorgensen2021bayesian, Nielsen2022}---we may interpret Eq.~\eqref{eq:jeffreys_prior} as encoding a principle of indifference to reparametrisations \cite{jaynes2003probability, rubio2023quantum} and thus, as a genuine ignorance prior. Secondly, since the Fisher information quantifies sensitivity in metrology, using Eq.~\eqref{eq:jeffreys_prior} presupposes that the unknown parameter likely lies within the sensitivity region of the sensor. This implication is arguably too strong, in general, for information geometry to serve as the \textit{default} approach to global quantum sensing. It can, however, be very useful for speeding up estimator convergence (see Sec.~\ref{sec:examples}). 

Another issue of the information-geometric approach is that the adaptive single-shot optimisation of the POVM and control parameter $y$ cannot be carried out in a consistent manner. Indeed, note that Eq.~\eqref{eq:jeffreys_prior} depends on the state $\rho_y(\theta)$ and POVM $M_y(x)$ implicitly, and may depend on the control parameter $y$ explicitly once we replace $p(x|\theta) \mapsto p(x|\theta, y)$ in the previous formulas. 
This modifies Eq.~\eqref{eq:jeffreys_prior} as
\begin{equation}
    p_{\mathrm{MI}}(\theta|y) \propto \sqrt{I_y(\theta)}.
    \label{eq:jeffreys_prior_control}
\end{equation}
Upon use of Eq.~\eqref{eq:symmetry-equation}, these dependencies are carried over further to the symmetry function for information geometry, 
\begin{equation}
    f_y(\theta) \propto \int^\theta dz \sqrt{I_y(z)},
    \label{eq:info-geo-f}
\end{equation}
and to the quadratic loss function in Eq.~\eqref{eq:quadratic}. 
Now, Eq.~\eqref{eq:info-geo-f} determines the optimiser required for the estimation strategies discussed in Secs.~\ref{sec:experiments-formalism} and \ref{sec:1shot-formulas}, and it should remain fixed throughout the experiment.  
Allowing $f$ to change between shots would make the definition of optimality itself vary with each run, which is conceptually inconsistent.
Consequently, the state, the applied POVM and the control parameter would all need to be fixed, precluding their adaptive updating. Thus, only the optimal multi-shot estimator in Eq.~\eqref{eq:opt-est-f} and its error in Eq.~\eqref{eq:opt-err-f} can be implemented when working in the information-geometric framework. This is why we conclude that information geometry is less widely applicable to global quantum sensing than relying on specific transformations, as advocated in Sec.~\ref{sec:specific-transformations}.

\subsubsection{Maximum ignorance from the space of quantum states}

Rather than working with likelihood functions $p(x\vert \theta)$, which implicitly refer to a fixed measurement procedure, one may look at the space of quantum states, where each point corresponds to a $\rho(\theta)$. There, the Bures metric is the natural choice \cite{Braunstein1994} to measure distance.
Furthermore, this metric is the only one \cite{lu2022, sommers2003} that reduces to the Fisher information metric in the classical case \cite{petz1996} and to the Fubini--Study metric for pure states \cite{Braunstein1994}. 
For this reason, it is widely used in quantum state estimation \cite{osipov2010}. 

The Bures metric can be expressed in terms of the 
\emph{quantum} Fisher information \cite{paris2009}, provided that the rank of the density does not change with respect to $\theta$ \cite{safranek2017}.  
Omitting a factor of four, the metric takes the form
\begin{equation}
    g(\theta) = \mathrm{Tr}[\rho(\theta)L(\theta)^2] \eqqcolon \mathcal{I}(\theta),
\end{equation}
where $L(\theta)$ is the symmetric logarithmic derivative \cite{helstrom1976quantum, holevo2011probabilistic} defined as the solution to
\begin{equation}
\frac{1}{2}\left[L(\theta)\rho(\theta)+\rho(\theta)L(\theta)\right] = \frac{d\rho(\theta)}{d\theta}.
\end{equation}
The quantum Fisher information $\mathcal{I}(\theta)$ can also be obtained directly by maximising the Fisher information in Eq.~\eqref{eq:fisher_info_metric} over all POVMs \cite{Braunstein1994}.

Following the reasoning in Sec.~\ref{sec:geometry_likelihood}, \emph{mutatis} \emph{mutandis}, and replacing $\rho(\theta) \mapsto \rho_y(\theta)$ in this new derivation, we obtain the ignorance prior
\begin{equation}
    p_{\mathrm{MI}}(\theta|y) \propto \sqrt{\mathcal{I}_{y}(\theta)},
\end{equation}
which generalises Jeffreys's general rule to the quantum regime \cite{slater1995,jorgensen2021bayesian}, and the associated symmetry function
\begin{equation}
    f_y(\theta) \propto \int^\theta dz \sqrt{\mathcal{I}_{ y}(z)},
    \label{eq:symmetry_function_geometry}
\end{equation}
which continues to depend on the state $\rho_y(\theta)$ implicitly, but no longer on a POVM. 
As such, one can inform the optimal estimator in Eq.~\eqref{eq:opt-est-f} by Eq.~\eqref{eq:symmetry_function_geometry} to post-process outcomes, as demonstrated by \citet{jorgensen2021bayesian} in quantum thermometry, as well as to solve the Bayesian Lyapunov equation in Eq.~\eqref{eq:lyapunov} that identifies the associated optimal POVM in Eq.~\eqref{eq:opt-povm} for each shot. 
The latter possibility remains to be fully explored, but will be illustrated in the next section.

\section{Case studies}\label{sec:examples}

In Sec.~\ref{sec:symmetry-section} we have discussed important differences between the transformation-informed and geometry-informed approaches for constructing optimal estimation frameworks. 
Through a series of case studies, in this section we demonstrate that, in practice, these two approaches can yield comparable performance when both can be reliably applied.
Conversely, we identify scenarios where only the transformation-informed method is feasible or where it enables the simplest estimation protocol.

\bigskip

\subsection{Estimating rates with an exponential distribution}

We first consider a classic example where the estimators and errors can be derived analytically, in order to demonstrate the logic of both frameworks.
Let a system decay at a unknown rate $\Theta$.
Suppose we measure the time $t$ it takes for its state to change. 
Here, $1/\Theta$ defines the time scale against which the duration $t$ is measured, through the product $\Theta\,t$.
Given a hypothesis $\theta$ for $\Theta$, this product is invariant under transformations 
\begin{equation}
\begin{cases}
\theta' = \gamma \theta \\
t' = t/\gamma. 
\end{cases}
\label{eq:scale-time}
\end{equation}
Hence, maximum ignorance is represented by Jeffreys's prior, $p_{\mathrm{MI}}^{\mathrm{T}}(\theta) \propto 1/\theta$, where $\mathrm{T}$ denotes the transformation-informed prescription, and the symmetry function can thus be chosen as $f_{\mathrm{T}}(\theta) = \log(\theta/\theta_u)$, where $\theta_u$ neutralises the dimension of time in the logarithmic function (cf. Sec.~\ref{sec:specific-transformations}). 

If the measurement is repeated $\mu$ times one generates the vector of outcomes $\boldsymbol{t} = (t_1, \dots, t_{\mu})$; assuming there is no control parameter, the optimal estimator in Eq.~\eqref{eq:opt-est-f} and its error in Eq.~\eqref{eq:opt-err-f} take the form
\begin{subequations}
\label{eq:optimal_strategy_rates}
\begin{align} \label{eq:optimal-est-rates}
    \tilde{\vartheta}(\boldsymbol{t}) 
    &= \theta_u \exp\left[ \int d\theta\,p(\theta|\boldsymbol{t}) \log\left( \frac{\theta}{\theta_u} \right) \right], \\
    \Delta \tilde{\vartheta}(\boldsymbol{t}) 
    &= \tilde{\vartheta}(\boldsymbol{t}) 
    \left\lbrace- \log^2\left[\frac{\tilde{\vartheta}(\boldsymbol{t})}{\theta_u}\right] + \int d\theta\,p(\theta|\boldsymbol{t})\log^2\left(\frac{\theta}{\theta_u}\right) 
    \right\rbrace^{\frac{1}{2}},
\end{align}
\end{subequations}
where
\begin{equation}
    p(\theta|\boldsymbol{t}) \propto p(\theta) \prod_{k=1}^{\mu} p(t_k|\theta).
    \label{eq:posterior-time}
\end{equation}

Some comments are in order. 
First, Eqs.~\eqref{eq:optimal_strategy_rates} follow from $\Theta$ being a rate, irrespective of the likelihood function $p(t|\theta)$ used to describe the measurement. 
Therefore, a specific statistical model is not always required to formulate symmetry-informed estimation. 
More importantly, the estimator in Eq.~\eqref{eq:optimal-est-rates} is  positive definite by construction, regardless of the initial prior knowledge.
This contrasts with the use of the square loss in Eq.~\eqref{eq:location-losses}, as is commonly done, since the corresponding optimal estimator, the posterior mean, does \emph{not} respect scale invariance and can, in principle, be negative, thus requiring a manual truncation of the prior.
Note that similar considerations apply in global thermometry \cite{rubio2021global, mehboudi2021fundamental, boeyens2021uninformed, rubio2023quantum, chang2024global}, where the equations for the optimal strategy are analogous to Eqs.~\eqref{eq:optimal_strategy_rates}.

We now consider the standard case of a Poisson process. The likelihood of detecting a state change at some time $t$ given a rate $\theta$ is then an exponential distribution, \begin{equation}
    p(t|\theta) = \theta \exp(-\theta t),
    \label{eq:exponential-dist}
\end{equation}
for which Eqs.~\eqref{eq:optimal_strategy_rates} can be analytically calculated.
This model is invariant under transformations in Eq.~\eqref{eq:scale-time} and yields the posterior
\begin{equation}
    p(\theta|\boldsymbol{t}) \propto \theta^{\mu-1}\exp(-\mu \theta \bar{t}),
    \label{eq:unnormalised-posterior}
\end{equation}
where $\bar{t} = \sum_{k=1}^\mu t_k/\mu$.
We assume the most conservative scenario---maximum ignorance---by setting $p(\theta) = p_{\mathrm{MI}}(\theta)$, $\theta \in (0, \infty)$.
While Jeffreys's prior cannot be normalised over this range, the following integrals over the unnormalised posterior in Eq.~\eqref{eq:unnormalised-posterior} converge:
\begin{widetext}
\begin{subequations}
\label{eq:improper_integrals}
\begin{align}
    \int_0^\infty d\theta\,\theta^{\mu-1}\exp(-\mu \theta \bar{t}) 
    &= \frac{\Gamma(\mu)}{\mu^\mu \bar{t}^\mu}, \\
    \int_0^\infty d\theta\,\theta^{\mu-1}\exp(-\mu \theta \bar{t})\log\left(\frac{\theta}{\theta_u}\right)  
    &= \frac{\Gamma(\mu)}{\mu^\mu \bar{t}^\mu} \left[\psi^{(0)}(\mu) - \log(\mu \theta_u \bar{t})\right], \\
    \int_0^\infty d\theta\,\theta^{\mu-1}\exp(-\mu \theta \bar{t}) \log^2\left(\frac{\theta}{\theta_u}\right) 
    &= \frac{\Gamma(\mu)}{\mu^\mu \bar{t}^\mu} 
    \left\lbrace\left[\psi^{(0)}(\mu) - \log(\mu \theta_u \bar{t})\right]^2 + \psi^{(1)}(\mu)\right\rbrace,
\end{align}
\end{subequations}
\end{widetext}
where $\psi^{(n)}(z)$ is the polygamma function of order $n$. 
By combining these integrals with Eqs.~\eqref{eq:optimal_strategy_rates}, we  arrive at the optimal estimator and error
\begin{equation}
    \tilde{\vartheta}(\boldsymbol{t}) \pm \Delta \tilde{\vartheta}(\boldsymbol{t}) = \frac{\mathrm{exp}[\psi^{(0)}(\mu)]}{\mu\,\bar{t}}\left[1 \pm \sqrt{\psi^{(1)}(\mu)}\right].
    \label{eq:rates}
\end{equation}
This provides a complete solution to rate estimation based on waiting-time measurements for any Poisson process without additional control parameters to optimise over.
The solution can be applied, for instance, to the estimation of biochemical rates in single-molecule experiments \cite{subramanian2021sensing, eerqing2021comparing, mpofu2022measuring, mpofu2022experimental}.
In addition, since $\psi^{(0)} (\mu) \approx \log(\mu)$ when $\mu \gg 1$, we asymptotically recover the intuitive result $\tilde{\vartheta}(\boldsymbol{t}) \rightarrow 1/\bar{t}$ as $\mu \rightarrow \infty$, which is the corresponding maximum likelihood estimator \cite{kay1993fundamentals}.

We could alternatively use information geometry to construct a prior. 
The Fisher information for the likelihood in Eq.~\eqref{eq:exponential-dist} is given by 
\begin{align}
    I(\theta) &= \int_0^{\infty} \frac{dt}{p(t|\theta)} \left[\frac{\partial p(t|\theta)}{\partial \theta} \right]^2
    \nonumber \\
    &= \frac{1}{\theta} \int_0^\infty dt\,\exp(-\theta t) (1-\theta t)^2 = \frac{1}{\theta^2}. 
\end{align}
This results in the same prior, $p_{\mathrm{MI}}^{\mathrm{G}}(\theta) \propto \sqrt{I(\theta)} = 1/\theta$, 
and hence in the optimal strategy in Eq.~\eqref{eq:rates}. Here, the superindex $\mathrm{G}$ stands for geometry-based.
We thus see that there are problems for which the transformation- and geometry-informed approaches yield identical results.
Another example of this phenomenon is found in quantum thermometry with thermal states of constant heat capacity, as discussed by \citet{jorgensen2021bayesian}.

\subsection{Estimating coherence under a depolarising channel}\label{sec:coherence}

We next consider a scenario in which the two approaches yield different ignorance priors---and thus distinct symmetry functions---and demonstrate that both still provide correct estimates despite their differing initial assumptions.
Let
\begin{equation}
    \ket{\psi(\theta)} = \sqrt{1-\theta}\ket{0} + \sqrt{\theta}\ket{1}
    \label{eq:qubit}
\end{equation}
be a qubit state with hypothesis $ \theta \in (1-a, a)$ for the unknown coherence parameter $\Theta$, where $ a \in (1/2, 1)$.
Suppose that the state above undergoes a depolarising channel, thus transforming into \cite{bertlmann2023modern}
\begin{equation}
    \rho_{\lambda}(\theta) = (1-\lambda) \ketbra{\psi(\theta)} + \frac{\lambda}{2}\sigma_0,
    \label{eq:coherence-state}
\end{equation}
where $\sigma_0 = \ketbra{0}{0} + \ketbra{1}{1}$ and $\lambda$ quantifies the degree of noise.
Given Eq.~\eqref{eq:coherence-state}, we choose $\lambda = 1/10$, for the sake of example, and get 
\begin{equation}
    \rho(\theta) = \frac{1}{10} \left[ 9\ketbra{\psi(\theta)} + \frac{\sigma_0}{2} \right].
    \label{eq:coherence-state-numerical}
\end{equation}
We shall address two problems; namely, a comparison of single-shot relative errors, which provides insight into the fundamental precision achievable in principle by each approach, and the post-processing of simulated measurement data based on the state in Eq.~\eqref{eq:coherence-state-numerical}.

\subsubsection{Single-shot analysis}

For the first problem, we define the single-shot relative error
\begin{align}
    \varepsilon \coloneqq \frac{ \Delta \tilde{\vartheta}_p^2 - \langle \bar{\mathcal{L}} \rangle_{\mathrm{opt}}}{\Delta \tilde{\vartheta}_p^2}
    = \frac{\Delta \mathcal{S}^2}{\Delta \tilde{\vartheta}_p^2}.
    \label{eq:intrinsic-gain}
\end{align}
Here, $\Delta \tilde{\vartheta}_p^2$ is the optimal information loss prior to performing a measurement, given as
\begin{equation}
    \Delta \tilde{\vartheta}_p^2 = \int d\theta\, p(\theta) f(\theta)^2 - \left[ \int d\theta \,p(\theta) f(\theta) \right]^2,
\end{equation}
while the minimum loss $\langle \bar{\mathcal{L}} \rangle_{\mathrm{opt}}$ is given by Eq.~\eqref{eq:minimum-loss}.
Furthermore, the variance of the operator $\mathcal{S}$ with respect to the average state $\rho_{0}$ (Eq.~\eqref{eq:state-moments}) is
\begin{equation}
   \Delta \mathcal{S}^2 = \mathcal{G} - \mathrm{Tr}(\rho_{0} \mathcal{S})^2,
\end{equation}
where $\mathcal{G}$ is given by Eq.~\eqref{eq:precision_gain} and we have used the identity $\mathrm{Tr}(\rho_{0} \mathcal{S}) = \int d\theta\, p(\theta) f(\theta)$.
Eq.~\eqref{eq:intrinsic-gain} provides an \emph{intrinsic} measure of precision gain, thus enabling a meaningful comparison between different estimation frameworks.\footnote{
In practice, once a specific framework is selected, the \emph{absolute} quantifier $\mathcal{G}$ is a sufficient and simpler choice  \cite{Glatthard2022, overton2024five}.
}

\begin{figure}[b]
    \includegraphics[trim={0cm 0cm 0cm 0cm},clip,width=\linewidth]{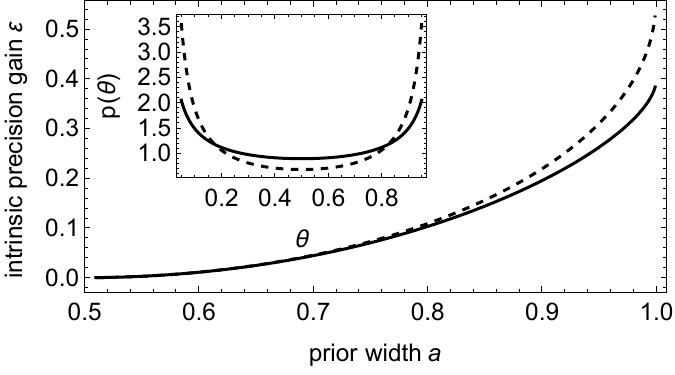}
    \caption{Intrinsic precision gain in Eq.~\eqref{eq:intrinsic-gain} for the estimation of a coherence parameter in a depolarising channel with noise rate $\lambda = 0.1$, using geometry- (solid) and transformation-informed (dashed) frameworks.
    A measurement within the geometry-informed approach provides less information---i.e., it renders a lower $\varepsilon$---because, being more informed initially, it leaves less room for precision improvement.
    The inset shows the prior for both approaches with $a=0.95$}.
\label{fig:coherence-intrinsic}
\end{figure}

We evaluate Eq.~\eqref{eq:intrinsic-gain} numerically for two symmetry functions. 
For the transformation-informed approach, here referred to as weight estimation, $p_{\mathrm{MI}}^{\mathrm{T}}(\theta) \propto 1/[\theta(1-\theta)]$ and hence $f_{\mathrm{T}}(\theta) = 2 \arctanh{(2\theta - 1)}$ because $\Theta$ determines the weight distribution of the states $\ket{0}$ and $\ket{1}$ (cf. Eqs.~\eqref{eq:weight-prior} and \eqref{eq:hyperbolic_symmetry_function}).
For information geometry, since the quantum Fisher information for Eq.~\eqref{eq:coherence-state} evaluates to
\begin{equation}
    \mathcal{I}(\theta) = \frac{81}{100\,\theta(1-\theta)},
\end{equation} maximum ignorance is represented in this framework as 
\begin{equation}
    p_{\mathrm{MI}}^{\mathrm{G}}(\theta) \propto \sqrt{\mathcal{I}_{}(\theta)} \propto \frac{1}{\sqrt{\theta(1-\theta)}}
\end{equation}
and so the corresponding symmetry function is
\begin{equation}
    f_{\mathrm{G}}(\theta) = c_1\,\arctan{\left( \sqrt{\frac{\theta}{1-\theta}} \right)} + c_2,
\end{equation}
as per Eq.~\eqref{eq:symmetry-equation}.
We choose $c_1 = 2$ and $c_2 = 0$. 
To enable the most conservative quantification of the relative precision gain, we select the ignorance prior intrinsic to each approach, i.e., $p(\theta) = p_{\mathrm{MI}}^{\mathrm{T}}(\theta)$ and $p(\theta) = p_{\mathrm{MI}}^{\mathrm{G}}(\theta)$, respectively.

The results are shown in Fig.~\ref{fig:coherence-intrinsic}. Eq.~\eqref{eq:intrinsic-gain} is plotted as a function of the prior width $a$ for weight estimation (dashed) and information geometry (solid).
For both approaches, a wider prior range (larger $a$) corresponds to a larger relative error $\varepsilon$.
This is expected, since a wider range indicates greater prior uncertainty, and therefore a greater potential to improve our knowledge by performing a measurement.
More importantly, weight estimation yields the largest $\varepsilon$ for fixed width $a$, which can be explained by observing the priors in the inset of Fig.~\ref{fig:coherence-intrinsic}; since $p_{\mathrm{MI}}^{\mathrm{T}}(\theta)$ approaches the boundaries of the hypothesis range asymptotically faster than $p_{\mathrm{MI}}^{\mathrm{G}}(\theta)$, the corresponding prior error is larger. 
This suggests that the information-geometric framework can be initially more informed about the unknown parameter than the framework derived from the invariance under a single transformation. Hence, one measurement outcome provides comparatively more information in the transformation-informed scenario.
Naturally, both approaches become equivalent as $a \rightarrow 1/2$, at which point both priors converge to the same Dirac delta.

\subsubsection{Data analysis in global quantum sensing}

We now address the problem of post-processing a sample of simulated data, $\boldsymbol{s} = (s_1, \dots, s_\mu)$, generated by performing the optimal POVM arising from the Lyapunov equation in Eq.~\eqref{eq:lyapunov} on the state in Eq.~\eqref{eq:coherence-state} a total of $\mu$ times.
The numerical solution of Eq.~\eqref{eq:lyapunov} for the state in Eq.~\eqref{eq:coherence-state} reveals that, for both symmetry functions, $f_{\mathrm{T}}(\theta)$ and $f_{\mathrm{G}}(\theta)$, the optimal measurement strategy consists in projecting onto the computational basis, $\lbrace \ketbra{0}, \ketbra{1} \rbrace$.
Using the Born rule, this renders the single-shot likelihood model
\begin{equation}
    p(s|\theta) = \frac{1}{10}\left\lbrace 9 [1-s - (-1)^s\theta] + \frac{1}{2} \right\rbrace,
\end{equation}
with $s = \{0, 1\}$. 
Consequently, the posterior for $\mu$ repetitions of this measurement is given as
\begin{equation}
    p(\theta|\boldsymbol{s}) \propto p(\theta) \prod_{i=1}^\mu\left\lbrace 9 [1-s_i - (-1)^{s_i}\theta] + \frac{1}{2} \right\rbrace,
    \label{eq:posterior-coherence}
\end{equation}
where we continue to use either $p(\theta) = p^\mathrm{T}_{\mathrm{MI}}(\theta)$ or $p(\theta) = p^\mathrm{G}_{\mathrm{MI}}(\theta)$.
Inserting Eq.~\eqref{eq:posterior-coherence} into Eqs.~\eqref{eq:opt-est-f} and \eqref{eq:opt-err-f} for each symmetry function further leads to the optimal estimators and errors 
\begin{widetext}
\begin{subequations}
\label{eq:estimator-weight}
\begin{align}
    \tilde{\vartheta}_{\mathrm{T}}(\boldsymbol{s}) &= \frac{1}{2} \left\lbrace 1 + \mathrm{tanh}\left[ \int d\theta \,p(\theta|\boldsymbol{s})\,\mathrm{artanh}(2\theta - 1) \right] \right\rbrace, \\
    \Delta \tilde{\vartheta}_{\mathrm{T}}(\boldsymbol{s}) &= 
    2 \tilde{\vartheta}_{\mathrm{T}}(\boldsymbol{s})[1 + \tilde{\vartheta}_{\mathrm{T}}(\boldsymbol{s})] \left\lbrace-\mathrm{artanh}^2[2\tilde{\vartheta}_{\mathrm{T}}(\boldsymbol{s}) - 1] + \int d\theta\,p^\mathrm{T}_{\mathrm{MI}}(\theta)\,\mathrm{artanh}^2(2\theta - 1)\right\rbrace^{\frac{1}{2}}
\end{align}
\end{subequations}
and 
\begin{subequations}
\label{eq:estimator-coherence-geometry}
\begin{align}
    \tilde{\vartheta}_{\mathrm{G}}(\boldsymbol{s}) &= \mathrm{tan}^2\left[ \int d\theta \,p(\theta|\boldsymbol{s})\,\arctan{\left(\sqrt{\frac{\theta}{1 - \theta}}\right)} \right] \left\lbrace1 + \mathrm{tan}^2\left[\int d\theta \,p(\theta|\boldsymbol{s})\,\arctan{\left(\sqrt{\frac{\theta}{1 - \theta}}\right)} \right] \right\rbrace^{-1}, \\
    \Delta \tilde{\vartheta}_{\mathrm{G}}(\boldsymbol{s}) &= 
    2\left(\lbrace\tilde{\vartheta}_{\mathrm{G}}(\boldsymbol{s})[1 + \tilde{\vartheta}_{\mathrm{G}}(\boldsymbol{s})] \rbrace \left[
    -\arctan^2{\left[\sqrt{\frac{\tilde{\vartheta}_{\mathrm{G}}(\boldsymbol{s})}{1 - \tilde{\vartheta}_{\mathrm{G}}(\boldsymbol{s})}}\,\right]} + \int d\theta\,p_{\mathrm{MI}}^\mathrm{G}(\theta)\,\arctan^2{\left(\sqrt{\frac{\theta}{1 - \theta}}\right)} \right]\right)^{\frac{1}{2}}. 
\end{align}
\end{subequations}
\end{widetext} 

Instead of focusing on the parameter $\theta$ itself, we consider estimating a more widely used coherence quantifier, based on the $l1$-norm \cite{baumgratz2014quantifying, winter2016operational, streltsov2017colloqium, chitambar2019quantum, wu2021experimental, ares2024unification}, 
\begin{equation}
    \zeta(\theta) = \sum_{i\neq j} |\rho_{ij}(\theta)| = \frac{9}{5}\sqrt{\theta(1-\theta)},
    \label{eq:coherence-quantifier}
\end{equation}
which is a function of the parameter \cite{proctor2018multiparameter, rubio2020networks, gross2021one}.
By inserting the optimal estimator and error in Eqs.~(\ref{eq:estimator-weight}, \ref{eq:estimator-coherence-geometry}) into Eq.~\eqref{eq:coherence-quantifier}, the estimator and propagated error, $\tilde{\zeta}(\boldsymbol{s}) \pm \Delta \tilde{\zeta}(\boldsymbol{s})$, are  
\begin{subequations}
\label{eq:estimation_coherence}
\begin{align}
    \tilde{\zeta}(\boldsymbol{s}) &= \frac{9}{5} \sqrt{\tilde{\vartheta} (\boldsymbol{s})[1-\tilde{\vartheta} (\boldsymbol{s})]}, \\
    \Delta \tilde{\zeta}(\boldsymbol{s}) &= \frac{9 |1-2\tilde{\vartheta} (\boldsymbol{s})|}{10 \sqrt{\tilde{\vartheta} (\boldsymbol{s})[1-\tilde{\vartheta} (\boldsymbol{s})]}} \Delta \tilde{\vartheta} (\boldsymbol{s}).
\end{align}
\end{subequations}
We then repeat this procedure $m$ times and employ the noise-to-signal ratio (NSR) $ \mathrm{var}(\tilde{\zeta})/\langle \tilde{\zeta} \rangle^2$, where
\begin{equation}
    \mathrm{var}(\tilde{\zeta}) = \langle \tilde{\zeta}^2 \rangle - \langle \tilde{\zeta} \rangle ^2,\hspace{0.4em}\text{with}\hspace{0.4em}
    \langle \tilde{\zeta} \rangle = \frac{1}{m}\sum_{i=1}^m \tilde{\zeta}_{i},
\end{equation}
as an empirical metric of estimator variability. 
This NSR is agnostic to the framework employed  \cite{Glatthard2022, overton2024five} and hence enables a meaningful comparison between the two. 

\begin{figure}[t]
    \includegraphics[trim={-0.5cm 0cm 0cm -0.4cm},clip,width=\linewidth]{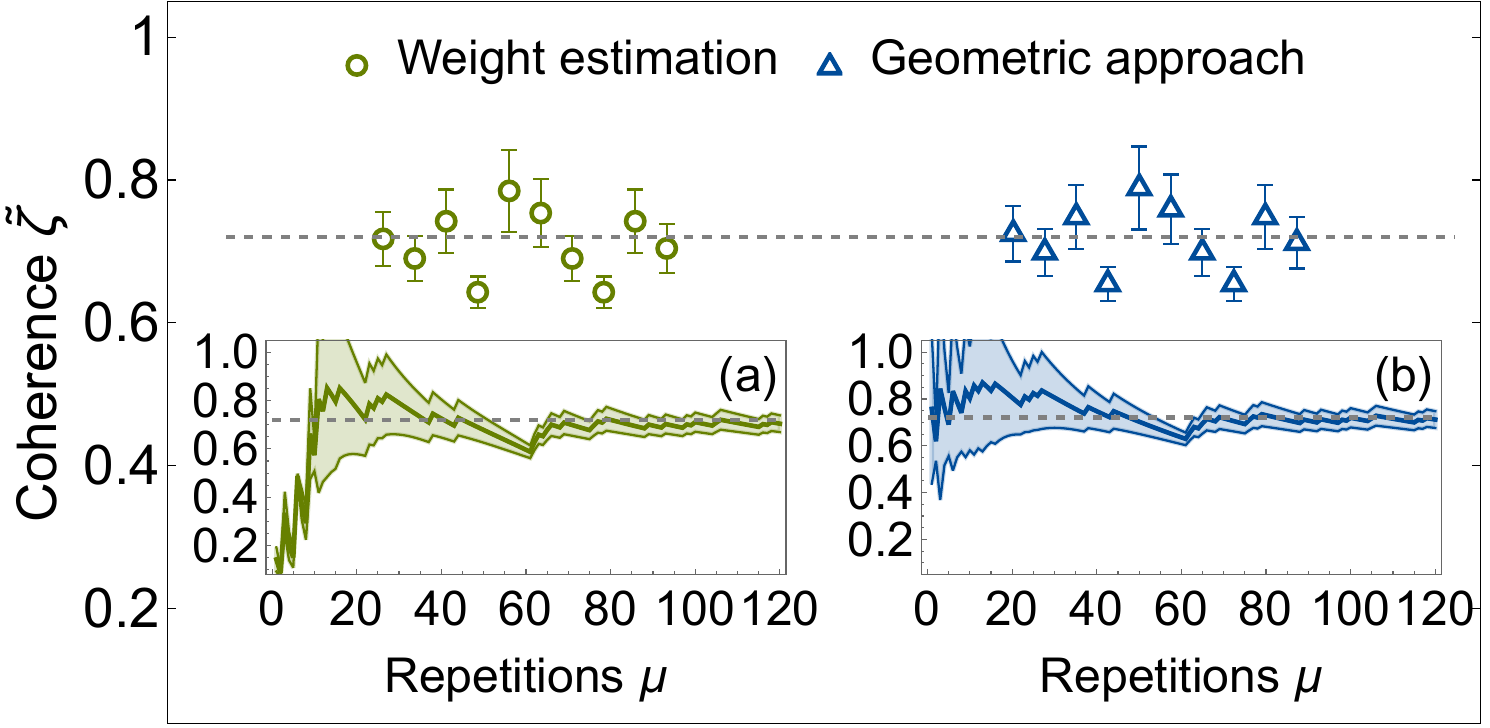}
    \includegraphics[trim={-0.5cm 0cm 0cm -0.3cm},clip,width=\linewidth]{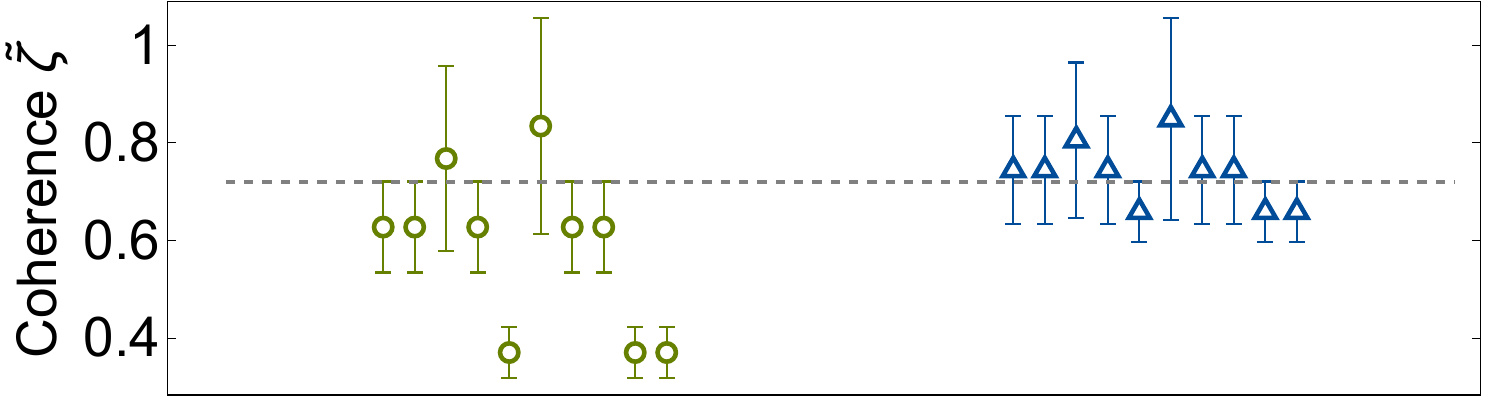}
    \caption{Variability of coherence estimates simulated in a depolarising channel with noise rate $\lambda=1/10$ using the transformation- (green circles) and geometry-based (blue triangles) approaches for $\mu=120$ (top panel) and $\mu = 20$ (bottom panel) repetitions of the single-shot optimal measurement rendered by Eq.~\eqref{eq:lyapunov}. 
    The prior width in all cases is $a=1-10^{-5}$. 
    The results obtained from $m = 10$ different realisations are shown, on which both estimation strategies are applied. 
    The underlying true coherence is set to $\zeta = 0.72$ (dashed grey). 
    Insets (a), (b) show the typical convergence of estimates with the number of measurements. 
    We see that the different strategies can give different results for a small dataset, but become similar for a large number of data.}
\label{fig:coherence-scatter}
\end{figure} 

We simulated $\mu = 120$ shots with true coherence $\zeta = 0.72$ and prior width $a=1-10^{-5}$, and we repeated this procedure $m = 10$ times.
The results are shown in Fig.~\ref{fig:coherence-scatter}, top panel. 
As can be seen, both approaches are practically equivalent even when based on different principles. 
This is further confirmed by comparing their NSRs, which evaluate to $0.38$ and $0.34$ for the transformation- and geometry-based frameworks, respectively. 

However, in the regime of limited data \cite{Teklu_2009, lumino2018experimental, rubio2018quantum, Bavaresco2024, meyer2023quantum, Cimini_2024}---low $\mu$---the differences become apparent. 
The insets in Fig.~\ref{fig:coherence-scatter} illustrate the progression of the estimate $\tilde{\zeta}$ in a single simulation as the posterior is updated with new data.  
While, from $\mu \sim 60$, both sequences display a similar convergence towards the true value,  weight estimation (inset (a)) yields generally worse estimates than the geometric approach (inset (b)) for very low $\mu$.
This is another consequence of using different ignorance priors.
To quantitatively assess these differences, we performed $m = 10$ simulations of $\mu = 20$ repetitions each. 
The result is shown in Fig.~\ref{fig:coherence-scatter}, bottom panel, with corresponding NSRs $0.66$ (geometric approach) and $7.08$ (transformation-based approach). 
As in the single-shot analysis above, this $11$-fold difference may be interpreted as a consequence of information geometry being based on the state-informed prior $p_{\mathrm{MI}}^\mathrm{G}(\theta)$, which enforces a stronger constraint and can, as discussed, be considered more informative than its transformation-based counterpart, $p_{\mathrm{MI}}^\mathrm{T}(\theta)$. 

By combining this result with the previous single-shot analysis, we conclude that, if the prior $p_{\mathrm{MI}}^\mathrm{G}(\theta)$ can be reasonably justified in a given experimental setup, the related expressions are computationally efficient, and the state space is fixed, information geometry may speed up estimator convergence in practice. 

\subsection{Lifetime of atomic states} \label{sec:lifetime}

Our final example is a two-level atom, initially prepared in the superposition $\ket{\psi(\eta)}=\sqrt{1-\eta}\ket{g}+\sqrt{\eta}\ket{e}$, that decays to the ground state $\ket{g}$, owing to spontaneous photon emission. The task here is to estimate the inverse spontaneous emission rate, i.e., the lifetime $\Theta$, by probing the atom after some time $t$. 
Assuming that the state evolves according to the standard quantum-optical master equation, the corresponding density matrix is \cite{2009barnett}
\begin{align}
    \rho_t(\theta) =&\,[1-\eta\,\exp(-t/\theta)] \ketbra{g}{g} + \eta\,\exp(-t/\theta)\ketbra{e}{e} \nonumber \\
    &+\sqrt{\eta(1-\eta)\,\exp(-t/\theta)}(\ketbra{e}{g}+\ketbra{g}{e}),
    \label{eq:state-atom}
\end{align}
where $t$ is the elapsed time and $\theta \in (1/b, b)$ is a hypothesis for the unknown lifetime of the excited state, with prior width $b \in (1, \infty)$.
Since the state is a function of the ratio $\theta/t$, in this section we use the notation $ \rho_t(\theta) \coloneqq \rho(\theta/t)$

Beyond providing additional evidence for our conclusions in Sec.~\ref{sec:coherence}, the motivation for considering this example is twofold.  
First, it extends the analysis in Ref.~\cite{rubio2023quantum}, where this problem was previously addressed through scale estimation, from the perspective of information geometry.
Second, in contrast to both Sec.~\ref{sec:coherence} and Ref.~\cite{rubio2023quantum}, here we include state optimisation.

We start by finding the relevant symmetry functions. 
Given that lifetimes are scale parameters, it holds that $p_{\mathrm{MI}}^{\mathrm{T}}(\theta) \propto 1/\theta$ and $f_{\mathrm{T}}(\theta) = \log(\theta/\theta_u)$, where we recall that $\theta_u$ is an arbitrary positive constant. 
A convenient choice is $\theta_u = t$, which can be made without loss of generality because the framework of scale estimation is independent of the value of $\theta_u$ \cite{rubio2023quantum}.
On the other hand, to apply information geometry, we calculate the quantum Fisher information for Eq.~\eqref{eq:state-atom}, which leads to
\begin{equation}
    \mathcal{I}_{\eta, t}\left(\theta\right) = \frac{\eta\,t^2 \exp(-t/\theta)[1+(\eta - 1)\exp(-t/\theta)]}{\theta^4 [1-\exp(-t/\theta)]}.
    \label{eq:fisher-atom}
\end{equation}

At this point we encounter a difficulty. 
The square root of Eq.~\eqref{eq:fisher-atom} cannot be analytically integrated over $\theta$ for an \textit{arbitrary} value of $\eta$, meaning that the corresponding symmetry function also cannot be defined analytically.
Hence, it is not generally possible to write down the expressions for information loss, optimal estimators and error bars.
While a fully numerical approach is feasible, it can quickly become computationally cumbersome, especially if the parameter range is broad. This may be detrimental to the experimental implementation of adaptive protocols, where optimisation must be carried out in real time, as the measurement data are collected \cite{mehboudi2021fundamental, Glatthard2022, overton2024five}. 
In contrast, the transformation-informed approach is straightforward in this case.

Luckily, as we show below, $\eta = 1$---the atom is initialised in its excited state---is the optimal preparation. 
Selecting this value greatly simplifies Eq.~\eqref{eq:fisher-atom}, yielding
\begin{equation}
    \mathcal{I}_{t}(\theta) = \left\lbrace \frac{\theta^4}{t^2} \left[\exp\left(\frac{t}{\theta}\right)-1\right] \right\rbrace^{-1}
    \label{eq:fisher-atom-simpler}.
\end{equation}
The corresponding ignorance prior is then
\begin{equation}
    p_{\mathrm{MI}}^{\mathrm{G}}(\theta|t)  \propto \sqrt{\mathcal{I}_{t}(\theta)} = \frac{t}{\theta^2}\left[ \exp\left(\frac{t}{\theta}\right)-1 \right]^{-\frac{1}{2}},
\end{equation}
and so, by virtue of Eq.~\eqref{eq:symmetry-equation}, the geometry-informed symmetry function is given by
\begin{equation}
    f_{\mathrm{G}}\left(\frac{\theta}{t}\right) = c_1 \arctan{\left[\sqrt{\exp\left(\frac{t}{\theta}\right) - 1}\, \right]} + c_2.
\end{equation}
We choose $c_1 = 2$, $c_2 = 0$, and $p(\theta) = p_{\mathrm{MI}}^{\mathrm{T}}(\theta)$ or $p(\theta|t) = p_{\mathrm{MI}}^{\mathrm{G}}(\theta|t)$ depending on the framework under analysis. 

\begin{figure}[b]
    \includegraphics[trim={0cm 0cm 0cm 0cm},clip,width=\linewidth]{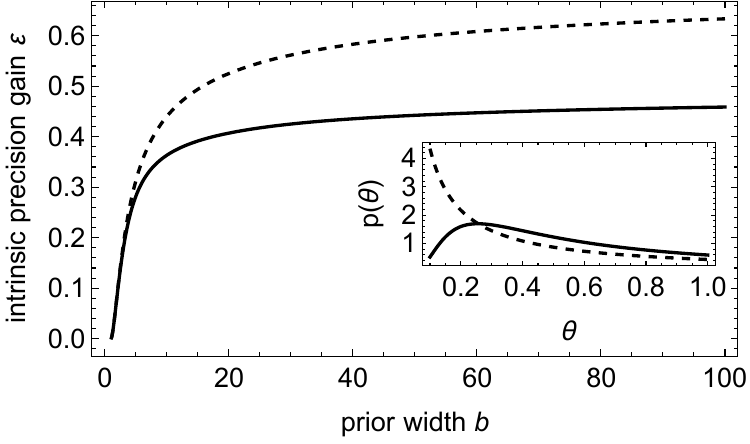}
    \caption{Intrinsic precision gain in Eq.~\eqref{eq:intrinsic-gain} in the estimation of the lifetime of an atomic state, shown for scale estimation (dashed) and information geometry (solid), as a function of the prior width $b$. 
    The inset shows the ignorance prior of each approach (here, we have set $t = 1\,$s for representation purposes).    
    We see that the prior of scale estimation is concentrated at low values of $\theta$, growing unbounded as $\theta$ is reduced, while the prior of information geometry is concentrated around an intermediate value.  
    This results, for wide prior ranges, in a more uninformed state of knowledge for scale estimation and, consequently, a higher precision gain due to the increased potential for information update.
    For small prior ranges, the intrinsic precision gain becomes the same for both approaches, reflecting that the prior densities approach a Dirac delta centred at $b=1$.
    }
\label{fig:lifetime-intrinsic}
\end{figure}

We compare these two frameworks following the procedures in Sec.~\ref{sec:coherence}.
First, the calculation of the intrinsic precision gain in Eq.~\eqref{eq:intrinsic-gain}, $\varepsilon$, as a function of the prior width, $b$, renders the numerical results shown in Fig.~\ref{fig:lifetime-intrinsic}.
The dashed line corresponds to scale estimation and the solid line to information geometry.  
As it can be seen, scale estimation generally produces in the largest relative error.
This is consistent with the unbounded growth of $ p_{\mathrm{MI}}^\mathrm{T}(\theta) $ as $ \theta \to 0 $, in contrast to the finite values of $ p_{\mathrm{MI}}^\mathrm{G}(\theta|t) $ in the same region, and further stresses the intuition drawn from the example in Sec.~\ref{sec:coherence} that the geometric approach is intrinsically more informed \textit{a priori}.
For $b \rightarrow 1$, both priors converge again to the same Dirac delta, thereby making the relative errors of both approaches identical.

As in the previous section, we shall also simulate $\mu$ shots of the optimal POVM determined by Lyapunov equation Eq.~\eqref{eq:lyapunov} and post-process the resulting outcomes, $\boldsymbol{s} = (s_1, \dots, s_\mu)$, into an optimal lifetime estimate. 
For $\eta = 1$, the state in Eq.~\eqref{eq:state-atom} becomes diagonal, so that the optimal POVM trivially consists in projecting onto the energy basis, $\lbrace \ketbra{g}, \ketbra{e} \rbrace$. 
The corresponding likelihood function is
\begin{equation}
    p(s|\theta,t) = \left[(1-s)+(-1)^{1-s} \exp\left(-\frac{t}{\theta}\right)\right],
\end{equation}
with $s = 0$ for the ground state and $s = 1$ for the excited state.
Repeating this measurement $\mu$ times, each after a time $t$, results in the multi-shot posterior $p(\theta | \boldsymbol{s}, t)$, defined as in Eq.~\eqref{eq:posterior-coherence}.
By inserting this posterior into Eqs.~\eqref{eq:opt-est-f} and \eqref{eq:opt-err-f} for each symmetry function, we find the optimal estimators and errors 
\begin{widetext}
\begin{subequations}
\label{eq:estimator-scale}
\begin{align}
    \frac{\tilde{\vartheta}_{\mathrm{T}}(\boldsymbol{s})}{t} &= \exp\left[\int d\theta\,p(\theta|\boldsymbol{s}) \log\left(\frac{\theta}{t}\right)\right],\\
    \frac{\Delta \tilde{\vartheta}_{\mathrm{T}}(\boldsymbol{s})}{\tilde{\vartheta}_{\mathrm{T}}(\boldsymbol{s})} &= \left\lbrace
    -\log^2\left[ \frac{\tilde{\vartheta}_{\mathrm{T}}(\boldsymbol{s})}{t} \right] + \int d\theta\,p^\mathrm{T}_{\mathrm{MI}}(\theta) \log^2\left( \frac{\theta}{t} \right)\right\rbrace^{\frac{1}{2}}
\end{align}
\end{subequations}
and
\begin{subequations}
\label{eq:estimator-atom-geometry}
\begin{align}
\begin{split}
    \frac{\tilde{\vartheta}_{\mathrm{G}}(\boldsymbol{s})}{t} =\,& \log^{-1} \left( 1 + \tan^2\left\lbrace\int d\theta\,p(\theta|\boldsymbol{s})\,\arctan{\left[\sqrt{\exp\left(\frac{t}{\theta}\right) -1}\,\right]} \right\rbrace \right), \\
    \frac{\Delta \tilde{\vartheta}_{\mathrm{G}}(\boldsymbol{s})}{\tilde{\vartheta}_{\mathrm{G}}(\boldsymbol{s})} =\,& 
    \frac{2 \tilde{\vartheta}_{\mathrm{G}}(\boldsymbol{s})}{t}
    \left(\left\lbrace\exp\left[\frac{t}{\tilde{\vartheta}_{\mathrm{G}}(\boldsymbol{s})}\right] -1\right\rbrace\left\lbrace
    -\arctan^2{\left[\sqrt{\exp\left(\frac{t}{\tilde{\vartheta}_{\mathrm{G}}(\boldsymbol{s})}\right) -1}\,\right]} \right. \right. \\
    &+ \left. \left.  \int d\theta\,p^\mathrm{G}_{\mathrm{MI}}(\theta|t)\,\arctan^2{\left[\sqrt{\exp\left(\frac{t}{\theta}\right) -1}\,\right]}\right\rbrace\right)^{\frac{1}{2}}.
    \end{split}
\end{align}
\end{subequations}
\end{widetext} 

\begin{figure}[t]
    \includegraphics[trim={-0.5cm 0cm 0cm -0.4cm},clip,width=\linewidth]{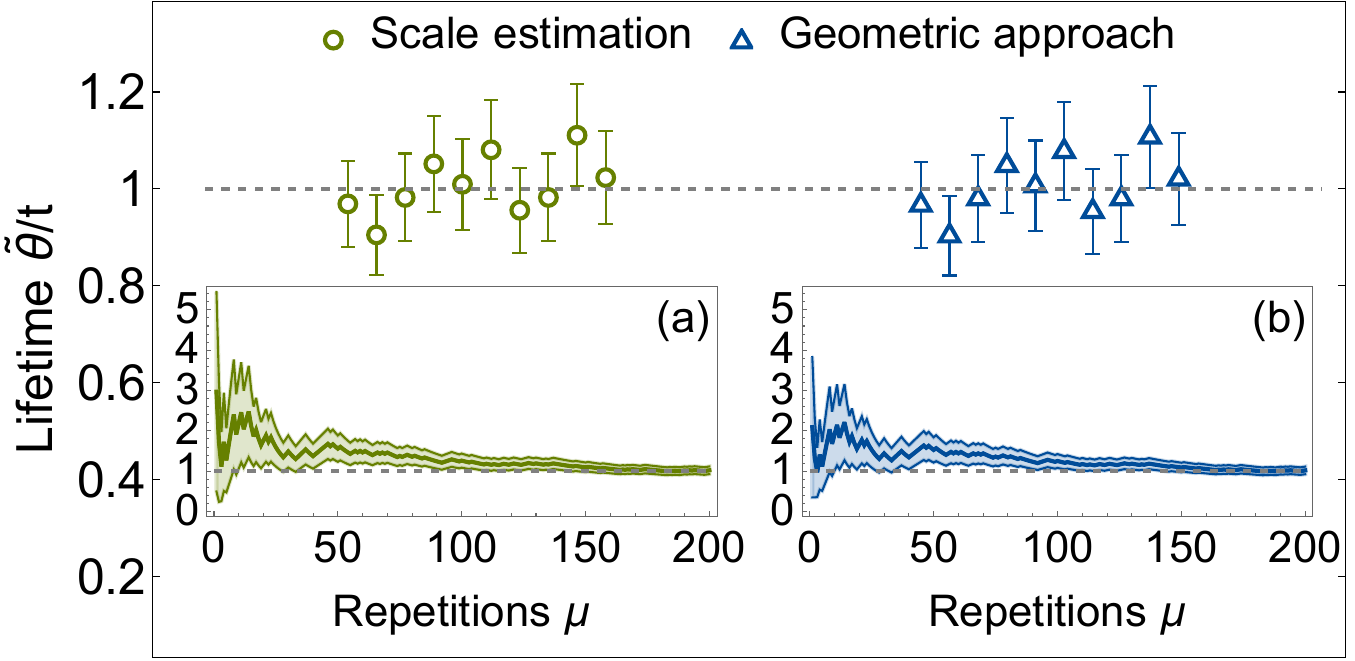}
    \includegraphics[trim={-0.5cm 0cm 0cm -0.4cm},clip,width=\linewidth]{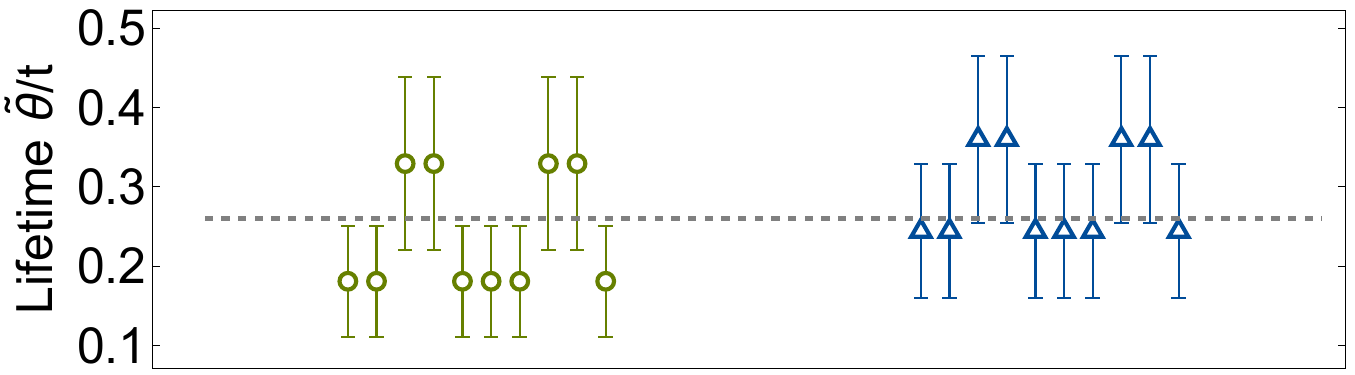}
    \caption{Variability of lifetime estimates for an atomic state for $\mu=200$ (top) and $\mu=20$ (bottom) measurements, using scale estimation (green circles) and information geometry (blue triangles). 
    The prior width in all cases is $b=10$. 
    The results obtained from $m=10$ different realisations are shown, on which both estimation strategies are applied. 
    The underlying true lifetime (dashed grey) is set to $\Theta/t = 1$ (top) and $\Theta/t = 0.26$ (bottom). 
    Insets (a), (b) show the typical convergence of estimates with the number of measurements. 
    As before, the different strategies can give different results for a small dataset, but become practically equivalent for larger numbers of data. 
    We note that scale estimation produces lower estimates for the small dataset due to the dominance of the corresponding prior in this regime, which grows unbounded as the lifetime tends to zero, as per the inset in Fig.~\ref{fig:lifetime-intrinsic}.
    }
\label{fig:lifetime}
\end{figure}

We simulated these protocols for $\mu = 200$ shots, prior width $b = 10$, and true lifetime $\Theta/t = 1$, repeating each procedure $m = 10$ times.  
The results are presented in Fig.~\ref{fig:lifetime}, top panel.  
As in Sec.~\ref{sec:coherence}, both methods perform practically equivalently, yielding NSRs of $0.34$ and $0.33$ for scale estimation and the geometric approach, respectively, with comparable error bars.  
A slight difference in precision appears for smaller lifetime values and low $\mu$; for instance, at $\Theta/t = 0.26$ and $\mu = 20$, Fig.~\ref{fig:lifetime}, bottom panel, shows that scale estimation is slightly less precise, with an NSR of $0.30$ vs. $0.20$.  
While this suggests potential for faster convergence with the more informed geometric approach, the advantage is less pronounced than in coherence estimation (Sec.~\ref{sec:coherence}).

So far, we have optimised estimators and POVMs.  
Additionally, we can optimise the state in Eq.~\eqref{eq:state-atom} over the coherence parameter $\eta$ by maximising the precision gain in Eq.~\eqref{eq:precision_gain}.  
However, beyond the difficulty of analytically computing the geometric symmetry function for an arbitrary $\eta$, we now face a second challenge: this function, which appears in the definition of precision gain, should remain fixed throughout a given experiment, yet it depends on the very control parameter we aim to optimise.
Equally problematic is that the geometric prior itself depends on $\eta$, since the initial prior in a Bayesian calculation should also remain fixed.
Formally, one could maximise the precision gain over $\eta$, but this is conceptually unclear, as both the loss function and prior would also change. 
Since these are chosen based on information and assumptions available \emph{a priori}, their optimisation is arguably meaningless once specified.  

\begin{figure}[t]
    \includegraphics[trim={0cm 0cm 0cm 0cm},clip,width=\linewidth]{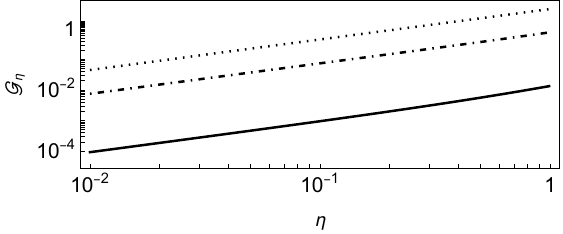}
    \caption{Precision gain quantifier in Eq.~\eqref{eq:precision_gain} as a function of the coherence parameter $\eta$ for the estimation of the lifetime of an atomic state, using scale estimation. 
    We performed the calculation for the prior widths $b=2$ (solid), $b=10$ (dot-dashed), and $b=100$ (dotted).  
    As can be seen, $\eta = 1$ leads to the maximum information gain per shot, thus yielding the optimal probe state.    
    }
\label{fig:lifetime-1shot}
\end{figure}

Scale estimation, on the other hand, presents no such difficulties.
Fig.~\ref{fig:lifetime-1shot} depicts the associated precision gain in Eq.~\eqref{eq:precision_gain} as a function of $\eta$ for different prior widths, revealing that the optimal state corresponds to $\eta = 1$.  
As seen before, the optimal POVM associated with this state is simply $\lbrace \ketbra{g}{g}, \ketbra{e}{e} \rbrace$, while the optimal estimator and empirical error are given by Eqs.~\eqref{eq:estimator-scale}. 
Furthermore, the fundamental precision limit for this strategy---given by Eq.~\eqref{eq:minimum-loss}--- evaluates to $0.99$. 

Finally, it is instructive to compare this performance with that of a suboptimal state.
Let $\eta = 1/2$, e.g.; the corresponding optimal POVM is $\lbrace \ketbra{\psi}{\psi}, |\psi^\perp\rangle \langle \psi^\perp| \rbrace$, with $\ket{\psi} = 0.44 \ket{g} + 0.90 \ket{e}$, while the minimum error turns out to be $1.52$. 
We observe that Bayesian optimisation of the POVM attempts to compensate for a suboptimal state by identifying a non-trivial projection, but this remains insufficient to achieve a precision close to that of the optimal strategy.

\section{Concluding remarks}\label{sec:remarks}

In this work, we have provided a cohesive foundation for \emph{global} quantum sensing and metrology using the framework of location-isomorphic estimation.
This has been achieved by unifying previous theoretical \cite{rubio2024first, rubio2023quantum, helstrom1976quantum} and experimental \cite{overton2024five, Glatthard2022} formulations with information geometry for Bayesian estimation \cite{jermyn2005invariant, jorgensen2021bayesian}.

In practice, optimal data post-processing is achieved via the methods in Sec.~\ref{sec:experiments-formalism}, while those in Sec.~\ref{sec:1shot-formulas} provide the means to design adaptive estimation protocols with optimal POVMs in each individual run. 
Furthermore, Eq.~\eqref{eq:symmetry-equation} serves as a fundamental equation guiding the selection of suitable loss functions, by incorporating symmetries or the geometry of the state space through ignorance priors. 
These priors can be determined in many cases of practical interest using the methods in Sec.~\ref{sec:symmetry-section}, as well as other well-known approaches in the literature~\cite{kass1996the, jaynes2003probability, linden2014bayesian}.

Owing to the Bayesian nature of this framework, it can be implemented in sensing platforms with any sample size or prior information~\cite{Glatthard2022, hewitt2024controlling, overton2024five}. 
Moreover, unlike in local estimation theory, the resulting estimators and POVMs are inherently independent of the true unknown parameter, thus making them substantially more practical.
These advantages are expected to carry over also to multiparameter scenarios \cite{demkowicz2020multiparameter}, where priors must reflect joint ignorance and potential correlations; this can be handled either through a system of functional equations \cite{jaynes2003probability, overton2024five} or by replacing the Fisher information with the determinant of its matrix version \cite{jeffreys1946invariant, kass1996the}.

We have also demonstrated the relative merits of enforcing invariance of the prior under specific parameter transformations versus employing information geometry.  
Through a series of case studies, including the estimation of exponential rates, quantum coherence, and atomic lifetimes, we found that information geometry provides a more systematic approach to computing loss functions.  
More importantly, it may enable faster convergence of estimates to the true value, consistent with the fact that geometric considerations lead to a framework best suited for scenarios where the parameter is known to lie within the sensitivity region of the sensor.

However, these potential gains come at a cost: the expressions for the associated loss functions, as well as estimators and errors, are generally more complex and not always analytically tractable.  
While a fully numerical approach is possible, it may be challenging for adaptive protocols requiring a rapid evaluation of the precision gain in Eq.~\eqref{eq:precision_gain} between runs, as in release-recapture thermometry \cite{Glatthard2022, hewitt2024controlling}.  
More critically, the potential dependence of the loss function on control parameters poses a significant challenge, as the gain quantifier in Eq.~\eqref{eq:precision_gain} would itself change at each step.

Furthermore, the requirement of a fixed state space as a prerequisite for information geometry makes state optimisation conceptually challenging.  
For instance, in the absence of control parameters, a parameterised state can be expressed as $\rho(\theta) = \Lambda_\theta (\rho_0)$, where \(\Lambda_\theta\) is a parameter-dependent channel and \(\rho_0\) is the probe state to be optimised.  
Here, the structure of the Bures metric is essentially determined by the channel.
However, this formulation does still not straightforwardly allow for probe optimisation, as the optimisation process may alter the state space itself.
Exploring ways to mitigate this problem is left for future work.

These difficulties contrast with the use of specific parameter transformations.
While it may not always be possible to identify a useful symmetry, this approach yields simpler and more general estimation strategies across a broad range of practical cases.  
These include, but are not limited to, locations \cite{helstrom1976quantum, jaynes2003probability}, scales \cite{rubio2023quantum}, weights \cite{rubio2024first}, and rotations \cite{demkowicz2015quantum, goldberg2021rotation}.  
Moreover, since these metrological frameworks are independent of the specific statistical model employed, they are readily applicable to machine learning protocols for quantum estimation \cite{lumino2018experimental, larocca2022group, perrier2024quantum, Cimini_2024, wakeham2024inference}.  

\section*{Acknowledgements}

We gratefully thank F. Albarelli, D. Branford, A. Luis, L. Ares, M. Mitchison, and the participants of the QUMINOS workshop for insightful discussions. 
We acknowledge the University of La Laguna (ULL)
and the Spanish Ministry of Universities for supporting the \textit{“DQDD Quantum Thermometry Program”}, during which this project started. LAC, JR, JG and EG acknowledge funding by the Ministerio de Ciencia e Innovaci\'{o}n and European
Union (FEDER) (PID2022-138269NB-I00). LAC is supported by a Ram\'{o}n y Cajal fellowship (RYC2021325804-I), funded by MCIN/AEI/10.13039/501100011033 and “NextGenerationEU”/PRTR.
JB acknowledges financial support from the House of Young Talents of Siegen.
JG acknowledges financial support from the Leverhulme Trust. 
JR acknowledges financial support from the Surrey Future Fellowship Programme. 

\bibliography{refs}

\begin{thebibliography}{118}%
\makeatletter
\providecommand \@ifxundefined [1]{%
 \@ifx{#1\undefined}
}%
\providecommand \@ifnum [1]{%
 \ifnum #1\expandafter \@firstoftwo
 \else \expandafter \@secondoftwo
 \fi
}%
\providecommand \@ifx [1]{%
 \ifx #1\expandafter \@firstoftwo
 \else \expandafter \@secondoftwo
 \fi
}%
\providecommand \natexlab [1]{#1}%
\providecommand \enquote  [1]{``#1''}%
\providecommand \bibnamefont  [1]{#1}%
\providecommand \bibfnamefont [1]{#1}%
\providecommand \citenamefont [1]{#1}%
\providecommand \href@noop [0]{\@secondoftwo}%
\providecommand \href [0]{\begingroup \@sanitize@url \@href}%
\providecommand \@href[1]{\@@startlink{#1}\@@href}%
\providecommand \@@href[1]{\endgroup#1\@@endlink}%
\providecommand \@sanitize@url [0]{\catcode `\\12\catcode `\$12\catcode
  `\&12\catcode `\#12\catcode `\^12\catcode `\_12\catcode `\%12\relax}%
\providecommand \@@startlink[1]{}%
\providecommand \@@endlink[0]{}%
\providecommand \url  [0]{\begingroup\@sanitize@url \@url }%
\providecommand \@url [1]{\endgroup\@href {#1}{\urlprefix }}%
\providecommand \urlprefix  [0]{URL }%
\providecommand \Eprint [0]{\href }%
\providecommand \doibase [0]{https://doi.org/}%
\providecommand \selectlanguage [0]{\@gobble}%
\providecommand \bibinfo  [0]{\@secondoftwo}%
\providecommand \bibfield  [0]{\@secondoftwo}%
\providecommand \translation [1]{[#1]}%
\providecommand \BibitemOpen [0]{}%
\providecommand \bibitemStop [0]{}%
\providecommand \bibitemNoStop [0]{.\EOS\space}%
\providecommand \EOS [0]{\spacefactor3000\relax}%
\providecommand \BibitemShut  [1]{\csname bibitem#1\endcsname}%
\let\auto@bib@innerbib\@empty
\bibitem [{\citenamefont {Degen}\ \emph
  {et~al.}(2017{\natexlab{a}})\citenamefont {Degen}, \citenamefont {Reinhard},\
  and\ \citenamefont {Cappellaro}}]{degen2017}%
  \BibitemOpen
  \bibfield  {author} {\bibinfo {author} {\bibfnamefont {C.~L.}\ \bibnamefont
  {Degen}}, \bibinfo {author} {\bibfnamefont {F.}~\bibnamefont {Reinhard}},\
  and\ \bibinfo {author} {\bibfnamefont {P.}~\bibnamefont {Cappellaro}},\
  }\bibfield  {title} {\bibinfo {title} {Quantum sensing},\ }\href
  {https://doi.org/10.1103/RevModPhys.89.035002} {\bibfield  {journal}
  {\bibinfo  {journal} {Rev. Mod. Phys.}\ }\textbf {\bibinfo {volume} {89}},\
  \bibinfo {pages} {035002} (\bibinfo {year} {2017}{\natexlab{a}})}\BibitemShut
  {NoStop}%
\bibitem [{\citenamefont {Doherty}\ \emph {et~al.}(2000)\citenamefont
  {Doherty}, \citenamefont {Habib}, \citenamefont {Jacobs}, \citenamefont
  {Mabuchi},\ and\ \citenamefont {Tan}}]{doherty2000}%
  \BibitemOpen
  \bibfield  {author} {\bibinfo {author} {\bibfnamefont {A.~C.}\ \bibnamefont
  {Doherty}}, \bibinfo {author} {\bibfnamefont {S.}~\bibnamefont {Habib}},
  \bibinfo {author} {\bibfnamefont {K.}~\bibnamefont {Jacobs}}, \bibinfo
  {author} {\bibfnamefont {H.}~\bibnamefont {Mabuchi}},\ and\ \bibinfo {author}
  {\bibfnamefont {S.~M.}\ \bibnamefont {Tan}},\ }\bibfield  {title} {\bibinfo
  {title} {Quantum feedback control and classical control theory},\ }\href
  {https://doi.org/10.1103/PhysRevA.62.012105} {\bibfield  {journal} {\bibinfo
  {journal} {Phys. Rev. A}\ }\textbf {\bibinfo {volume} {62}},\ \bibinfo
  {pages} {012105} (\bibinfo {year} {2000})}\BibitemShut {NoStop}%
\bibitem [{\citenamefont {Schnabel}\ \emph {et~al.}(2010)\citenamefont
  {Schnabel}, \citenamefont {Mavalvala}, \citenamefont {McClelland},\ and\
  \citenamefont {Lam}}]{schnabel_quantum_2010}%
  \BibitemOpen
  \bibfield  {author} {\bibinfo {author} {\bibfnamefont {R.}~\bibnamefont
  {Schnabel}}, \bibinfo {author} {\bibfnamefont {N.}~\bibnamefont {Mavalvala}},
  \bibinfo {author} {\bibfnamefont {D.~E.}\ \bibnamefont {McClelland}},\ and\
  \bibinfo {author} {\bibfnamefont {P.~K.}\ \bibnamefont {Lam}},\ }\bibfield
  {title} {\bibinfo {title} {Quantum metrology for gravitational wave
  astronomy},\ }\href {https://doi.org/10.1038/ncomms1122} {\bibfield
  {journal} {\bibinfo  {journal} {Nat. Commun.}\ }\textbf {\bibinfo {volume}
  {1}},\ \bibinfo {pages} {121} (\bibinfo {year} {2010})}\BibitemShut {NoStop}%
\bibitem [{\citenamefont {Belenchia}\ \emph {et~al.}(2022)\citenamefont
  {Belenchia}, \citenamefont {Carlesso}, \citenamefont {Ömer Bayraktar},
  \citenamefont {Dequal}, \citenamefont {Derkach}, \citenamefont {Gasbarri},
  \citenamefont {Herr}, \citenamefont {Li}, \citenamefont {Rademacher},
  \citenamefont {Sidhu}, \citenamefont {Oi}, \citenamefont {Seidel},
  \citenamefont {Kaltenbaek}, \citenamefont {Marquardt}, \citenamefont
  {Ulbricht}, \citenamefont {Usenko}, \citenamefont {Wörner}, \citenamefont
  {Xuereb}, \citenamefont {Paternostro},\ and\ \citenamefont
  {Bassi}}]{belenchia2021quantum}%
  \BibitemOpen
  \bibfield  {author} {\bibinfo {author} {\bibfnamefont {A.}~\bibnamefont
  {Belenchia}}, \bibinfo {author} {\bibfnamefont {M.}~\bibnamefont {Carlesso}},
  \bibinfo {author} {\bibnamefont {Ömer Bayraktar}}, \bibinfo {author}
  {\bibfnamefont {D.}~\bibnamefont {Dequal}}, \bibinfo {author} {\bibfnamefont
  {I.}~\bibnamefont {Derkach}}, \bibinfo {author} {\bibfnamefont
  {G.}~\bibnamefont {Gasbarri}}, \bibinfo {author} {\bibfnamefont
  {W.}~\bibnamefont {Herr}}, \bibinfo {author} {\bibfnamefont {Y.~L.}\
  \bibnamefont {Li}}, \bibinfo {author} {\bibfnamefont {M.}~\bibnamefont
  {Rademacher}}, \bibinfo {author} {\bibfnamefont {J.}~\bibnamefont {Sidhu}},
  \bibinfo {author} {\bibfnamefont {D.~K.}\ \bibnamefont {Oi}}, \bibinfo
  {author} {\bibfnamefont {S.~T.}\ \bibnamefont {Seidel}}, \bibinfo {author}
  {\bibfnamefont {R.}~\bibnamefont {Kaltenbaek}}, \bibinfo {author}
  {\bibfnamefont {C.}~\bibnamefont {Marquardt}}, \bibinfo {author}
  {\bibfnamefont {H.}~\bibnamefont {Ulbricht}}, \bibinfo {author}
  {\bibfnamefont {V.~C.}\ \bibnamefont {Usenko}}, \bibinfo {author}
  {\bibfnamefont {L.}~\bibnamefont {Wörner}}, \bibinfo {author} {\bibfnamefont
  {A.}~\bibnamefont {Xuereb}}, \bibinfo {author} {\bibfnamefont
  {M.}~\bibnamefont {Paternostro}},\ and\ \bibinfo {author} {\bibfnamefont
  {A.}~\bibnamefont {Bassi}},\ }\bibfield  {title} {\bibinfo {title} {Quantum
  physics in space},\ }\href
  {https://doi.org/https://doi.org/10.1016/j.physrep.2021.11.004} {\bibfield
  {journal} {\bibinfo  {journal} {Phys. Rep.}\ }\textbf {\bibinfo {volume}
  {951}},\ \bibinfo {pages} {1} (\bibinfo {year} {2022})},\ \bibinfo {note}
  {quantum Physics in Space}\BibitemShut {NoStop}%
\bibitem [{\citenamefont {Carney}\ \emph {et~al.}(2021)\citenamefont {Carney},
  \citenamefont {Krnjaic}, \citenamefont {Moore}, \citenamefont {Regal},
  \citenamefont {Afek}, \citenamefont {Bhave}, \citenamefont {Brubaker},
  \citenamefont {Corbitt}, \citenamefont {Cripe}, \citenamefont {Crisosto},
  \citenamefont {Geraci}, \citenamefont {Ghosh}, \citenamefont {Harris},
  \citenamefont {Hook}, \citenamefont {Kolb}, \citenamefont {Kunjummen},
  \citenamefont {Lang}, \citenamefont {Li}, \citenamefont {Lin}, \citenamefont
  {Liu}, \citenamefont {Lykken}, \citenamefont {Magrini}, \citenamefont
  {Manley}, \citenamefont {Matsumoto}, \citenamefont {Monte}, \citenamefont
  {Monteiro}, \citenamefont {Purdy}, \citenamefont {Riedel}, \citenamefont
  {Singh}, \citenamefont {Singh}, \citenamefont {Sinha}, \citenamefont
  {Taylor}, \citenamefont {Qin}, \citenamefont {Wilson},\ and\ \citenamefont
  {Zhao}}]{carney2021mechanical}%
  \BibitemOpen
  \bibfield  {author} {\bibinfo {author} {\bibfnamefont {D.}~\bibnamefont
  {Carney}}, \bibinfo {author} {\bibfnamefont {G.}~\bibnamefont {Krnjaic}},
  \bibinfo {author} {\bibfnamefont {D.~C.}\ \bibnamefont {Moore}}, \bibinfo
  {author} {\bibfnamefont {C.~A.}\ \bibnamefont {Regal}}, \bibinfo {author}
  {\bibfnamefont {G.}~\bibnamefont {Afek}}, \bibinfo {author} {\bibfnamefont
  {S.}~\bibnamefont {Bhave}}, \bibinfo {author} {\bibfnamefont
  {B.}~\bibnamefont {Brubaker}}, \bibinfo {author} {\bibfnamefont
  {T.}~\bibnamefont {Corbitt}}, \bibinfo {author} {\bibfnamefont
  {J.}~\bibnamefont {Cripe}}, \bibinfo {author} {\bibfnamefont
  {N.}~\bibnamefont {Crisosto}}, \bibinfo {author} {\bibfnamefont
  {A.}~\bibnamefont {Geraci}}, \bibinfo {author} {\bibfnamefont
  {S.}~\bibnamefont {Ghosh}}, \bibinfo {author} {\bibfnamefont {J.~G.~E.}\
  \bibnamefont {Harris}}, \bibinfo {author} {\bibfnamefont {A.}~\bibnamefont
  {Hook}}, \bibinfo {author} {\bibfnamefont {E.~W.}\ \bibnamefont {Kolb}},
  \bibinfo {author} {\bibfnamefont {J.}~\bibnamefont {Kunjummen}}, \bibinfo
  {author} {\bibfnamefont {R.~F.}\ \bibnamefont {Lang}}, \bibinfo {author}
  {\bibfnamefont {T.}~\bibnamefont {Li}}, \bibinfo {author} {\bibfnamefont
  {T.}~\bibnamefont {Lin}}, \bibinfo {author} {\bibfnamefont {Z.}~\bibnamefont
  {Liu}}, \bibinfo {author} {\bibfnamefont {J.}~\bibnamefont {Lykken}},
  \bibinfo {author} {\bibfnamefont {L.}~\bibnamefont {Magrini}}, \bibinfo
  {author} {\bibfnamefont {J.}~\bibnamefont {Manley}}, \bibinfo {author}
  {\bibfnamefont {N.}~\bibnamefont {Matsumoto}}, \bibinfo {author}
  {\bibfnamefont {A.}~\bibnamefont {Monte}}, \bibinfo {author} {\bibfnamefont
  {F.}~\bibnamefont {Monteiro}}, \bibinfo {author} {\bibfnamefont
  {T.}~\bibnamefont {Purdy}}, \bibinfo {author} {\bibfnamefont {C.~J.}\
  \bibnamefont {Riedel}}, \bibinfo {author} {\bibfnamefont {R.}~\bibnamefont
  {Singh}}, \bibinfo {author} {\bibfnamefont {S.}~\bibnamefont {Singh}},
  \bibinfo {author} {\bibfnamefont {K.}~\bibnamefont {Sinha}}, \bibinfo
  {author} {\bibfnamefont {J.~M.}\ \bibnamefont {Taylor}}, \bibinfo {author}
  {\bibfnamefont {J.}~\bibnamefont {Qin}}, \bibinfo {author} {\bibfnamefont
  {D.~J.}\ \bibnamefont {Wilson}},\ and\ \bibinfo {author} {\bibfnamefont
  {Y.}~\bibnamefont {Zhao}},\ }\bibfield  {title} {\bibinfo {title} {Mechanical
  quantum sensing in the search for dark matter},\ }\href
  {https://doi.org/10.1088/2058-9565/abcfcd} {\bibfield  {journal} {\bibinfo
  {journal} {Quantum Sci. Technol.}\ }\textbf {\bibinfo {volume} {6}},\
  \bibinfo {pages} {024002} (\bibinfo {year} {2021})}\BibitemShut {NoStop}%
\bibitem [{\citenamefont {Jackson~Kimball}\ \emph {et~al.}(2023)\citenamefont
  {Jackson~Kimball}, \citenamefont {Budker}, \citenamefont {Chupp},
  \citenamefont {Geraci}, \citenamefont {Kolkowitz}, \citenamefont {Singh},\
  and\ \citenamefont {Sushkov}}]{kimball2023}%
  \BibitemOpen
  \bibfield  {author} {\bibinfo {author} {\bibfnamefont {D.~F.}\ \bibnamefont
  {Jackson~Kimball}}, \bibinfo {author} {\bibfnamefont {D.}~\bibnamefont
  {Budker}}, \bibinfo {author} {\bibfnamefont {T.~E.}\ \bibnamefont {Chupp}},
  \bibinfo {author} {\bibfnamefont {A.~A.}\ \bibnamefont {Geraci}}, \bibinfo
  {author} {\bibfnamefont {S.}~\bibnamefont {Kolkowitz}}, \bibinfo {author}
  {\bibfnamefont {J.~T.}\ \bibnamefont {Singh}},\ and\ \bibinfo {author}
  {\bibfnamefont {A.~O.}\ \bibnamefont {Sushkov}},\ }\bibfield  {title}
  {\bibinfo {title} {Probing fundamental physics with spin-based quantum
  sensors},\ }\href {https://doi.org/10.1103/PhysRevA.108.010101} {\bibfield
  {journal} {\bibinfo  {journal} {Phys. Rev. A}\ }\textbf {\bibinfo {volume}
  {108}},\ \bibinfo {pages} {010101} (\bibinfo {year} {2023})}\BibitemShut
  {NoStop}%
\bibitem [{\citenamefont {Ye}\ and\ \citenamefont
  {Zoller}(2024)}]{ye2024essayquantumsensing}%
  \BibitemOpen
  \bibfield  {author} {\bibinfo {author} {\bibfnamefont {J.}~\bibnamefont
  {Ye}}\ and\ \bibinfo {author} {\bibfnamefont {P.}~\bibnamefont {Zoller}},\
  }\bibfield  {title} {\bibinfo {title} {Essay: Quantum sensing with atomic,
  molecular, and optical platforms for fundamental physics},\ }\href
  {https://doi.org/10.1103/PhysRevLett.132.190001} {\bibfield  {journal}
  {\bibinfo  {journal} {Phys. Rev. Lett.}\ }\textbf {\bibinfo {volume} {132}},\
  \bibinfo {pages} {190001} (\bibinfo {year} {2024})}\BibitemShut {NoStop}%
\bibitem [{\citenamefont {Bass}\ and\ \citenamefont
  {Doser}(2024)}]{bass2024quantum}%
  \BibitemOpen
  \bibfield  {author} {\bibinfo {author} {\bibfnamefont {S.~D.}\ \bibnamefont
  {Bass}}\ and\ \bibinfo {author} {\bibfnamefont {M.}~\bibnamefont {Doser}},\
  }\bibfield  {title} {\bibinfo {title} {Quantum sensing for particle
  physics},\ }\href {https://doi.org/doi.org/10.1038/s42254-024-00714-3}
  {\bibfield  {journal} {\bibinfo  {journal} {Nat. Rev. Phys.}\ ,\ \bibinfo
  {pages} {1}} (\bibinfo {year} {2024})}\BibitemShut {NoStop}%
\bibitem [{\citenamefont {DeMille}\ \emph {et~al.}(2024)\citenamefont
  {DeMille}, \citenamefont {Hutzler}, \citenamefont {Rey},\ and\ \citenamefont
  {Zelevinsky}}]{demille2024quantum}%
  \BibitemOpen
  \bibfield  {author} {\bibinfo {author} {\bibfnamefont {D.}~\bibnamefont
  {DeMille}}, \bibinfo {author} {\bibfnamefont {N.~R.}\ \bibnamefont
  {Hutzler}}, \bibinfo {author} {\bibfnamefont {A.~M.}\ \bibnamefont {Rey}},\
  and\ \bibinfo {author} {\bibfnamefont {T.}~\bibnamefont {Zelevinsky}},\
  }\bibfield  {title} {\bibinfo {title} {Quantum sensing and metrology for
  fundamental physics with molecules},\ }\href
  {https://doi.org/doi.org/10.1038/s41567-024-02499-9} {\bibfield  {journal}
  {\bibinfo  {journal} {Nat. Phys.}\ ,\ \bibinfo {pages} {19}} (\bibinfo {year}
  {2024})}\BibitemShut {NoStop}%
\bibitem [{\citenamefont {Xiao}\ \emph {et~al.}(1987)\citenamefont {Xiao},
  \citenamefont {Wu},\ and\ \citenamefont {Kimble}}]{xiao1987}%
  \BibitemOpen
  \bibfield  {author} {\bibinfo {author} {\bibfnamefont {M.}~\bibnamefont
  {Xiao}}, \bibinfo {author} {\bibfnamefont {L.-A.}\ \bibnamefont {Wu}},\ and\
  \bibinfo {author} {\bibfnamefont {H.~J.}\ \bibnamefont {Kimble}},\ }\bibfield
   {title} {\bibinfo {title} {Precision measurement beyond the shot-noise
  limit},\ }\href {https://doi.org/10.1103/PhysRevLett.59.278} {\bibfield
  {journal} {\bibinfo  {journal} {Phys. Rev. Lett.}\ }\textbf {\bibinfo
  {volume} {59}},\ \bibinfo {pages} {278} (\bibinfo {year} {1987})}\BibitemShut
  {NoStop}%
\bibitem [{\citenamefont {Giovannetti}\ \emph {et~al.}(2006)\citenamefont
  {Giovannetti}, \citenamefont {Lloyd},\ and\ \citenamefont
  {Maccone}}]{Giovanetti2006}%
  \BibitemOpen
  \bibfield  {author} {\bibinfo {author} {\bibfnamefont {V.}~\bibnamefont
  {Giovannetti}}, \bibinfo {author} {\bibfnamefont {S.}~\bibnamefont {Lloyd}},\
  and\ \bibinfo {author} {\bibfnamefont {L.}~\bibnamefont {Maccone}},\
  }\bibfield  {title} {\bibinfo {title} {Quantum metrology},\ }\href
  {https://doi.org/10.1103/PhysRevLett.96.010401} {\bibfield  {journal}
  {\bibinfo  {journal} {Phys. Rev. Lett.}\ }\textbf {\bibinfo {volume} {96}},\
  \bibinfo {pages} {010401} (\bibinfo {year} {2006})}\BibitemShut {NoStop}%
\bibitem [{\citenamefont {Dorner}\ \emph {et~al.}(2009)\citenamefont {Dorner},
  \citenamefont {Demkowicz-Dobrzanski}, \citenamefont {Smith}, \citenamefont
  {Lundeen}, \citenamefont {Wasilewski}, \citenamefont {Banaszek},\ and\
  \citenamefont {Walmsley}}]{Dorner2009}%
  \BibitemOpen
  \bibfield  {author} {\bibinfo {author} {\bibfnamefont {U.}~\bibnamefont
  {Dorner}}, \bibinfo {author} {\bibfnamefont {R.}~\bibnamefont
  {Demkowicz-Dobrzanski}}, \bibinfo {author} {\bibfnamefont {B.~J.}\
  \bibnamefont {Smith}}, \bibinfo {author} {\bibfnamefont {J.~S.}\ \bibnamefont
  {Lundeen}}, \bibinfo {author} {\bibfnamefont {W.}~\bibnamefont {Wasilewski}},
  \bibinfo {author} {\bibfnamefont {K.}~\bibnamefont {Banaszek}},\ and\
  \bibinfo {author} {\bibfnamefont {I.~A.}\ \bibnamefont {Walmsley}},\
  }\bibfield  {title} {\bibinfo {title} {Optimal quantum phase estimation},\
  }\href {https://doi.org/10.1103/PhysRevLett.102.040403} {\bibfield  {journal}
  {\bibinfo  {journal} {Phys. Rev. Lett.}\ }\textbf {\bibinfo {volume} {102}},\
  \bibinfo {pages} {040403} (\bibinfo {year} {2009})}\BibitemShut {NoStop}%
\bibitem [{\citenamefont {Demkowicz-Dobrza\ifmmode~\acute{n}\else
  \'{n}\fi{}ski}\ \emph {et~al.}(2015)\citenamefont
  {Demkowicz-Dobrza\ifmmode~\acute{n}\else \'{n}\fi{}ski}, \citenamefont
  {Jarzyna},\ and\ \citenamefont {Ko\l{}ody\ifmmode~\acute{n}\else
  \'{n}\fi{}ski}}]{demkowicz2015quantum}%
  \BibitemOpen
  \bibfield  {author} {\bibinfo {author} {\bibfnamefont {R.}~\bibnamefont
  {Demkowicz-Dobrza\ifmmode~\acute{n}\else \'{n}\fi{}ski}}, \bibinfo {author}
  {\bibfnamefont {M.}~\bibnamefont {Jarzyna}},\ and\ \bibinfo {author}
  {\bibfnamefont {J.}~\bibnamefont {Ko\l{}ody\ifmmode~\acute{n}\else
  \'{n}\fi{}ski}},\ }\bibfield  {title} {\bibinfo {title} {{Quantum Limits in
  Optical Interferometry}},\ }\href {https://doi.org/10.1016/bs.po.2015.02.003}
  {\bibfield  {journal} {\bibinfo  {journal} {Prog. Opt.}\ }\textbf {\bibinfo
  {volume} {60}},\ \bibinfo {pages} {345} (\bibinfo {year} {2015})}\BibitemShut
  {NoStop}%
\bibitem [{\citenamefont {Correa}\ \emph {et~al.}(2015)\citenamefont {Correa},
  \citenamefont {Mehboudi}, \citenamefont {Adesso},\ and\ \citenamefont
  {Sanpera}}]{Correa2015}%
  \BibitemOpen
  \bibfield  {author} {\bibinfo {author} {\bibfnamefont {L.~A.}\ \bibnamefont
  {Correa}}, \bibinfo {author} {\bibfnamefont {M.}~\bibnamefont {Mehboudi}},
  \bibinfo {author} {\bibfnamefont {G.}~\bibnamefont {Adesso}},\ and\ \bibinfo
  {author} {\bibfnamefont {A.}~\bibnamefont {Sanpera}},\ }\bibfield  {title}
  {\bibinfo {title} {Individual quantum probes for optimal thermometry},\
  }\href {https://doi.org/10.1103/PhysRevLett.114.220405} {\bibfield  {journal}
  {\bibinfo  {journal} {Phys. Rev. Lett.}\ }\textbf {\bibinfo {volume} {114}},\
  \bibinfo {pages} {220405} (\bibinfo {year} {2015})}\BibitemShut {NoStop}%
\bibitem [{\citenamefont {Wu}\ and\ \citenamefont {Shi}(2020)}]{Wu2020}%
  \BibitemOpen
  \bibfield  {author} {\bibinfo {author} {\bibfnamefont {W.}~\bibnamefont
  {Wu}}\ and\ \bibinfo {author} {\bibfnamefont {C.}~\bibnamefont {Shi}},\
  }\bibfield  {title} {\bibinfo {title} {Quantum parameter estimation in a
  dissipative environment},\ }\href
  {https://doi.org/10.1103/PhysRevA.102.032607} {\bibfield  {journal} {\bibinfo
   {journal} {Phys. Rev. A}\ }\textbf {\bibinfo {volume} {102}},\ \bibinfo
  {pages} {032607} (\bibinfo {year} {2020})}\BibitemShut {NoStop}%
\bibitem [{\citenamefont {Saleem}\ \emph {et~al.}(2023)\citenamefont {Saleem},
  \citenamefont {Shaji},\ and\ \citenamefont {Gray}}]{saleem2023}%
  \BibitemOpen
  \bibfield  {author} {\bibinfo {author} {\bibfnamefont {Z.~H.}\ \bibnamefont
  {Saleem}}, \bibinfo {author} {\bibfnamefont {A.}~\bibnamefont {Shaji}},\ and\
  \bibinfo {author} {\bibfnamefont {S.~K.}\ \bibnamefont {Gray}},\ }\bibfield
  {title} {\bibinfo {title} {Optimal time for sensing in open quantum
  systems},\ }\href {https://doi.org/10.1103/PhysRevA.108.022413} {\bibfield
  {journal} {\bibinfo  {journal} {Phys. Rev. A}\ }\textbf {\bibinfo {volume}
  {108}},\ \bibinfo {pages} {022413} (\bibinfo {year} {2023})}\BibitemShut
  {NoStop}%
\bibitem [{\citenamefont {Mirkhalaf}\ \emph {et~al.}(2024)\citenamefont
  {Mirkhalaf}, \citenamefont {Mehboudi}, \citenamefont {Qaleh},\ and\
  \citenamefont {Rahimi-Keshari}}]{Mirkhalaf2024}%
  \BibitemOpen
  \bibfield  {author} {\bibinfo {author} {\bibfnamefont {S.}~\bibnamefont
  {Mirkhalaf}}, \bibinfo {author} {\bibfnamefont {M.}~\bibnamefont {Mehboudi}},
  \bibinfo {author} {\bibfnamefont {Z.~N.}\ \bibnamefont {Qaleh}},\ and\
  \bibinfo {author} {\bibfnamefont {S.}~\bibnamefont {Rahimi-Keshari}},\
  }\bibfield  {title} {\bibinfo {title} {Operational significance of
  nonclassicality in nonequilibrium gaussian quantum thermometry},\ }\href
  {https://doi.org/10.1088/1367-2630/ad23a1} {\bibfield  {journal} {\bibinfo
  {journal} {New J. Phys.}\ }\textbf {\bibinfo {volume} {26}},\ \bibinfo
  {pages} {023046} (\bibinfo {year} {2024})}\BibitemShut {NoStop}%
\bibitem [{\citenamefont {Paris}(2009)}]{paris2009}%
  \BibitemOpen
  \bibfield  {author} {\bibinfo {author} {\bibfnamefont {M.~G.~A.}\
  \bibnamefont {Paris}},\ }\bibfield  {title} {\bibinfo {title} {Quantum
  estimation for quantum technology},\ }\href
  {https://doi.org/10.1142/S0219749909004839} {\bibfield  {journal} {\bibinfo
  {journal} {Int. J. Quantum Inf.}\ }\textbf {\bibinfo {volume} {07}},\
  \bibinfo {pages} {125} (\bibinfo {year} {2009})}\BibitemShut {NoStop}%
\bibitem [{\citenamefont {Teklu}\ \emph {et~al.}(2009)\citenamefont {Teklu},
  \citenamefont {Olivares},\ and\ \citenamefont {Paris}}]{Teklu_2009}%
  \BibitemOpen
  \bibfield  {author} {\bibinfo {author} {\bibfnamefont {B.}~\bibnamefont
  {Teklu}}, \bibinfo {author} {\bibfnamefont {S.}~\bibnamefont {Olivares}},\
  and\ \bibinfo {author} {\bibfnamefont {M.~G.~A.}\ \bibnamefont {Paris}},\
  }\bibfield  {title} {\bibinfo {title} {Bayesian estimation of one-parameter
  qubit gates},\ }\href {https://doi.org/10.1088/0953-4075/42/3/035502}
  {\bibfield  {journal} {\bibinfo  {journal} {J. Phys. B: At. Mol. Opt. Phys.}\
  }\textbf {\bibinfo {volume} {42}},\ \bibinfo {pages} {035502} (\bibinfo
  {year} {2009})}\BibitemShut {NoStop}%
\bibitem [{\citenamefont {Lumino}\ \emph {et~al.}(2018)\citenamefont {Lumino},
  \citenamefont {Polino}, \citenamefont {Rab}, \citenamefont {Milani},
  \citenamefont {Spagnolo}, \citenamefont {Wiebe},\ and\ \citenamefont
  {Sciarrino}}]{lumino2018experimental}%
  \BibitemOpen
  \bibfield  {author} {\bibinfo {author} {\bibfnamefont {A.}~\bibnamefont
  {Lumino}}, \bibinfo {author} {\bibfnamefont {E.}~\bibnamefont {Polino}},
  \bibinfo {author} {\bibfnamefont {A.~S.}\ \bibnamefont {Rab}}, \bibinfo
  {author} {\bibfnamefont {G.}~\bibnamefont {Milani}}, \bibinfo {author}
  {\bibfnamefont {N.}~\bibnamefont {Spagnolo}}, \bibinfo {author}
  {\bibfnamefont {N.}~\bibnamefont {Wiebe}},\ and\ \bibinfo {author}
  {\bibfnamefont {F.}~\bibnamefont {Sciarrino}},\ }\bibfield  {title} {\bibinfo
  {title} {Experimental phase estimation enhanced by machine learning},\ }\href
  {https://doi.org/10.1103/PhysRevApplied.10.044033} {\bibfield  {journal}
  {\bibinfo  {journal} {Phys. Rev. Appl.}\ }\textbf {\bibinfo {volume} {10}},\
  \bibinfo {pages} {044033} (\bibinfo {year} {2018})}\BibitemShut {NoStop}%
\bibitem [{\citenamefont {Rubio}\ and\ \citenamefont
  {Dunningham}(2019)}]{rubio2018quantum}%
  \BibitemOpen
  \bibfield  {author} {\bibinfo {author} {\bibfnamefont {J.}~\bibnamefont
  {Rubio}}\ and\ \bibinfo {author} {\bibfnamefont {J.~A.}\ \bibnamefont
  {Dunningham}},\ }\bibfield  {title} {\bibinfo {title} {Quantum metrology in
  the presence of limited data},\ }\href
  {https://doi.org/10.1088/1367-2630/ab098b} {\bibfield  {journal} {\bibinfo
  {journal} {New J. Phys.}\ }\textbf {\bibinfo {volume} {21}},\ \bibinfo
  {pages} {043037} (\bibinfo {year} {2019})}\BibitemShut {NoStop}%
\bibitem [{\citenamefont {Bavaresco}\ \emph
  {et~al.}(2024{\natexlab{a}})\citenamefont {Bavaresco}, \citenamefont
  {Lipka-Bartosik}, \citenamefont {Sekatski},\ and\ \citenamefont
  {Mehboudi}}]{Bavaresco2024}%
  \BibitemOpen
  \bibfield  {author} {\bibinfo {author} {\bibfnamefont {J.}~\bibnamefont
  {Bavaresco}}, \bibinfo {author} {\bibfnamefont {P.}~\bibnamefont
  {Lipka-Bartosik}}, \bibinfo {author} {\bibfnamefont {P.}~\bibnamefont
  {Sekatski}},\ and\ \bibinfo {author} {\bibfnamefont {M.}~\bibnamefont
  {Mehboudi}},\ }\bibfield  {title} {\bibinfo {title} {Designing optimal
  protocols in bayesian quantum parameter estimation with higher-order
  operations},\ }\href {https://doi.org/10.1103/PhysRevResearch.6.023305}
  {\bibfield  {journal} {\bibinfo  {journal} {Phys. Rev. Res.}\ }\textbf
  {\bibinfo {volume} {6}},\ \bibinfo {pages} {023305} (\bibinfo {year}
  {2024}{\natexlab{a}})}\BibitemShut {NoStop}%
\bibitem [{\citenamefont {Meyer}\ \emph {et~al.}(2025)\citenamefont {Meyer},
  \citenamefont {Khatri}, \citenamefont {Fran\c{c}a}, \citenamefont {Eisert},\
  and\ \citenamefont {Faist}}]{meyer2023quantum}%
  \BibitemOpen
  \bibfield  {author} {\bibinfo {author} {\bibfnamefont {J.~J.}\ \bibnamefont
  {Meyer}}, \bibinfo {author} {\bibfnamefont {S.}~\bibnamefont {Khatri}},
  \bibinfo {author} {\bibfnamefont {D.~S.}\ \bibnamefont {Fran\c{c}a}},
  \bibinfo {author} {\bibfnamefont {J.}~\bibnamefont {Eisert}},\ and\ \bibinfo
  {author} {\bibfnamefont {P.}~\bibnamefont {Faist}},\ }\bibfield  {title}
  {\bibinfo {title} {Quantum metrology in the finite-sample regime},\ }\href
  {https://doi.org/10.1103/qbn1-p6bq} {\bibfield  {journal} {\bibinfo
  {journal} {PRX Quantum}\ ,\ } (\bibinfo {year} {2025})}\BibitemShut {NoStop}%
\bibitem [{\citenamefont {Cimini}\ \emph {et~al.}(2024)\citenamefont {Cimini},
  \citenamefont {Polino}, \citenamefont {Valeri}, \citenamefont {Spagnolo},\
  and\ \citenamefont {Sciarrino}}]{Cimini_2024}%
  \BibitemOpen
  \bibfield  {author} {\bibinfo {author} {\bibfnamefont {V.}~\bibnamefont
  {Cimini}}, \bibinfo {author} {\bibfnamefont {E.}~\bibnamefont {Polino}},
  \bibinfo {author} {\bibfnamefont {M.}~\bibnamefont {Valeri}}, \bibinfo
  {author} {\bibfnamefont {N.}~\bibnamefont {Spagnolo}},\ and\ \bibinfo
  {author} {\bibfnamefont {F.}~\bibnamefont {Sciarrino}},\ }\bibfield  {title}
  {\bibinfo {title} {Benchmarking bayesian quantum estimation},\ }\href
  {https://doi.org/10.1088/2058-9565/ad48b3} {\bibfield  {journal} {\bibinfo
  {journal} {Quantum Sci. Technol.}\ }\textbf {\bibinfo {volume} {9}},\
  \bibinfo {pages} {035035} (\bibinfo {year} {2024})}\BibitemShut {NoStop}%
\bibitem [{\citenamefont {Oliver~Buchmueller}\ and\ \citenamefont
  {Schneider}(2023)}]{buchmueller2023largescale}%
  \BibitemOpen
  \bibfield  {author} {\bibinfo {author} {\bibfnamefont {J.~E.}\ \bibnamefont
  {Oliver~Buchmueller}}\ and\ \bibinfo {author} {\bibfnamefont
  {U.}~\bibnamefont {Schneider}},\ }\bibfield  {title} {\bibinfo {title}
  {Large-scale atom interferometry for fundamental physics},\ }\href
  {https://doi.org/10.1080/00107514.2023.2239008} {\bibfield  {journal}
  {\bibinfo  {journal} {Contemp. Phys.}\ }\textbf {\bibinfo {volume} {64}},\
  \bibinfo {pages} {93} (\bibinfo {year} {2023})}\BibitemShut {NoStop}%
\bibitem [{\citenamefont {Bose}\ \emph {et~al.}(2025)\citenamefont {Bose},
  \citenamefont {Fuentes}, \citenamefont {Geraci}, \citenamefont {Khan},
  \citenamefont {Qvarfort}, \citenamefont {Rademacher}, \citenamefont {Rashid},
  \citenamefont {Toro\ifmmode~\check{s}\else \v{s}\fi{}}, \citenamefont
  {Ulbricht},\ and\ \citenamefont
  {Wanjura}}]{bose2024massivequantumsystemsinterfaces}%
  \BibitemOpen
  \bibfield  {author} {\bibinfo {author} {\bibfnamefont {S.}~\bibnamefont
  {Bose}}, \bibinfo {author} {\bibfnamefont {I.}~\bibnamefont {Fuentes}},
  \bibinfo {author} {\bibfnamefont {A.~A.}\ \bibnamefont {Geraci}}, \bibinfo
  {author} {\bibfnamefont {S.~M.}\ \bibnamefont {Khan}}, \bibinfo {author}
  {\bibfnamefont {S.}~\bibnamefont {Qvarfort}}, \bibinfo {author}
  {\bibfnamefont {M.}~\bibnamefont {Rademacher}}, \bibinfo {author}
  {\bibfnamefont {M.}~\bibnamefont {Rashid}}, \bibinfo {author} {\bibfnamefont
  {M.}~\bibnamefont {Toro\ifmmode~\check{s}\else \v{s}\fi{}}}, \bibinfo
  {author} {\bibfnamefont {H.}~\bibnamefont {Ulbricht}},\ and\ \bibinfo
  {author} {\bibfnamefont {C.~C.}\ \bibnamefont {Wanjura}},\ }\bibfield
  {title} {\bibinfo {title} {Massive quantum systems as interfaces of quantum
  mechanics and gravity},\ }\href
  {https://doi.org/10.1103/RevModPhys.97.015003} {\bibfield  {journal}
  {\bibinfo  {journal} {Rev. Mod. Phys.}\ }\textbf {\bibinfo {volume} {97}},\
  \bibinfo {pages} {015003} (\bibinfo {year} {2025})}\BibitemShut {NoStop}%
\bibitem [{\citenamefont {Braun}\ \emph {et~al.}(2025)\citenamefont {Braun},
  \citenamefont {Marchese}, \citenamefont {Nimmrichter}, \citenamefont
  {Qvarfort}, \citenamefont {Rätzel},\ and\ \citenamefont
  {Ulbricht}}]{braun2025metrologygravitationaleffectsmechanical}%
  \BibitemOpen
  \bibfield  {author} {\bibinfo {author} {\bibfnamefont {D.}~\bibnamefont
  {Braun}}, \bibinfo {author} {\bibfnamefont {M.~M.}\ \bibnamefont {Marchese}},
  \bibinfo {author} {\bibfnamefont {S.}~\bibnamefont {Nimmrichter}}, \bibinfo
  {author} {\bibfnamefont {S.}~\bibnamefont {Qvarfort}}, \bibinfo {author}
  {\bibfnamefont {D.}~\bibnamefont {Rätzel}},\ and\ \bibinfo {author}
  {\bibfnamefont {H.}~\bibnamefont {Ulbricht}},\ }\href
  {https://arxiv.org/abs/2501.18274} {\bibinfo {title} {Metrology of
  gravitational effects with mechanical quantum systems}} (\bibinfo {year}
  {2025}),\ \Eprint {https://arxiv.org/abs/2501.18274} {arXiv:2501.18274
  [quant-ph]} \BibitemShut {NoStop}%
\bibitem [{\citenamefont {Sherrill}\ \emph {et~al.}(2023)\citenamefont
  {Sherrill}, \citenamefont {Parsons}, \citenamefont {Baynham}, \citenamefont
  {Bowden}, \citenamefont {Anne~Curtis}, \citenamefont {Hendricks},
  \citenamefont {Hill}, \citenamefont {Hobson}, \citenamefont {Margolis},
  \citenamefont {Robertson}, \citenamefont {Schioppo}, \citenamefont
  {Szymaniec}, \citenamefont {Tofful}, \citenamefont {Tunesi}, \citenamefont
  {Godun},\ and\ \citenamefont {Calmet}}]{sherrill2023}%
  \BibitemOpen
  \bibfield  {author} {\bibinfo {author} {\bibfnamefont {N.}~\bibnamefont
  {Sherrill}}, \bibinfo {author} {\bibfnamefont {A.~O.}\ \bibnamefont
  {Parsons}}, \bibinfo {author} {\bibfnamefont {C.~F.~A.}\ \bibnamefont
  {Baynham}}, \bibinfo {author} {\bibfnamefont {W.}~\bibnamefont {Bowden}},
  \bibinfo {author} {\bibfnamefont {E.}~\bibnamefont {Anne~Curtis}}, \bibinfo
  {author} {\bibfnamefont {R.}~\bibnamefont {Hendricks}}, \bibinfo {author}
  {\bibfnamefont {I.~R.}\ \bibnamefont {Hill}}, \bibinfo {author}
  {\bibfnamefont {R.}~\bibnamefont {Hobson}}, \bibinfo {author} {\bibfnamefont
  {H.~S.}\ \bibnamefont {Margolis}}, \bibinfo {author} {\bibfnamefont {B.~I.}\
  \bibnamefont {Robertson}}, \bibinfo {author} {\bibfnamefont {M.}~\bibnamefont
  {Schioppo}}, \bibinfo {author} {\bibfnamefont {K.}~\bibnamefont {Szymaniec}},
  \bibinfo {author} {\bibfnamefont {A.}~\bibnamefont {Tofful}}, \bibinfo
  {author} {\bibfnamefont {J.}~\bibnamefont {Tunesi}}, \bibinfo {author}
  {\bibfnamefont {R.~M.}\ \bibnamefont {Godun}},\ and\ \bibinfo {author}
  {\bibfnamefont {X.}~\bibnamefont {Calmet}},\ }\bibfield  {title} {\bibinfo
  {title} {Analysis of atomic-clock data to constrain variations of fundamental
  constants},\ }\href {https://doi.org/10.1088/1367-2630/aceff6} {\bibfield
  {journal} {\bibinfo  {journal} {New J. Phys.}\ }\textbf {\bibinfo {volume}
  {25}},\ \bibinfo {pages} {093012} (\bibinfo {year} {2023})}\BibitemShut
  {NoStop}%
\bibitem [{\citenamefont {Amaral}\ \emph {et~al.}(2024)\citenamefont {Amaral},
  \citenamefont {Jain}, \citenamefont {Amin},\ and\ \citenamefont
  {Tunnell}}]{amaral2024vector}%
  \BibitemOpen
  \bibfield  {author} {\bibinfo {author} {\bibfnamefont {D.~W.}\ \bibnamefont
  {Amaral}}, \bibinfo {author} {\bibfnamefont {M.}~\bibnamefont {Jain}},
  \bibinfo {author} {\bibfnamefont {M.~A.}\ \bibnamefont {Amin}},\ and\
  \bibinfo {author} {\bibfnamefont {C.}~\bibnamefont {Tunnell}},\ }\bibfield
  {title} {\bibinfo {title} {Vector wave dark matter and terrestrial quantum
  sensors},\ }\href {https://doi.org/10.1088/1475-7516/2024/06/050} {\bibfield
  {journal} {\bibinfo  {journal} {J. Cosmol. Astropart. Phys.}\ }\textbf
  {\bibinfo {volume} {2024}}\bibinfo  {number} { (06)},\ \bibinfo {pages}
  {050}}\BibitemShut {NoStop}%
\bibitem [{\citenamefont {Amaral}\ \emph {et~al.}(2025)\citenamefont {Amaral},
  \citenamefont {Uitenbroek}, \citenamefont {Oosterkamp},\ and\ \citenamefont
  {Tunnell}}]{amaral2024first}%
  \BibitemOpen
\bibfield  {number} {  }\bibfield  {author} {\bibinfo {author} {\bibfnamefont
  {D.~W.~P.}\ \bibnamefont {Amaral}}, \bibinfo {author} {\bibfnamefont {D.~G.}\
  \bibnamefont {Uitenbroek}}, \bibinfo {author} {\bibfnamefont {T.~H.}\
  \bibnamefont {Oosterkamp}},\ and\ \bibinfo {author} {\bibfnamefont {C.~D.}\
  \bibnamefont {Tunnell}},\ }\bibfield  {title} {\bibinfo {title} {First search
  for ultralight dark matter using a magnetically levitated particle},\ }\href
  {https://doi.org/10.1103/PhysRevLett.134.251001} {\bibfield  {journal}
  {\bibinfo  {journal} {Phys. Rev. Lett.}\ }\textbf {\bibinfo {volume} {134}},\
  \bibinfo {pages} {251001} (\bibinfo {year} {2025})}\BibitemShut {NoStop}%
\bibitem [{\citenamefont {Ralph}\ \emph {et~al.}(2018)\citenamefont {Ralph},
  \citenamefont {Toro\ifmmode~\check{s}\else \v{s}\fi{}}, \citenamefont
  {Maskell}, \citenamefont {Jacobs}, \citenamefont {Rashid}, \citenamefont
  {Setter},\ and\ \citenamefont {Ulbricht}}]{ralph2018dynamical}%
  \BibitemOpen
  \bibfield  {author} {\bibinfo {author} {\bibfnamefont {J.~F.}\ \bibnamefont
  {Ralph}}, \bibinfo {author} {\bibfnamefont {M.}~\bibnamefont
  {Toro\ifmmode~\check{s}\else \v{s}\fi{}}}, \bibinfo {author} {\bibfnamefont
  {S.}~\bibnamefont {Maskell}}, \bibinfo {author} {\bibfnamefont
  {K.}~\bibnamefont {Jacobs}}, \bibinfo {author} {\bibfnamefont
  {M.}~\bibnamefont {Rashid}}, \bibinfo {author} {\bibfnamefont {A.~J.}\
  \bibnamefont {Setter}},\ and\ \bibinfo {author} {\bibfnamefont
  {H.}~\bibnamefont {Ulbricht}},\ }\bibfield  {title} {\bibinfo {title}
  {Dynamical model selection near the quantum-classical boundary},\ }\href
  {https://doi.org/10.1103/PhysRevA.98.010102} {\bibfield  {journal} {\bibinfo
  {journal} {Phys. Rev. A}\ }\textbf {\bibinfo {volume} {98}},\ \bibinfo
  {pages} {010102} (\bibinfo {year} {2018})}\BibitemShut {NoStop}%
\bibitem [{\citenamefont {Schrinski}\ \emph {et~al.}(2019)\citenamefont
  {Schrinski}, \citenamefont {Nimmrichter}, \citenamefont {Stickler},\ and\
  \citenamefont {Hornberger}}]{schrinski2020macroscopicity}%
  \BibitemOpen
  \bibfield  {author} {\bibinfo {author} {\bibfnamefont {B.}~\bibnamefont
  {Schrinski}}, \bibinfo {author} {\bibfnamefont {S.}~\bibnamefont
  {Nimmrichter}}, \bibinfo {author} {\bibfnamefont {B.~A.}\ \bibnamefont
  {Stickler}},\ and\ \bibinfo {author} {\bibfnamefont {K.}~\bibnamefont
  {Hornberger}},\ }\bibfield  {title} {\bibinfo {title} {Macroscopicity of
  quantum mechanical superposition tests via hypothesis falsification},\ }\href
  {https://doi.org/10.1103/PhysRevA.100.032111} {\bibfield  {journal} {\bibinfo
   {journal} {Phys. Rev. A}\ }\textbf {\bibinfo {volume} {100}},\ \bibinfo
  {pages} {032111} (\bibinfo {year} {2019})}\BibitemShut {NoStop}%
\bibitem [{\citenamefont {Schrinski}\ \emph {et~al.}(2020)\citenamefont
  {Schrinski}, \citenamefont {Nimmrichter},\ and\ \citenamefont
  {Hornberger}}]{schrinski2020quantumclassical}%
  \BibitemOpen
  \bibfield  {author} {\bibinfo {author} {\bibfnamefont {B.}~\bibnamefont
  {Schrinski}}, \bibinfo {author} {\bibfnamefont {S.}~\bibnamefont
  {Nimmrichter}},\ and\ \bibinfo {author} {\bibfnamefont {K.}~\bibnamefont
  {Hornberger}},\ }\bibfield  {title} {\bibinfo {title} {Quantum-classical
  hypothesis tests in macroscopic matter-wave interferometry},\ }\href
  {https://doi.org/10.1103/PhysRevResearch.2.033034} {\bibfield  {journal}
  {\bibinfo  {journal} {Phys. Rev. Res.}\ }\textbf {\bibinfo {volume} {2}},\
  \bibinfo {pages} {033034} (\bibinfo {year} {2020})}\BibitemShut {NoStop}%
\bibitem [{\citenamefont {Schrinski}\ \emph
  {et~al.}(2023{\natexlab{a}})\citenamefont {Schrinski}, \citenamefont
  {Haslinger}, \citenamefont {Schmiedmayer}, \citenamefont {Hornberger},\ and\
  \citenamefont {Nimmrichter}}]{schrinski2023testing}%
  \BibitemOpen
  \bibfield  {author} {\bibinfo {author} {\bibfnamefont {B.}~\bibnamefont
  {Schrinski}}, \bibinfo {author} {\bibfnamefont {P.}~\bibnamefont
  {Haslinger}}, \bibinfo {author} {\bibfnamefont {J.}~\bibnamefont
  {Schmiedmayer}}, \bibinfo {author} {\bibfnamefont {K.}~\bibnamefont
  {Hornberger}},\ and\ \bibinfo {author} {\bibfnamefont {S.}~\bibnamefont
  {Nimmrichter}},\ }\bibfield  {title} {\bibinfo {title} {Testing collapse
  models with bose-einstein-condensate interferometry},\ }\href
  {https://doi.org/10.1103/PhysRevA.107.043320} {\bibfield  {journal} {\bibinfo
   {journal} {Phys. Rev. A}\ }\textbf {\bibinfo {volume} {107}},\ \bibinfo
  {pages} {043320} (\bibinfo {year} {2023}{\natexlab{a}})}\BibitemShut
  {NoStop}%
\bibitem [{\citenamefont {Schrinski}\ \emph
  {et~al.}(2023{\natexlab{b}})\citenamefont {Schrinski}, \citenamefont {Yang},
  \citenamefont {von L\"upke}, \citenamefont {Bild}, \citenamefont {Chu},
  \citenamefont {Hornberger}, \citenamefont {Nimmrichter},\ and\ \citenamefont
  {Fadel}}]{schrinski2023macroscopic}%
  \BibitemOpen
  \bibfield  {author} {\bibinfo {author} {\bibfnamefont {B.}~\bibnamefont
  {Schrinski}}, \bibinfo {author} {\bibfnamefont {Y.}~\bibnamefont {Yang}},
  \bibinfo {author} {\bibfnamefont {U.}~\bibnamefont {von L\"upke}}, \bibinfo
  {author} {\bibfnamefont {M.}~\bibnamefont {Bild}}, \bibinfo {author}
  {\bibfnamefont {Y.}~\bibnamefont {Chu}}, \bibinfo {author} {\bibfnamefont
  {K.}~\bibnamefont {Hornberger}}, \bibinfo {author} {\bibfnamefont
  {S.}~\bibnamefont {Nimmrichter}},\ and\ \bibinfo {author} {\bibfnamefont
  {M.}~\bibnamefont {Fadel}},\ }\bibfield  {title} {\bibinfo {title}
  {Macroscopic quantum test with bulk acoustic wave resonators},\ }\href
  {https://doi.org/10.1103/PhysRevLett.130.133604} {\bibfield  {journal}
  {\bibinfo  {journal} {Phys. Rev. Lett.}\ }\textbf {\bibinfo {volume} {130}},\
  \bibinfo {pages} {133604} (\bibinfo {year} {2023}{\natexlab{b}})}\BibitemShut
  {NoStop}%
\bibitem [{\citenamefont {Laing}\ and\ \citenamefont
  {Bateman}(2024)}]{laing2024bayesian}%
  \BibitemOpen
  \bibfield  {author} {\bibinfo {author} {\bibfnamefont {S.}~\bibnamefont
  {Laing}}\ and\ \bibinfo {author} {\bibfnamefont {J.}~\bibnamefont
  {Bateman}},\ }\bibfield  {title} {\bibinfo {title} {Bayesian inference for
  near-field interferometric tests of collapse models},\ }\href
  {https://doi.org/10.1103/PhysRevA.110.012214} {\bibfield  {journal} {\bibinfo
   {journal} {Phys. Rev. A}\ }\textbf {\bibinfo {volume} {110}},\ \bibinfo
  {pages} {012214} (\bibinfo {year} {2024})}\BibitemShut {NoStop}%
\bibitem [{\citenamefont {Jaynes}(2003)}]{jaynes2003probability}%
  \BibitemOpen
  \bibfield  {author} {\bibinfo {author} {\bibfnamefont {E.~T.}\ \bibnamefont
  {Jaynes}},\ }\href {https://doi.org/10.1017/CBO9780511790423} {\emph
  {\bibinfo {title} {{Probability Theory: The Logic of Science}}}}\ (\bibinfo
  {publisher} {Cambridge University Press},\ \bibinfo {year}
  {2003})\BibitemShut {NoStop}%
\bibitem [{\citenamefont {von Toussaint}(2011)}]{toussaint2011bayesian}%
  \BibitemOpen
  \bibfield  {author} {\bibinfo {author} {\bibfnamefont {U.}~\bibnamefont {von
  Toussaint}},\ }\bibfield  {title} {\bibinfo {title} {Bayesian inference in
  physics},\ }\href {https://doi.org/10.1103/RevModPhys.83.943} {\bibfield
  {journal} {\bibinfo  {journal} {Rev. Mod. Phys.}\ }\textbf {\bibinfo {volume}
  {83}},\ \bibinfo {pages} {943} (\bibinfo {year} {2011})}\BibitemShut
  {NoStop}%
\bibitem [{\citenamefont {Helstrom}(1976)}]{helstrom1976quantum}%
  \BibitemOpen
  \bibfield  {author} {\bibinfo {author} {\bibfnamefont {C.~W.}\ \bibnamefont
  {Helstrom}},\ }\href@noop {} {\emph {\bibinfo {title} {{Quantum Detection and
  Estimation Theory}}}}\ (\bibinfo  {publisher} {Academic Press},\ \bibinfo
  {address} {New York},\ \bibinfo {year} {1976})\BibitemShut {NoStop}%
\bibitem [{\citenamefont {Tsang}(2012)}]{tsang2012zivzakai}%
  \BibitemOpen
  \bibfield  {author} {\bibinfo {author} {\bibfnamefont {M.}~\bibnamefont
  {Tsang}},\ }\bibfield  {title} {\bibinfo {title} {{Ziv-Zakai Error Bounds for
  Quantum Parameter Estimation}},\ }\href
  {https://doi.org/10.1103/PhysRevLett.108.230401} {\bibfield  {journal}
  {\bibinfo  {journal} {Phys. Rev. Lett.}\ }\textbf {\bibinfo {volume} {108}},\
  \bibinfo {pages} {230401} (\bibinfo {year} {2012})}\BibitemShut {NoStop}%
\bibitem [{\citenamefont {Lu}\ and\ \citenamefont
  {Tsang}(2016)}]{tsang2016quantum}%
  \BibitemOpen
  \bibfield  {author} {\bibinfo {author} {\bibfnamefont {X.-M.}\ \bibnamefont
  {Lu}}\ and\ \bibinfo {author} {\bibfnamefont {M.}~\bibnamefont {Tsang}},\
  }\bibfield  {title} {\bibinfo {title} {{Quantum Weiss-Weinstein bounds for
  quantum metrology}},\ }\href {http://stacks.iop.org/2058-9565/1/i=1/a=015002}
  {\bibfield  {journal} {\bibinfo  {journal} {Quantum Sci. Technol.}\ }\textbf
  {\bibinfo {volume} {1}},\ \bibinfo {pages} {015002} (\bibinfo {year}
  {2016})}\BibitemShut {NoStop}%
\bibitem [{\citenamefont {Mehboudi}\ \emph {et~al.}(2022)\citenamefont
  {Mehboudi}, \citenamefont {J\o{}rgensen}, \citenamefont {Seah}, \citenamefont
  {Brask}, \citenamefont {Ko\l{}ody\ifmmode~\acute{n}\else \'{n}\fi{}ski},\
  and\ \citenamefont {Perarnau-Llobet}}]{mehboudi2021fundamental}%
  \BibitemOpen
  \bibfield  {author} {\bibinfo {author} {\bibfnamefont {M.}~\bibnamefont
  {Mehboudi}}, \bibinfo {author} {\bibfnamefont {M.~R.}\ \bibnamefont
  {J\o{}rgensen}}, \bibinfo {author} {\bibfnamefont {S.}~\bibnamefont {Seah}},
  \bibinfo {author} {\bibfnamefont {J.~B.}\ \bibnamefont {Brask}}, \bibinfo
  {author} {\bibfnamefont {J.}~\bibnamefont {Ko\l{}ody\ifmmode~\acute{n}\else
  \'{n}\fi{}ski}},\ and\ \bibinfo {author} {\bibfnamefont {M.}~\bibnamefont
  {Perarnau-Llobet}},\ }\bibfield  {title} {\bibinfo {title} {{Fundamental
  Limits in Bayesian Thermometry and Attainability via Adaptive Strategies}},\
  }\href {https://doi.org/10.1103/PhysRevLett.128.130502} {\bibfield  {journal}
  {\bibinfo  {journal} {Phys. Rev. Lett.}\ }\textbf {\bibinfo {volume} {128}},\
  \bibinfo {pages} {130502} (\bibinfo {year} {2022})}\BibitemShut {NoStop}%
\bibitem [{\citenamefont {Li}\ \emph {et~al.}(2018{\natexlab{a}})\citenamefont
  {Li}, \citenamefont {Pezz\`{e}}, \citenamefont {Gessner}, \citenamefont
  {Ren}, \citenamefont {Li},\ and\ \citenamefont
  {Smerzi}}]{yan2018frequentist}%
  \BibitemOpen
  \bibfield  {author} {\bibinfo {author} {\bibfnamefont {Y.}~\bibnamefont
  {Li}}, \bibinfo {author} {\bibfnamefont {L.}~\bibnamefont {Pezz\`{e}}},
  \bibinfo {author} {\bibfnamefont {M.}~\bibnamefont {Gessner}}, \bibinfo
  {author} {\bibfnamefont {Z.}~\bibnamefont {Ren}}, \bibinfo {author}
  {\bibfnamefont {W.}~\bibnamefont {Li}},\ and\ \bibinfo {author}
  {\bibfnamefont {A.}~\bibnamefont {Smerzi}},\ }\bibfield  {title} {\bibinfo
  {title} {{Frequentist and Bayesian Quantum Phase Estimation}},\ }\href
  {https://doi.org/10.3390/e20090628} {\bibfield  {journal} {\bibinfo
  {journal} {Entropy}\ }\textbf {\bibinfo {volume} {20}},\ \bibinfo {pages}
  {628} (\bibinfo {year} {2018}{\natexlab{a}})}\BibitemShut {NoStop}%
\bibitem [{\citenamefont {Gessner}\ and\ \citenamefont
  {Smerzi}(2023)}]{gessner2023hierarchies}%
  \BibitemOpen
  \bibfield  {author} {\bibinfo {author} {\bibfnamefont {M.}~\bibnamefont
  {Gessner}}\ and\ \bibinfo {author} {\bibfnamefont {A.}~\bibnamefont
  {Smerzi}},\ }\bibfield  {title} {\bibinfo {title} {{Hierarchies of
  Frequentist Bounds for Quantum Metrology: From Cram\'er-Rao to Barankin}},\
  }\href {https://doi.org/10.1103/PhysRevLett.130.260801} {\bibfield  {journal}
  {\bibinfo  {journal} {Phys. Rev. Lett.}\ }\textbf {\bibinfo {volume} {130}},\
  \bibinfo {pages} {260801} (\bibinfo {year} {2023})}\BibitemShut {NoStop}%
\bibitem [{\citenamefont {Rubio}\ \emph {et~al.}(2018)\citenamefont {Rubio},
  \citenamefont {Knott},\ and\ \citenamefont {Dunningham}}]{rubio2018non}%
  \BibitemOpen
  \bibfield  {author} {\bibinfo {author} {\bibfnamefont {J.}~\bibnamefont
  {Rubio}}, \bibinfo {author} {\bibfnamefont {P.}~\bibnamefont {Knott}},\ and\
  \bibinfo {author} {\bibfnamefont {J.}~\bibnamefont {Dunningham}},\ }\bibfield
   {title} {\bibinfo {title} {{Non-asymptotic analysis of quantum metrology
  protocols beyond the Cramér–Rao bound}},\ }\href
  {https://doi.org/10.1088/2399-6528/aaa234} {\bibfield  {journal} {\bibinfo
  {journal} {J. Phys. Commun.}\ }\textbf {\bibinfo {volume} {2}},\ \bibinfo
  {pages} {015027} (\bibinfo {year} {2018})}\BibitemShut {NoStop}%
\bibitem [{\citenamefont {Salmon}\ \emph {et~al.}(2023)\citenamefont {Salmon},
  \citenamefont {Strelchuk},\ and\ \citenamefont
  {Arvidsson-Shukur}}]{salmon2023only}%
  \BibitemOpen
  \bibfield  {author} {\bibinfo {author} {\bibfnamefont {W.}~\bibnamefont
  {Salmon}}, \bibinfo {author} {\bibfnamefont {S.}~\bibnamefont {Strelchuk}},\
  and\ \bibinfo {author} {\bibfnamefont {D.}~\bibnamefont {Arvidsson-Shukur}},\
  }\bibfield  {title} {\bibinfo {title} {Only {C}lassical {P}arameterised
  {S}tates have {O}ptimal {M}easurements under {L}east {S}quares {L}oss},\
  }\href {https://doi.org/10.22331/q-2023-05-11-998} {\bibfield  {journal}
  {\bibinfo  {journal} {{Quantum}}\ }\textbf {\bibinfo {volume} {7}},\ \bibinfo
  {pages} {998} (\bibinfo {year} {2023})}\BibitemShut {NoStop}%
\bibitem [{\citenamefont {Law}\ and\ \citenamefont
  {Bern\'ad}(2025)}]{dennis2024}%
  \BibitemOpen
  \bibfield  {author} {\bibinfo {author} {\bibfnamefont {C.~K.~D.}\
  \bibnamefont {Law}}\ and\ \bibinfo {author} {\bibfnamefont {J.~Z.}\
  \bibnamefont {Bern\'ad}},\ }\bibfield  {title} {\bibinfo {title} {Bayesian
  and frequentist estimators for the transition frequency of a driven two-level
  quantum system},\ }\href {https://doi.org/10.1103/PhysRevA.111.042218}
  {\bibfield  {journal} {\bibinfo  {journal} {Phys. Rev. A}\ }\textbf {\bibinfo
  {volume} {111}},\ \bibinfo {pages} {042218} (\bibinfo {year}
  {2025})}\BibitemShut {NoStop}%
\bibitem [{\citenamefont {Bavaresco}\ \emph
  {et~al.}(2024{\natexlab{b}})\citenamefont {Bavaresco}, \citenamefont
  {Lipka-Bartosik}, \citenamefont {Sekatski},\ and\ \citenamefont
  {Mehboudi}}]{bavaresco2024designing}%
  \BibitemOpen
  \bibfield  {author} {\bibinfo {author} {\bibfnamefont {J.}~\bibnamefont
  {Bavaresco}}, \bibinfo {author} {\bibfnamefont {P.}~\bibnamefont
  {Lipka-Bartosik}}, \bibinfo {author} {\bibfnamefont {P.}~\bibnamefont
  {Sekatski}},\ and\ \bibinfo {author} {\bibfnamefont {M.}~\bibnamefont
  {Mehboudi}},\ }\bibfield  {title} {\bibinfo {title} {{Designing optimal
  protocols in Bayesian quantum parameter estimation with higher-order
  operations}},\ }\href {https://doi.org/10.1103/PhysRevResearch.6.023305}
  {\bibfield  {journal} {\bibinfo  {journal} {Phys. Rev. Res.}\ }\textbf
  {\bibinfo {volume} {6}},\ \bibinfo {pages} {023305} (\bibinfo {year}
  {2024}{\natexlab{b}})}\BibitemShut {NoStop}%
\bibitem [{\citenamefont {Kurdziałek}\ \emph {et~al.}(2025)\citenamefont
  {Kurdziałek}, \citenamefont {Dulian}, \citenamefont {Majsak}, \citenamefont
  {Chakraborty},\ and\ \citenamefont
  {Demkowicz-Dobrzański}}]{kurdzialek2024quantum}%
  \BibitemOpen
  \bibfield  {author} {\bibinfo {author} {\bibfnamefont {S.}~\bibnamefont
  {Kurdziałek}}, \bibinfo {author} {\bibfnamefont {P.}~\bibnamefont {Dulian}},
  \bibinfo {author} {\bibfnamefont {J.}~\bibnamefont {Majsak}}, \bibinfo
  {author} {\bibfnamefont {S.}~\bibnamefont {Chakraborty}},\ and\ \bibinfo
  {author} {\bibfnamefont {R.}~\bibnamefont {Demkowicz-Dobrzański}},\
  }\bibfield  {title} {\bibinfo {title} {Quantum metrology using quantum combs
  and tensor network formalism},\ }\href
  {https://doi.org/10.1088/1367-2630/ada8d1} {\bibfield  {journal} {\bibinfo
  {journal} {New J. Phys.}\ }\textbf {\bibinfo {volume} {27}},\ \bibinfo
  {pages} {013019} (\bibinfo {year} {2025})}\BibitemShut {NoStop}%
\bibitem [{\citenamefont {Rubio}(2024)}]{rubio2024first}%
  \BibitemOpen
  \bibfield  {author} {\bibinfo {author} {\bibfnamefont {J.}~\bibnamefont
  {Rubio}},\ }\bibfield  {title} {\bibinfo {title} {First-principles
  construction of symmetry-informed quantum metrologies},\ }\href
  {https://doi.org/10.1103/PhysRevA.110.L030401} {\bibfield  {journal}
  {\bibinfo  {journal} {Phys. Rev. A}\ }\textbf {\bibinfo {volume} {110}},\
  \bibinfo {pages} {L030401} (\bibinfo {year} {2024})}\BibitemShut {NoStop}%
\bibitem [{\citenamefont {Overton}\ \emph {et~al.}(2024)\citenamefont
  {Overton}, \citenamefont {Rubio}, \citenamefont {Cooper}, \citenamefont
  {Baldolini}, \citenamefont {Johnson}, \citenamefont {Anders},\ and\
  \citenamefont {Hackermüller}}]{overton2024five}%
  \BibitemOpen
  \bibfield  {author} {\bibinfo {author} {\bibfnamefont {M.}~\bibnamefont
  {Overton}}, \bibinfo {author} {\bibfnamefont {J.}~\bibnamefont {Rubio}},
  \bibinfo {author} {\bibfnamefont {N.}~\bibnamefont {Cooper}}, \bibinfo
  {author} {\bibfnamefont {D.}~\bibnamefont {Baldolini}}, \bibinfo {author}
  {\bibfnamefont {D.}~\bibnamefont {Johnson}}, \bibinfo {author} {\bibfnamefont
  {J.}~\bibnamefont {Anders}},\ and\ \bibinfo {author} {\bibfnamefont
  {L.}~\bibnamefont {Hackermüller}},\ }\href
  {https://arxiv.org/abs/2410.10615} {\bibinfo {title} {{Substantial precision
  enhancements via adaptive symmetry-informed Bayesian metrology}}},\ \bibinfo
  {howpublished} {arXiv:2410.10615} (\bibinfo {year} {2024})\BibitemShut
  {NoStop}%
\bibitem [{\citenamefont {Linden}\ \emph {et~al.}(2014)\citenamefont {Linden},
  \citenamefont {Dose},\ and\ \citenamefont {Toussaint}}]{linden2014bayesian}%
  \BibitemOpen
  \bibfield  {author} {\bibinfo {author} {\bibfnamefont {W.~v.~d.}\
  \bibnamefont {Linden}}, \bibinfo {author} {\bibfnamefont {V.}~\bibnamefont
  {Dose}},\ and\ \bibinfo {author} {\bibfnamefont {U.~v.}\ \bibnamefont
  {Toussaint}},\ }\href@noop {} {\emph {\bibinfo {title} {Bayesian Probability
  Theory: Applications in the Physical Sciences}}}\ (\bibinfo  {publisher}
  {Cambridge University Press},\ \bibinfo {year} {2014})\BibitemShut {NoStop}%
\bibitem [{\citenamefont {Personick}(1971)}]{personick1971application}%
  \BibitemOpen
  \bibfield  {author} {\bibinfo {author} {\bibfnamefont {S.}~\bibnamefont
  {Personick}},\ }\bibfield  {title} {\bibinfo {title} {Application of quantum
  estimation theory to analog communication over quantum channels},\ }\href
  {https://doi.org/10.1109/TIT.1971.1054643} {\bibfield  {journal} {\bibinfo
  {journal} {IEEE Trans. Inf. Theory}\ }\textbf {\bibinfo {volume} {17}},\
  \bibinfo {pages} {240} (\bibinfo {year} {1971})}\BibitemShut {NoStop}%
\bibitem [{\citenamefont {Rubio}\ \emph {et~al.}(2021)\citenamefont {Rubio},
  \citenamefont {Anders},\ and\ \citenamefont {Correa}}]{rubio2021global}%
  \BibitemOpen
  \bibfield  {author} {\bibinfo {author} {\bibfnamefont {J.}~\bibnamefont
  {Rubio}}, \bibinfo {author} {\bibfnamefont {J.}~\bibnamefont {Anders}},\ and\
  \bibinfo {author} {\bibfnamefont {L.~A.}\ \bibnamefont {Correa}},\ }\bibfield
   {title} {\bibinfo {title} {Global quantum thermometry},\ }\href
  {https://doi.org/10.1103/PhysRevLett.127.190402} {\bibfield  {journal}
  {\bibinfo  {journal} {Phys. Rev. Lett.}\ }\textbf {\bibinfo {volume} {127}},\
  \bibinfo {pages} {190402} (\bibinfo {year} {2021})}\BibitemShut {NoStop}%
\bibitem [{\citenamefont {Boeyens}\ \emph {et~al.}(2021)\citenamefont
  {Boeyens}, \citenamefont {Seah},\ and\ \citenamefont
  {Nimmrichter}}]{boeyens2021uninformed}%
  \BibitemOpen
  \bibfield  {author} {\bibinfo {author} {\bibfnamefont {J.}~\bibnamefont
  {Boeyens}}, \bibinfo {author} {\bibfnamefont {S.}~\bibnamefont {Seah}},\ and\
  \bibinfo {author} {\bibfnamefont {S.}~\bibnamefont {Nimmrichter}},\
  }\bibfield  {title} {\bibinfo {title} {Uninformed bayesian quantum
  thermometry},\ }\href {https://doi.org/10.1103/PhysRevA.104.052214}
  {\bibfield  {journal} {\bibinfo  {journal} {Phys. Rev. A}\ }\textbf {\bibinfo
  {volume} {104}},\ \bibinfo {pages} {052214} (\bibinfo {year}
  {2021})}\BibitemShut {NoStop}%
\bibitem [{\citenamefont {Chang}\ \emph {et~al.}(2024)\citenamefont {Chang},
  \citenamefont {Yan}, \citenamefont {Wang}, \citenamefont {Ye}, \citenamefont
  {Rao}, \citenamefont {Zhang}, \citenamefont {Huang}, \citenamefont {Luo},
  \citenamefont {Chen}, \citenamefont {Ma},\ and\ \citenamefont
  {Gao}}]{chang2024global}%
  \BibitemOpen
  \bibfield  {author} {\bibinfo {author} {\bibfnamefont {S.}~\bibnamefont
  {Chang}}, \bibinfo {author} {\bibfnamefont {Y.}~\bibnamefont {Yan}}, \bibinfo
  {author} {\bibfnamefont {L.}~\bibnamefont {Wang}}, \bibinfo {author}
  {\bibfnamefont {W.}~\bibnamefont {Ye}}, \bibinfo {author} {\bibfnamefont
  {X.}~\bibnamefont {Rao}}, \bibinfo {author} {\bibfnamefont {H.}~\bibnamefont
  {Zhang}}, \bibinfo {author} {\bibfnamefont {L.}~\bibnamefont {Huang}},
  \bibinfo {author} {\bibfnamefont {M.}~\bibnamefont {Luo}}, \bibinfo {author}
  {\bibfnamefont {Y.}~\bibnamefont {Chen}}, \bibinfo {author} {\bibfnamefont
  {Q.}~\bibnamefont {Ma}},\ and\ \bibinfo {author} {\bibfnamefont
  {S.}~\bibnamefont {Gao}},\ }\bibfield  {title} {\bibinfo {title} {Global
  quantum thermometry based on the optimal biased bound},\ }\href
  {https://doi.org/10.1103/PhysRevResearch.6.043171} {\bibfield  {journal}
  {\bibinfo  {journal} {Phys. Rev. Res.}\ }\textbf {\bibinfo {volume} {6}},\
  \bibinfo {pages} {043171} (\bibinfo {year} {2024})}\BibitemShut {NoStop}%
\bibitem [{\citenamefont {Mok}\ \emph {et~al.}(2021)\citenamefont {Mok},
  \citenamefont {Bharti}, \citenamefont {Kwek},\ and\ \citenamefont
  {Bayat}}]{mok2021optimal}%
  \BibitemOpen
  \bibfield  {author} {\bibinfo {author} {\bibfnamefont {W.-K.}\ \bibnamefont
  {Mok}}, \bibinfo {author} {\bibfnamefont {K.}~\bibnamefont {Bharti}},
  \bibinfo {author} {\bibfnamefont {L.-C.}\ \bibnamefont {Kwek}},\ and\
  \bibinfo {author} {\bibfnamefont {A.}~\bibnamefont {Bayat}},\ }\bibfield
  {title} {\bibinfo {title} {Optimal probes for global quantum thermometry},\
  }\href {https://doi.org/https://doi.org/10.1038/s42005-021-00572-w}
  {\bibfield  {journal} {\bibinfo  {journal} {Commun. Phys.}\ }\textbf
  {\bibinfo {volume} {4}},\ \bibinfo {pages} {62} (\bibinfo {year}
  {2021})}\BibitemShut {NoStop}%
\bibitem [{\citenamefont {Alves}\ and\ \citenamefont
  {Landi}(2022)}]{alves2022}%
  \BibitemOpen
  \bibfield  {author} {\bibinfo {author} {\bibfnamefont {G.~O.}\ \bibnamefont
  {Alves}}\ and\ \bibinfo {author} {\bibfnamefont {G.~T.}\ \bibnamefont
  {Landi}},\ }\bibfield  {title} {\bibinfo {title} {Bayesian estimation for
  collisional thermometry},\ }\href
  {https://doi.org/10.1103/PhysRevA.105.012212} {\bibfield  {journal} {\bibinfo
   {journal} {Phys. Rev. A}\ }\textbf {\bibinfo {volume} {105}},\ \bibinfo
  {pages} {012212} (\bibinfo {year} {2022})}\BibitemShut {NoStop}%
\bibitem [{\citenamefont {Mukhopadhyay}\ \emph {et~al.}(2025)\citenamefont
  {Mukhopadhyay}, \citenamefont {Montenegro},\ and\ \citenamefont
  {Bayat}}]{mukhopadhyay2025current}%
  \BibitemOpen
  \bibfield  {author} {\bibinfo {author} {\bibfnamefont {C.}~\bibnamefont
  {Mukhopadhyay}}, \bibinfo {author} {\bibfnamefont {V.}~\bibnamefont
  {Montenegro}},\ and\ \bibinfo {author} {\bibfnamefont {A.}~\bibnamefont
  {Bayat}},\ }\bibfield  {title} {\bibinfo {title} {Current trends in global
  quantum metrology},\ }\href {https://doi.org/10.1088/1751-8121/adb112}
  {\bibfield  {journal} {\bibinfo  {journal} {J. Phys. A: Math. Theor.}\
  }\textbf {\bibinfo {volume} {58}},\ \bibinfo {pages} {063001} (\bibinfo
  {year} {2025})}\BibitemShut {NoStop}%
\bibitem [{\citenamefont {Rubio}(2022)}]{rubio2023quantum}%
  \BibitemOpen
  \bibfield  {author} {\bibinfo {author} {\bibfnamefont {J.}~\bibnamefont
  {Rubio}},\ }\bibfield  {title} {\bibinfo {title} {Quantum scale estimation},\
  }\href {https://doi.org/10.1088/2058-9565/aca04b} {\bibfield  {journal}
  {\bibinfo  {journal} {Quantum Sci. Technol.}\ }\textbf {\bibinfo {volume}
  {8}},\ \bibinfo {pages} {015009} (\bibinfo {year} {2022})}\BibitemShut
  {NoStop}%
\bibitem [{\citenamefont {J\o{}rgensen}\ \emph {et~al.}(2022)\citenamefont
  {J\o{}rgensen}, \citenamefont {Ko\l{}ody\ifmmode~\acute{n}\else
  \'{n}\fi{}ski}, \citenamefont {Mehboudi}, \citenamefont {Perarnau-Llobet},\
  and\ \citenamefont {Brask}}]{jorgensen2021bayesian}%
  \BibitemOpen
  \bibfield  {author} {\bibinfo {author} {\bibfnamefont {M.~R.}\ \bibnamefont
  {J\o{}rgensen}}, \bibinfo {author} {\bibfnamefont {J.}~\bibnamefont
  {Ko\l{}ody\ifmmode~\acute{n}\else \'{n}\fi{}ski}}, \bibinfo {author}
  {\bibfnamefont {M.}~\bibnamefont {Mehboudi}}, \bibinfo {author}
  {\bibfnamefont {M.}~\bibnamefont {Perarnau-Llobet}},\ and\ \bibinfo {author}
  {\bibfnamefont {J.~B.}\ \bibnamefont {Brask}},\ }\bibfield  {title} {\bibinfo
  {title} {Bayesian quantum thermometry based on thermodynamic length},\ }\href
  {https://doi.org/10.1103/PhysRevA.105.042601} {\bibfield  {journal} {\bibinfo
   {journal} {Phys. Rev. A}\ }\textbf {\bibinfo {volume} {105}},\ \bibinfo
  {pages} {042601} (\bibinfo {year} {2022})}\BibitemShut {NoStop}%
\bibitem [{\citenamefont {T.~Volkoff}(2025)}]{volkoff2025}%
  \BibitemOpen
  \bibfield  {author} {\bibinfo {author} {\bibfnamefont {G.~G.}\ \bibnamefont
  {T.~Volkoff}},\ }\href {https://arxiv.org/abs/2501.18673} {\bibinfo {title}
  {{Length scale estimation of excited quantum oscillators}}},\ \bibinfo
  {howpublished} {arXiv:2501.18673} (\bibinfo {year} {2025})\BibitemShut
  {NoStop}%
\bibitem [{\citenamefont {Glatthard}\ \emph {et~al.}(2022)\citenamefont
  {Glatthard}, \citenamefont {Rubio}, \citenamefont {Sawant}, \citenamefont
  {Hewitt}, \citenamefont {Barontini},\ and\ \citenamefont
  {Correa}}]{Glatthard2022}%
  \BibitemOpen
  \bibfield  {author} {\bibinfo {author} {\bibfnamefont {J.}~\bibnamefont
  {Glatthard}}, \bibinfo {author} {\bibfnamefont {J.}~\bibnamefont {Rubio}},
  \bibinfo {author} {\bibfnamefont {R.}~\bibnamefont {Sawant}}, \bibinfo
  {author} {\bibfnamefont {T.}~\bibnamefont {Hewitt}}, \bibinfo {author}
  {\bibfnamefont {G.}~\bibnamefont {Barontini}},\ and\ \bibinfo {author}
  {\bibfnamefont {L.~A.}\ \bibnamefont {Correa}},\ }\bibfield  {title}
  {\bibinfo {title} {Optimal cold atom thermometry using adaptive bayesian
  strategies},\ }\href {https://doi.org/10.1103/PRXQuantum.3.040330} {\bibfield
   {journal} {\bibinfo  {journal} {PRX Quantum}\ }\textbf {\bibinfo {volume}
  {3}},\ \bibinfo {pages} {040330} (\bibinfo {year} {2022})}\BibitemShut
  {NoStop}%
\bibitem [{\citenamefont {Hewitt}\ \emph {et~al.}(2024)\citenamefont {Hewitt},
  \citenamefont {Bertheas}, \citenamefont {Jain}, \citenamefont {Nishida},\
  and\ \citenamefont {Barontini}}]{hewitt2024controlling}%
  \BibitemOpen
  \bibfield  {author} {\bibinfo {author} {\bibfnamefont {T.}~\bibnamefont
  {Hewitt}}, \bibinfo {author} {\bibfnamefont {T.}~\bibnamefont {Bertheas}},
  \bibinfo {author} {\bibfnamefont {M.}~\bibnamefont {Jain}}, \bibinfo {author}
  {\bibfnamefont {Y.}~\bibnamefont {Nishida}},\ and\ \bibinfo {author}
  {\bibfnamefont {G.}~\bibnamefont {Barontini}},\ }\bibfield  {title} {\bibinfo
  {title} {Controlling the interactions in a cold atom quantum impurity
  system},\ }\href {https://doi.org/10.1088/2058-9565/ad4c91} {\bibfield
  {journal} {\bibinfo  {journal} {Quantum Sci. Technol.}\ }\textbf {\bibinfo
  {volume} {9}},\ \bibinfo {pages} {035039} (\bibinfo {year}
  {2024})}\BibitemShut {NoStop}%
\bibitem [{\citenamefont {Kass}\ and\ \citenamefont
  {Wasserman}(1996)}]{kass1996the}%
  \BibitemOpen
  \bibfield  {author} {\bibinfo {author} {\bibfnamefont {R.~E.}\ \bibnamefont
  {Kass}}\ and\ \bibinfo {author} {\bibfnamefont {L.}~\bibnamefont
  {Wasserman}},\ }\bibfield  {title} {\bibinfo {title} {{The Selection of Prior
  Distributions by Formal Rules}},\ }\href
  {https://doi.org/10.1080/01621459.1996.10477003} {\bibfield  {journal}
  {\bibinfo  {journal} {J. Am. Stat. Assoc.}\ }\textbf {\bibinfo {volume}
  {91}},\ \bibinfo {pages} {1343} (\bibinfo {year} {1996})}\BibitemShut
  {NoStop}%
\bibitem [{\citenamefont {Jaynes}(1968)}]{jaynes1968prior}%
  \BibitemOpen
  \bibfield  {author} {\bibinfo {author} {\bibfnamefont {E.~T.}\ \bibnamefont
  {Jaynes}},\ }\bibfield  {title} {\bibinfo {title} {{Prior Probabilities}},\
  }\href {https://ieeexplore.ieee.org/document/4082152} {\bibfield  {journal}
  {\bibinfo  {journal} {IEEE Trans. Cybern.}\ }\textbf {\bibinfo {volume}
  {4}},\ \bibinfo {pages} {227} (\bibinfo {year} {1968})}\BibitemShut {NoStop}%
\bibitem [{\citenamefont {Harney}(2003)}]{harney_bayesian_2003}%
  \BibitemOpen
  \bibfield  {author} {\bibinfo {author} {\bibfnamefont {H.~L.}\ \bibnamefont
  {Harney}},\ }\href {https://doi.org/10.1007/978-3-662-06006-3} {\emph
  {\bibinfo {title} {Bayesian {Inference}}}}\ (\bibinfo  {publisher}
  {Springer},\ \bibinfo {address} {Berlin, Heidelberg},\ \bibinfo {year}
  {2003})\BibitemShut {NoStop}%
\bibitem [{\citenamefont {Eaton}\ and\ \citenamefont
  {Sudderth}(2004)}]{eatonsudderth2004}%
  \BibitemOpen
  \bibfield  {author} {\bibinfo {author} {\bibfnamefont {M.~L.}\ \bibnamefont
  {Eaton}}\ and\ \bibinfo {author} {\bibfnamefont {W.~D.}\ \bibnamefont
  {Sudderth}},\ }\bibfield  {title} {\bibinfo {title} {Properties of right haar
  predictive inference},\ }\href {http://www.jstor.org/stable/25053376}
  {\bibfield  {journal} {\bibinfo  {journal} {Sankhyā: Indian J. Stat.
  (2003-2007)}\ }\textbf {\bibinfo {volume} {66}},\ \bibinfo {pages} {487}
  (\bibinfo {year} {2004})}\BibitemShut {NoStop}%
\bibitem [{\citenamefont {Tanaka}(2012)}]{tanaka2012}%
  \BibitemOpen
  \bibfield  {author} {\bibinfo {author} {\bibfnamefont {F.}~\bibnamefont
  {Tanaka}},\ }\bibfield  {title} {\bibinfo {title} {Noninformative prior in
  the quantum statistical model of pure states},\ }\href
  {https://doi.org/10.1103/PhysRevA.85.062305} {\bibfield  {journal} {\bibinfo
  {journal} {Phys. Rev. A}\ }\textbf {\bibinfo {volume} {85}},\ \bibinfo
  {pages} {062305} (\bibinfo {year} {2012})}\BibitemShut {NoStop}%
\bibitem [{\citenamefont {Demkowicz-Dobrzański}\ \emph
  {et~al.}(2020)\citenamefont {Demkowicz-Dobrzański}, \citenamefont
  {Górecki},\ and\ \citenamefont {Guţă}}]{demkowicz2020multi}%
  \BibitemOpen
  \bibfield  {author} {\bibinfo {author} {\bibfnamefont {R.}~\bibnamefont
  {Demkowicz-Dobrzański}}, \bibinfo {author} {\bibfnamefont {W.}~\bibnamefont
  {Górecki}},\ and\ \bibinfo {author} {\bibfnamefont {M.}~\bibnamefont
  {Guţă}},\ }\bibfield  {title} {\bibinfo {title} {Multi-parameter estimation
  beyond quantum fisher information},\ }\href
  {https://doi.org/10.1088/1751-8121/ab8ef3} {\bibfield  {journal} {\bibinfo
  {journal} {J. Phys. A: Math. Theor.}\ }\textbf {\bibinfo {volume} {53}},\
  \bibinfo {pages} {363001} (\bibinfo {year} {2020})}\BibitemShut {NoStop}%
\bibitem [{\citenamefont {Brody}\ and\ \citenamefont
  {Hughston}(1996)}]{brody1996geometry}%
  \BibitemOpen
  \bibfield  {author} {\bibinfo {author} {\bibfnamefont {D.~C.}\ \bibnamefont
  {Brody}}\ and\ \bibinfo {author} {\bibfnamefont {L.~P.}\ \bibnamefont
  {Hughston}},\ }\bibfield  {title} {\bibinfo {title} {Geometry of quantum
  statistical inference},\ }\href {https://doi.org/10.1103/PhysRevLett.77.2851}
  {\bibfield  {journal} {\bibinfo  {journal} {Phys. Rev. Lett.}\ }\textbf
  {\bibinfo {volume} {77}},\ \bibinfo {pages} {2851} (\bibinfo {year}
  {1996})}\BibitemShut {NoStop}%
\bibitem [{\citenamefont {Amari}(2016)}]{amari2016}%
  \BibitemOpen
  \bibfield  {author} {\bibinfo {author} {\bibfnamefont {S.-I.}\ \bibnamefont
  {Amari}},\ }\href@noop {} {\emph {\bibinfo {title} {Information Geometry and
  Its Applications}}},\ \bibinfo {edition} {1st}\ ed.,\ Applied mathematical
  sciences\ (\bibinfo  {publisher} {Springer},\ \bibinfo {address} {Tokyo,
  Japan},\ \bibinfo {year} {2016})\BibitemShut {NoStop}%
\bibitem [{\citenamefont {Goldberg}\ \emph
  {et~al.}(2021{\natexlab{a}})\citenamefont {Goldberg}, \citenamefont
  {S\'anchez-Soto},\ and\ \citenamefont {Ferretti}}]{goldberg2021}%
  \BibitemOpen
  \bibfield  {author} {\bibinfo {author} {\bibfnamefont {A.~Z.}\ \bibnamefont
  {Goldberg}}, \bibinfo {author} {\bibfnamefont {L.~L.}\ \bibnamefont
  {S\'anchez-Soto}},\ and\ \bibinfo {author} {\bibfnamefont {H.}~\bibnamefont
  {Ferretti}},\ }\bibfield  {title} {\bibinfo {title} {Intrinsic sensitivity
  limits for multiparameter quantum metrology},\ }\href
  {https://doi.org/10.1103/PhysRevLett.127.110501} {\bibfield  {journal}
  {\bibinfo  {journal} {Phys. Rev. Lett.}\ }\textbf {\bibinfo {volume} {127}},\
  \bibinfo {pages} {110501} (\bibinfo {year} {2021}{\natexlab{a}})}\BibitemShut
  {NoStop}%
\bibitem [{\citenamefont {Jermyn}(2005)}]{jermyn2005invariant}%
  \BibitemOpen
  \bibfield  {author} {\bibinfo {author} {\bibfnamefont {I.~H.}\ \bibnamefont
  {Jermyn}},\ }\bibfield  {title} {\bibinfo {title} {{Invariant Bayesian
  estimation on manifolds}},\ }\href
  {https://doi.org/10.1214/009053604000001273} {\bibfield  {journal} {\bibinfo
  {journal} {Ann. Stat.}\ }\textbf {\bibinfo {volume} {33}},\ \bibinfo {pages}
  {583 } (\bibinfo {year} {2005})}\BibitemShut {NoStop}%
\bibitem [{\citenamefont {Snoussi}(2007)}]{snoussi2007}%
  \BibitemOpen
  \bibfield  {author} {\bibinfo {author} {\bibfnamefont {H.}~\bibnamefont
  {Snoussi}},\ }\bibfield  {title} {\bibinfo {title} {{Bayesian information
  geometry: Application to prior selection on statistical manifolds}},\ }in\
  \href {https://doi.org/https://doi.org/10.1016/S1076-5670(06)46003-1} {\emph
  {\bibinfo {booktitle} {Advances in Imaging and Electron Physics Volume
  146}}}\ (\bibinfo  {publisher} {Elsevier},\ \bibinfo {year} {2007})\ pp.\
  \bibinfo {pages} {163--207}\BibitemShut {NoStop}%
\bibitem [{\citenamefont {Jeffreys}(1946)}]{jeffreys1946invariant}%
  \BibitemOpen
  \bibfield  {author} {\bibinfo {author} {\bibfnamefont {H.}~\bibnamefont
  {Jeffreys}},\ }\bibfield  {title} {\bibinfo {title} {An invariant form for
  the prior probability in estimation problems},\ }\href
  {https://doi.org/10.1098/rspa.1946.0056} {\bibfield  {journal} {\bibinfo
  {journal} {Proc. R. Soc. Lond. A.}\ }\textbf {\bibinfo {volume} {186}},\
  \bibinfo {pages} {453} (\bibinfo {year} {1946})}\BibitemShut {NoStop}%
\bibitem [{\citenamefont {Slater}(1995)}]{slater1995}%
  \BibitemOpen
  \bibfield  {author} {\bibinfo {author} {\bibfnamefont {P.~B.}\ \bibnamefont
  {Slater}},\ }\bibfield  {title} {\bibinfo {title} {{Quantum coin-tossing in a
  Bayesian Jeffreys framework}},\ }\href
  {https://doi.org/https://doi.org/10.1016/0375-9601(95)00601-X} {\bibfield
  {journal} {\bibinfo  {journal} {Phys. Lett. A.}\ }\textbf {\bibinfo {volume}
  {206}},\ \bibinfo {pages} {66} (\bibinfo {year} {1995})}\BibitemShut
  {NoStop}%
\bibitem [{\citenamefont {Kwek}\ \emph {et~al.}(1999)\citenamefont {Kwek},
  \citenamefont {Oh},\ and\ \citenamefont {Wang}}]{kwek1999quantum}%
  \BibitemOpen
  \bibfield  {author} {\bibinfo {author} {\bibfnamefont {L.}~\bibnamefont
  {Kwek}}, \bibinfo {author} {\bibfnamefont {C.}~\bibnamefont {Oh}},\ and\
  \bibinfo {author} {\bibfnamefont {X.-B.}\ \bibnamefont {Wang}},\ }\bibfield
  {title} {\bibinfo {title} {{Quantum Jeffreys prior for displaced squeezed
  thermal states}},\ }\href {https://doi.org/10.1088/0305-4470/32/37/310}
  {\bibfield  {journal} {\bibinfo  {journal} {J. Phys. A: Math. Gen.}\ }\textbf
  {\bibinfo {volume} {32}},\ \bibinfo {pages} {6613} (\bibinfo {year}
  {1999})}\BibitemShut {NoStop}%
\bibitem [{\citenamefont {Li}\ \emph {et~al.}(2018{\natexlab{b}})\citenamefont
  {Li}, \citenamefont {Pezzè}, \citenamefont {Gessner}, \citenamefont {Ren},
  \citenamefont {Li},\ and\ \citenamefont {Smerzi}}]{li2018}%
  \BibitemOpen
  \bibfield  {author} {\bibinfo {author} {\bibfnamefont {Y.}~\bibnamefont
  {Li}}, \bibinfo {author} {\bibfnamefont {L.}~\bibnamefont {Pezzè}}, \bibinfo
  {author} {\bibfnamefont {M.}~\bibnamefont {Gessner}}, \bibinfo {author}
  {\bibfnamefont {Z.}~\bibnamefont {Ren}}, \bibinfo {author} {\bibfnamefont
  {W.}~\bibnamefont {Li}},\ and\ \bibinfo {author} {\bibfnamefont
  {A.}~\bibnamefont {Smerzi}},\ }\bibfield  {title} {\bibinfo {title}
  {{Frequentist and Bayesian Quantum Phase Estimation}},\ }\bibfield  {journal}
  {\bibinfo  {journal} {Entropy}\ }\textbf {\bibinfo {volume} {20}},\ \href
  {https://doi.org/10.3390/e20090628} {10.3390/e20090628} (\bibinfo {year}
  {2018}{\natexlab{b}})\BibitemShut {NoStop}%
\bibitem [{\citenamefont {Dunningham}(2006)}]{dunningham2006using}%
  \BibitemOpen
  \bibfield  {author} {\bibinfo {author} {\bibfnamefont {J.~A.}\ \bibnamefont
  {Dunningham}},\ }\bibfield  {title} {\bibinfo {title} {Using quantum theory
  to improve measurement precision},\ }\href
  {https://doi.org/10.1080/00107510601009871} {\bibfield  {journal} {\bibinfo
  {journal} {Contemp. Phys.}\ }\textbf {\bibinfo {volume} {47}},\ \bibinfo
  {pages} {257} (\bibinfo {year} {2006})}\BibitemShut {NoStop}%
\bibitem [{\citenamefont {Degen}\ \emph
  {et~al.}(2017{\natexlab{b}})\citenamefont {Degen}, \citenamefont {Reinhard},\
  and\ \citenamefont {Cappellaro}}]{degen2017quantum}%
  \BibitemOpen
  \bibfield  {author} {\bibinfo {author} {\bibfnamefont {C.~L.}\ \bibnamefont
  {Degen}}, \bibinfo {author} {\bibfnamefont {F.}~\bibnamefont {Reinhard}},\
  and\ \bibinfo {author} {\bibfnamefont {P.}~\bibnamefont {Cappellaro}},\
  }\bibfield  {title} {\bibinfo {title} {Quantum sensing},\ }\href
  {https://doi.org/10.1103/RevModPhys.89.035002} {\bibfield  {journal}
  {\bibinfo  {journal} {{Rev. Mod. Phys.}}\ }\textbf {\bibinfo {volume} {89}},\
  \bibinfo {pages} {035002} (\bibinfo {year} {2017}{\natexlab{b}})}\BibitemShut
  {NoStop}%
\bibitem [{\citenamefont {V.~Giovannetti}(2004)}]{giovannetti2004quantum}%
  \BibitemOpen
  \bibfield  {author} {\bibinfo {author} {\bibfnamefont {L.~M.}\ \bibnamefont
  {V.~Giovannetti}, \bibfnamefont {S.~Lloyd}},\ }\bibfield  {title} {\bibinfo
  {title} {{Quantum-Enhanced Measurements: Beating the Standard Quantum
  Limit}},\ }\href {https://doi.org/http://dx.doi.org/10.1126/science.1104149}
  {\bibfield  {journal} {\bibinfo  {journal} {Phys. Rev. A}\ }\textbf {\bibinfo
  {volume} {306}},\ \bibinfo {pages} {1330} (\bibinfo {year}
  {2004})}\BibitemShut {NoStop}%
\bibitem [{\citenamefont {Holevo}(2011)}]{holevo2011probabilistic}%
  \BibitemOpen
  \bibfield  {author} {\bibinfo {author} {\bibfnamefont {A.~S.}\ \bibnamefont
  {Holevo}},\ }\href@noop {} {\emph {\bibinfo {title} {{Probabilistic and
  Statistical Aspects of Quantum Theory}}}}\ (\bibinfo  {publisher} {Edizioni
  della Normale, Springer Basel},\ \bibinfo {year} {2011})\BibitemShut
  {NoStop}%
\bibitem [{\citenamefont {Crowley}\ \emph {et~al.}(2014)\citenamefont
  {Crowley}, \citenamefont {Datta}, \citenamefont {Barbieri},\ and\
  \citenamefont {Walmsley}}]{crowley2014tradeoff}%
  \BibitemOpen
  \bibfield  {author} {\bibinfo {author} {\bibfnamefont {P.~J.~D.}\
  \bibnamefont {Crowley}}, \bibinfo {author} {\bibfnamefont {A.}~\bibnamefont
  {Datta}}, \bibinfo {author} {\bibfnamefont {M.}~\bibnamefont {Barbieri}},\
  and\ \bibinfo {author} {\bibfnamefont {I.~A.}\ \bibnamefont {Walmsley}},\
  }\bibfield  {title} {\bibinfo {title} {Tradeoff in simultaneous
  quantum-limited phase and loss estimation in interferometry},\ }\href
  {https://doi.org/10.1103/PhysRevA.89.023845} {\bibfield  {journal} {\bibinfo
  {journal} {Phys. Rev. A}\ }\textbf {\bibinfo {volume} {89}},\ \bibinfo
  {pages} {023845} (\bibinfo {year} {2014})}\BibitemShut {NoStop}%
\bibitem [{\citenamefont {Branford}\ and\ \citenamefont
  {Rubio}(2021)}]{branford2021average}%
  \BibitemOpen
  \bibfield  {author} {\bibinfo {author} {\bibfnamefont {D.}~\bibnamefont
  {Branford}}\ and\ \bibinfo {author} {\bibfnamefont {J.}~\bibnamefont
  {Rubio}},\ }\bibfield  {title} {\bibinfo {title} {Average number is an
  insufficient metric for interferometry},\ }\href
  {https://doi.org/10.1088/1367-2630/ac3571} {\bibfield  {journal} {\bibinfo
  {journal} {New J. Phys.}\ }\textbf {\bibinfo {volume} {23}},\ \bibinfo
  {pages} {123041} (\bibinfo {year} {2021})}\BibitemShut {NoStop}%
\bibitem [{\citenamefont {Smith}\ \emph {et~al.}(2024)\citenamefont {Smith},
  \citenamefont {Barnes},\ and\ \citenamefont
  {Arvidsson-Shukur}}]{smith2024adaptive}%
  \BibitemOpen
  \bibfield  {author} {\bibinfo {author} {\bibfnamefont {J.~G.}\ \bibnamefont
  {Smith}}, \bibinfo {author} {\bibfnamefont {C.~H.~W.}\ \bibnamefont
  {Barnes}},\ and\ \bibinfo {author} {\bibfnamefont {D.~R.~M.}\ \bibnamefont
  {Arvidsson-Shukur}},\ }\bibfield  {title} {\bibinfo {title} {{Adaptive
  Bayesian quantum algorithm for phase estimation}},\ }\href
  {https://doi.org/10.1103/PhysRevA.109.042412} {\bibfield  {journal} {\bibinfo
   {journal} {Phys. Rev. A}\ }\textbf {\bibinfo {volume} {109}},\ \bibinfo
  {pages} {042412} (\bibinfo {year} {2024})}\BibitemShut {NoStop}%
\bibitem [{\citenamefont {Calin}\ and\ \citenamefont
  {Udri{\c{s}}te}(2014)}]{calin2014geometric}%
  \BibitemOpen
  \bibfield  {author} {\bibinfo {author} {\bibfnamefont {O.}~\bibnamefont
  {Calin}}\ and\ \bibinfo {author} {\bibfnamefont {C.}~\bibnamefont
  {Udri{\c{s}}te}},\ }\href@noop {} {\emph {\bibinfo {title} {Geometric
  modeling in probability and statistics}}},\ Vol.\ \bibinfo {volume} {121}\
  (\bibinfo  {publisher} {Springer},\ \bibinfo {year} {2014})\BibitemShut
  {NoStop}%
\bibitem [{\citenamefont {Nielsen}(2022)}]{Nielsen2022}%
  \BibitemOpen
  \bibfield  {author} {\bibinfo {author} {\bibfnamefont {F.}~\bibnamefont
  {Nielsen}},\ }\bibfield  {title} {\bibinfo {title} {The many faces of
  information geometry},\ }\href {https://doi.org/10.1090/noti2403} {\bibfield
  {journal} {\bibinfo  {journal} {Notices Amer. Math. Soc.}\ }\textbf {\bibinfo
  {volume} {69}},\ \bibinfo {pages} {36} (\bibinfo {year} {2022})}\BibitemShut
  {NoStop}%
\bibitem [{\citenamefont {Rao}(1945)}]{rao_information_1945}%
  \BibitemOpen
  \bibfield  {author} {\bibinfo {author} {\bibfnamefont {C.~R.}\ \bibnamefont
  {Rao}},\ }\bibfield  {title} {\bibinfo {title} {Information and accuracy
  attainable in the estimation of statistical parameters},\ }\href
  {https://www.ias.ac.in/article/fulltext/reso/020/01/0076-0090} {\bibfield
  {journal} {\bibinfo  {journal} {Bull. Calcutta Math. Soc.}\ }\textbf
  {\bibinfo {volume} {37}},\ \bibinfo {pages} {81} (\bibinfo {year}
  {1945})}\BibitemShut {NoStop}%
\bibitem [{\citenamefont {Amari}\ and\ \citenamefont
  {Nagaoka}(2007)}]{amari2007}%
  \BibitemOpen
  \bibfield  {author} {\bibinfo {author} {\bibfnamefont {S.-I.}\ \bibnamefont
  {Amari}}\ and\ \bibinfo {author} {\bibfnamefont {H.}~\bibnamefont
  {Nagaoka}},\ }\href@noop {} {\emph {\bibinfo {title} {Methods of information
  geometry}}},\ edited by\ \bibinfo {editor} {\bibfnamefont {S.-I.}\
  \bibnamefont {Amari}}\ and\ \bibinfo {editor} {\bibfnamefont
  {H.}~\bibnamefont {Nagaoka}},\ Translations of mathematical monographs\
  (\bibinfo  {publisher} {American Mathematical Society},\ \bibinfo {address}
  {Providence, RI},\ \bibinfo {year} {2007})\BibitemShut {NoStop}%
\bibitem [{\citenamefont {Eaton}\ and\ \citenamefont
  {Sudderth}(2010)}]{eatonsudderth2010}%
  \BibitemOpen
  \bibfield  {author} {\bibinfo {author} {\bibfnamefont {M.~L.}\ \bibnamefont
  {Eaton}}\ and\ \bibinfo {author} {\bibfnamefont {W.~D.}\ \bibnamefont
  {Sudderth}},\ }\bibfield  {title} {\bibinfo {title} {Invariance of posterior
  distributions under reparametrization},\ }\href
  {http://www.jstor.org/stable/41941457} {\bibfield  {journal} {\bibinfo
  {journal} {Sankhyā: Indian J. Stat., Series A (2008-)}\ }\textbf {\bibinfo
  {volume} {72}},\ \bibinfo {pages} {101} (\bibinfo {year} {2010})}\BibitemShut
  {NoStop}%
\bibitem [{\citenamefont {Braunstein}\ and\ \citenamefont
  {Caves}(1994)}]{Braunstein1994}%
  \BibitemOpen
  \bibfield  {author} {\bibinfo {author} {\bibfnamefont {S.~L.}\ \bibnamefont
  {Braunstein}}\ and\ \bibinfo {author} {\bibfnamefont {C.~M.}\ \bibnamefont
  {Caves}},\ }\bibfield  {title} {\bibinfo {title} {Statistical distance and
  the geometry of quantum states},\ }\href
  {https://doi.org/10.1103/PhysRevLett.72.3439} {\bibfield  {journal} {\bibinfo
   {journal} {Phys. Rev. Lett.}\ }\textbf {\bibinfo {volume} {72}},\ \bibinfo
  {pages} {3439} (\bibinfo {year} {1994})}\BibitemShut {NoStop}%
\bibitem [{\citenamefont {Lu}\ \emph {et~al.}(2022)\citenamefont {Lu},
  \citenamefont {Myilswamy}, \citenamefont {Bennink}, \citenamefont {Seshadri},
  \citenamefont {Alshaykh}, \citenamefont {Liu}, \citenamefont {Kippenberg},
  \citenamefont {Leaird}, \citenamefont {Weiner},\ and\ \citenamefont
  {Lukens}}]{lu2022}%
  \BibitemOpen
  \bibfield  {author} {\bibinfo {author} {\bibfnamefont {H.-H.}\ \bibnamefont
  {Lu}}, \bibinfo {author} {\bibfnamefont {K.~V.}\ \bibnamefont {Myilswamy}},
  \bibinfo {author} {\bibfnamefont {R.~S.}\ \bibnamefont {Bennink}}, \bibinfo
  {author} {\bibfnamefont {S.}~\bibnamefont {Seshadri}}, \bibinfo {author}
  {\bibfnamefont {M.~S.}\ \bibnamefont {Alshaykh}}, \bibinfo {author}
  {\bibfnamefont {J.}~\bibnamefont {Liu}}, \bibinfo {author} {\bibfnamefont
  {T.~J.}\ \bibnamefont {Kippenberg}}, \bibinfo {author} {\bibfnamefont
  {D.~E.}\ \bibnamefont {Leaird}}, \bibinfo {author} {\bibfnamefont {A.~M.}\
  \bibnamefont {Weiner}},\ and\ \bibinfo {author} {\bibfnamefont {J.~M.}\
  \bibnamefont {Lukens}},\ }\bibfield  {title} {\bibinfo {title} {Bayesian
  tomography of high-dimensional on-chip biphoton frequency combs with
  randomized measurements},\ }\href@noop {} {\bibfield  {journal} {\bibinfo
  {journal} {Nat. Commun.}\ }\textbf {\bibinfo {volume} {13}},\ \bibinfo
  {pages} {4338} (\bibinfo {year} {2022})}\BibitemShut {NoStop}%
\bibitem [{\citenamefont {Sommers}\ and\ \citenamefont
  {Zyczkowski}(2003)}]{sommers2003}%
  \BibitemOpen
  \bibfield  {author} {\bibinfo {author} {\bibfnamefont {H.-J.~r.}\
  \bibnamefont {Sommers}}\ and\ \bibinfo {author} {\bibfnamefont
  {K.}~\bibnamefont {Zyczkowski}},\ }\bibfield  {title} {\bibinfo {title}
  {Bures volume of the set of mixed quantum states},\ }\href
  {https://doi.org/10.1088/0305-4470/36/39/308} {\bibfield  {journal} {\bibinfo
   {journal} {Journal of Physics A: Mathematical and General}\ }\textbf
  {\bibinfo {volume} {36}},\ \bibinfo {pages} {10083–10100} (\bibinfo {year}
  {2003})}\BibitemShut {NoStop}%
\bibitem [{\citenamefont {Petz}\ and\ \citenamefont {Sudár}(1996)}]{petz1996}%
  \BibitemOpen
  \bibfield  {author} {\bibinfo {author} {\bibfnamefont {D.}~\bibnamefont
  {Petz}}\ and\ \bibinfo {author} {\bibfnamefont {C.}~\bibnamefont {Sudár}},\
  }\bibfield  {title} {\bibinfo {title} {Geometries of quantum states},\
  }\href@noop {} {\bibfield  {journal} {\bibinfo  {journal} {Journal of
  Mathematical Physics}\ }\textbf {\bibinfo {volume} {37}},\ \bibinfo {pages}
  {2662} (\bibinfo {year} {1996})}\BibitemShut {NoStop}%
\bibitem [{\citenamefont {Al~Osipov}\ \emph {et~al.}(2010)\citenamefont
  {Al~Osipov}, \citenamefont {Sommers},\ and\ \citenamefont
  {{\.Z}yczkowski}}]{osipov2010}%
  \BibitemOpen
  \bibfield  {author} {\bibinfo {author} {\bibfnamefont {V.}~\bibnamefont
  {Al~Osipov}}, \bibinfo {author} {\bibfnamefont {H.-J.}\ \bibnamefont
  {Sommers}},\ and\ \bibinfo {author} {\bibfnamefont {K.}~\bibnamefont
  {{\.Z}yczkowski}},\ }\bibfield  {title} {\bibinfo {title} {Random bures mixed
  states and the distribution of their purity},\ }\href@noop {} {\bibfield
  {journal} {\bibinfo  {journal} {J. Phys. A Math. Theor.}\ }\textbf {\bibinfo
  {volume} {43}},\ \bibinfo {pages} {055302} (\bibinfo {year}
  {2010})}\BibitemShut {NoStop}%
\bibitem [{\citenamefont {Šafránek}(2017)}]{safranek2017}%
  \BibitemOpen
  \bibfield  {author} {\bibinfo {author} {\bibfnamefont {D.}~\bibnamefont
  {Šafránek}},\ }\bibfield  {title} {\bibinfo {title} {Discontinuities of the
  quantum fisher information and the bures metric},\ }\bibfield  {journal}
  {\bibinfo  {journal} {Phys. Rev. A}\ }\textbf {\bibinfo {volume} {95}},\
  \href {https://doi.org/10.1103/physreva.95.052320}
  {10.1103/physreva.95.052320} (\bibinfo {year} {2017})\BibitemShut {NoStop}%
\bibitem [{\citenamefont {Subramanian}\ \emph {et~al.}(2021)\citenamefont
  {Subramanian}, \citenamefont {Jones}, \citenamefont {Frustaci}, \citenamefont
  {Winter}, \citenamefont {van~der Kamp}, \citenamefont {Arcus}, \citenamefont
  {Pudney},\ and\ \citenamefont {Vollmer}}]{subramanian2021sensing}%
  \BibitemOpen
  \bibfield  {author} {\bibinfo {author} {\bibfnamefont {S.}~\bibnamefont
  {Subramanian}}, \bibinfo {author} {\bibfnamefont {H.~B.}\ \bibnamefont
  {Jones}}, \bibinfo {author} {\bibfnamefont {S.}~\bibnamefont {Frustaci}},
  \bibinfo {author} {\bibfnamefont {S.}~\bibnamefont {Winter}}, \bibinfo
  {author} {\bibfnamefont {M.~W.}\ \bibnamefont {van~der Kamp}}, \bibinfo
  {author} {\bibfnamefont {V.~L.}\ \bibnamefont {Arcus}}, \bibinfo {author}
  {\bibfnamefont {C.~R.}\ \bibnamefont {Pudney}},\ and\ \bibinfo {author}
  {\bibfnamefont {F.}~\bibnamefont {Vollmer}},\ }\bibfield  {title} {\bibinfo
  {title} {Sensing enzyme activation heat capacity at the single-molecule level
  using gold-nanorod-based optical whispering gallery modes},\ }\href
  {https://doi.org/doi.org/10.1021/acsanm.1c00176} {\bibfield  {journal}
  {\bibinfo  {journal} {ACS Appl. Nano Mater.}\ }\textbf {\bibinfo {volume}
  {4}},\ \bibinfo {pages} {4576} (\bibinfo {year} {2021})}\BibitemShut
  {NoStop}%
\bibitem [{\citenamefont {Eerqing}\ \emph {et~al.}(2021)\citenamefont
  {Eerqing}, \citenamefont {Subramanian}, \citenamefont {Rubio}, \citenamefont
  {Lutz}, \citenamefont {Wu}, \citenamefont {Anders}, \citenamefont {Soeller},\
  and\ \citenamefont {Vollmer}}]{eerqing2021comparing}%
  \BibitemOpen
  \bibfield  {author} {\bibinfo {author} {\bibfnamefont {N.}~\bibnamefont
  {Eerqing}}, \bibinfo {author} {\bibfnamefont {S.}~\bibnamefont
  {Subramanian}}, \bibinfo {author} {\bibfnamefont {J.}~\bibnamefont {Rubio}},
  \bibinfo {author} {\bibfnamefont {T.}~\bibnamefont {Lutz}}, \bibinfo {author}
  {\bibfnamefont {H.-Y.}\ \bibnamefont {Wu}}, \bibinfo {author} {\bibfnamefont
  {J.}~\bibnamefont {Anders}}, \bibinfo {author} {\bibfnamefont
  {C.}~\bibnamefont {Soeller}},\ and\ \bibinfo {author} {\bibfnamefont
  {F.}~\bibnamefont {Vollmer}},\ }\bibfield  {title} {\bibinfo {title}
  {Comparing transient oligonucleotide hybridization kinetics using dna-paint
  and optoplasmonic single-molecule sensing on gold nanorods},\ }\href
  {https://doi.org/10.1021/acsphotonics.1c01179} {\bibfield  {journal}
  {\bibinfo  {journal} {ACS photonics}\ }\textbf {\bibinfo {volume} {8}},\
  \bibinfo {pages} {2882} (\bibinfo {year} {2021})}\BibitemShut {NoStop}%
\bibitem [{\citenamefont {Mpofu}\ \emph
  {et~al.}(2022{\natexlab{a}})\citenamefont {Mpofu}, \citenamefont {Lee},
  \citenamefont {Maguire}, \citenamefont {Kruger},\ and\ \citenamefont
  {Tame}}]{mpofu2022measuring}%
  \BibitemOpen
  \bibfield  {author} {\bibinfo {author} {\bibfnamefont {K.~T.}\ \bibnamefont
  {Mpofu}}, \bibinfo {author} {\bibfnamefont {C.}~\bibnamefont {Lee}}, \bibinfo
  {author} {\bibfnamefont {G.~E.~M.}\ \bibnamefont {Maguire}}, \bibinfo
  {author} {\bibfnamefont {H.~G.}\ \bibnamefont {Kruger}},\ and\ \bibinfo
  {author} {\bibfnamefont {M.~S.}\ \bibnamefont {Tame}},\ }\bibfield  {title}
  {\bibinfo {title} {Measuring kinetic parameters using quantum plasmonic
  sensing},\ }\href {https://doi.org/10.1103/PhysRevA.105.032619} {\bibfield
  {journal} {\bibinfo  {journal} {Phys. Rev. A}\ }\textbf {\bibinfo {volume}
  {105}},\ \bibinfo {pages} {032619} (\bibinfo {year}
  {2022}{\natexlab{a}})}\BibitemShut {NoStop}%
\bibitem [{\citenamefont {Mpofu}\ \emph
  {et~al.}(2022{\natexlab{b}})\citenamefont {Mpofu}, \citenamefont {Lee},
  \citenamefont {Maguire}, \citenamefont {Kruger},\ and\ \citenamefont
  {Tame}}]{mpofu2022experimental}%
  \BibitemOpen
  \bibfield  {author} {\bibinfo {author} {\bibfnamefont {K.~T.}\ \bibnamefont
  {Mpofu}}, \bibinfo {author} {\bibfnamefont {C.}~\bibnamefont {Lee}}, \bibinfo
  {author} {\bibfnamefont {G.~E.~M.}\ \bibnamefont {Maguire}}, \bibinfo
  {author} {\bibfnamefont {H.~G.}\ \bibnamefont {Kruger}},\ and\ \bibinfo
  {author} {\bibfnamefont {M.~S.}\ \bibnamefont {Tame}},\ }\bibfield  {title}
  {\bibinfo {title} {Experimental measurement of kinetic parameters using
  quantum plasmonic sensing},\ }\href {https://doi.org/10.1063/5.0079896}
  {\bibfield  {journal} {\bibinfo  {journal} {J. Appl. Phys.}\ }\textbf
  {\bibinfo {volume} {131}},\ \bibinfo {pages} {084402} (\bibinfo {year}
  {2022}{\natexlab{b}})}\BibitemShut {NoStop}%
\bibitem [{\citenamefont {Kay}(1993)}]{kay1993fundamentals}%
  \BibitemOpen
  \bibfield  {author} {\bibinfo {author} {\bibfnamefont {S.~M.}\ \bibnamefont
  {Kay}},\ }\href@noop {} {\emph {\bibinfo {title} {{Fundamentals of
  Statistical Signal Processing: Estimation Theory}}}}\ (\bibinfo  {publisher}
  {Prentice Hall},\ \bibinfo {year} {1993})\BibitemShut {NoStop}%
\bibitem [{\citenamefont {Bertlmann}\ and\ \citenamefont
  {Friis}(2023)}]{bertlmann2023modern}%
  \BibitemOpen
  \bibfield  {author} {\bibinfo {author} {\bibfnamefont {R.}~\bibnamefont
  {Bertlmann}}\ and\ \bibinfo {author} {\bibfnamefont {N.}~\bibnamefont
  {Friis}},\ }\href@noop {} {\emph {\bibinfo {title} {Modern Quantum Theory:
  From Quantum Mechanics to Entanglement and Quantum Information}}}\ (\bibinfo
  {publisher} {Oxford University Press},\ \bibinfo {year} {2023})\BibitemShut
  {NoStop}%
\bibitem [{\citenamefont {Baumgratz}\ \emph {et~al.}(2014)\citenamefont
  {Baumgratz}, \citenamefont {Cramer},\ and\ \citenamefont
  {Plenio}}]{baumgratz2014quantifying}%
  \BibitemOpen
  \bibfield  {author} {\bibinfo {author} {\bibfnamefont {T.}~\bibnamefont
  {Baumgratz}}, \bibinfo {author} {\bibfnamefont {M.}~\bibnamefont {Cramer}},\
  and\ \bibinfo {author} {\bibfnamefont {M.~B.}\ \bibnamefont {Plenio}},\
  }\bibfield  {title} {\bibinfo {title} {Quantifying coherence},\ }\href
  {https://doi.org/10.1103/PhysRevLett.113.140401} {\bibfield  {journal}
  {\bibinfo  {journal} {Phys. Rev. Lett.}\ }\textbf {\bibinfo {volume} {113}},\
  \bibinfo {pages} {140401} (\bibinfo {year} {2014})}\BibitemShut {NoStop}%
\bibitem [{\citenamefont {Winter}\ and\ \citenamefont
  {Yang}(2016)}]{winter2016operational}%
  \BibitemOpen
  \bibfield  {author} {\bibinfo {author} {\bibfnamefont {A.}~\bibnamefont
  {Winter}}\ and\ \bibinfo {author} {\bibfnamefont {D.}~\bibnamefont {Yang}},\
  }\bibfield  {title} {\bibinfo {title} {Operational resource theory of
  coherence},\ }\href {https://doi.org/10.1103/PhysRevLett.116.120404}
  {\bibfield  {journal} {\bibinfo  {journal} {Phys. Rev. Lett.}\ }\textbf
  {\bibinfo {volume} {116}},\ \bibinfo {pages} {120404} (\bibinfo {year}
  {2016})}\BibitemShut {NoStop}%
\bibitem [{\citenamefont {Streltsov}\ \emph {et~al.}(2017)\citenamefont
  {Streltsov}, \citenamefont {Adesso},\ and\ \citenamefont
  {Plenio}}]{streltsov2017colloqium}%
  \BibitemOpen
  \bibfield  {author} {\bibinfo {author} {\bibfnamefont {A.}~\bibnamefont
  {Streltsov}}, \bibinfo {author} {\bibfnamefont {G.}~\bibnamefont {Adesso}},\
  and\ \bibinfo {author} {\bibfnamefont {M.~B.}\ \bibnamefont {Plenio}},\
  }\bibfield  {title} {\bibinfo {title} {Colloquium: Quantum coherence as a
  resource},\ }\href {https://doi.org/10.1103/RevModPhys.89.041003} {\bibfield
  {journal} {\bibinfo  {journal} {Rev. Mod. Phys.}\ }\textbf {\bibinfo {volume}
  {89}},\ \bibinfo {pages} {041003} (\bibinfo {year} {2017})}\BibitemShut
  {NoStop}%
\bibitem [{\citenamefont {Chitambar}\ and\ \citenamefont
  {Gour}(2019)}]{chitambar2019quantum}%
  \BibitemOpen
  \bibfield  {author} {\bibinfo {author} {\bibfnamefont {E.}~\bibnamefont
  {Chitambar}}\ and\ \bibinfo {author} {\bibfnamefont {G.}~\bibnamefont
  {Gour}},\ }\bibfield  {title} {\bibinfo {title} {Quantum resource theories},\
  }\href {https://doi.org/10.1103/RevModPhys.91.025001} {\bibfield  {journal}
  {\bibinfo  {journal} {Rev. Mod. Phys.}\ }\textbf {\bibinfo {volume} {91}},\
  \bibinfo {pages} {025001} (\bibinfo {year} {2019})}\BibitemShut {NoStop}%
\bibitem [{\citenamefont {Wu}\ \emph {et~al.}(2021)\citenamefont {Wu},
  \citenamefont {Streltsov}, \citenamefont {Regula}, \citenamefont {Xiang},
  \citenamefont {Li},\ and\ \citenamefont {Guo}}]{wu2021experimental}%
  \BibitemOpen
  \bibfield  {author} {\bibinfo {author} {\bibfnamefont {K.-D.}\ \bibnamefont
  {Wu}}, \bibinfo {author} {\bibfnamefont {A.}~\bibnamefont {Streltsov}},
  \bibinfo {author} {\bibfnamefont {B.}~\bibnamefont {Regula}}, \bibinfo
  {author} {\bibfnamefont {G.-Y.}\ \bibnamefont {Xiang}}, \bibinfo {author}
  {\bibfnamefont {C.-F.}\ \bibnamefont {Li}},\ and\ \bibinfo {author}
  {\bibfnamefont {G.-C.}\ \bibnamefont {Guo}},\ }\bibfield  {title} {\bibinfo
  {title} {Experimental progress on quantum coherence: Detection,
  quantification, and manipulation},\ }\href
  {https://doi.org/https://doi.org/10.1002/qute.202100040} {\bibfield
  {journal} {\bibinfo  {journal} {Adv. Quantum Technol.}\ }\textbf {\bibinfo
  {volume} {4}},\ \bibinfo {pages} {2100040} (\bibinfo {year}
  {2021})}\BibitemShut {NoStop}%
\bibitem [{\citenamefont {Ares~Santos}(2024)}]{ares2024unification}%
  \BibitemOpen
  \bibfield  {author} {\bibinfo {author} {\bibfnamefont {L.}~\bibnamefont
  {Ares~Santos}},\ }\emph {\bibinfo {title} {Unification of nonclassical
  signatures under the concept of quantum coherence}},\ \href
  {https://docta.ucm.es/entities/publication/436af6e1-b4a3-47c3-a61d-c57db3b51b84}
  {Ph.D. thesis},\ \bibinfo  {school} {Universidad Complutense de Madrid}
  (\bibinfo {year} {2024})\BibitemShut {NoStop}%
\bibitem [{\citenamefont {Proctor}\ \emph {et~al.}(2018)\citenamefont
  {Proctor}, \citenamefont {Knott},\ and\ \citenamefont
  {Dunningham}}]{proctor2018multiparameter}%
  \BibitemOpen
  \bibfield  {author} {\bibinfo {author} {\bibfnamefont {T.~J.}\ \bibnamefont
  {Proctor}}, \bibinfo {author} {\bibfnamefont {P.~A.}\ \bibnamefont {Knott}},\
  and\ \bibinfo {author} {\bibfnamefont {J.~A.}\ \bibnamefont {Dunningham}},\
  }\bibfield  {title} {\bibinfo {title} {Multiparameter estimation in networked
  quantum sensors},\ }\href {https://doi.org/10.1103/PhysRevLett.120.080501}
  {\bibfield  {journal} {\bibinfo  {journal} {Phys. Rev. Lett.}\ }\textbf
  {\bibinfo {volume} {120}},\ \bibinfo {pages} {080501} (\bibinfo {year}
  {2018})}\BibitemShut {NoStop}%
\bibitem [{\citenamefont {Rubio}\ \emph {et~al.}(2020)\citenamefont {Rubio},
  \citenamefont {Knott}, \citenamefont {Proctor},\ and\ \citenamefont
  {Dunningham}}]{rubio2020networks}%
  \BibitemOpen
  \bibfield  {author} {\bibinfo {author} {\bibfnamefont {J.}~\bibnamefont
  {Rubio}}, \bibinfo {author} {\bibfnamefont {P.~A.}\ \bibnamefont {Knott}},
  \bibinfo {author} {\bibfnamefont {T.~J.}\ \bibnamefont {Proctor}},\ and\
  \bibinfo {author} {\bibfnamefont {J.~A.}\ \bibnamefont {Dunningham}},\
  }\bibfield  {title} {\bibinfo {title} {Quantum sensing networks for the
  estimation of linear functions},\ }\href
  {https://doi.org/10.1088/1751-8121/ab9d46} {\bibfield  {journal} {\bibinfo
  {journal} {J. Phys. A: Math. Theor.}\ }\textbf {\bibinfo {volume} {53}},\
  \bibinfo {pages} {344001} (\bibinfo {year} {2020})}\BibitemShut {NoStop}%
\bibitem [{\citenamefont {Gross}\ and\ \citenamefont
  {Caves}(2020)}]{gross2021one}%
  \BibitemOpen
  \bibfield  {author} {\bibinfo {author} {\bibfnamefont {J.~A.}\ \bibnamefont
  {Gross}}\ and\ \bibinfo {author} {\bibfnamefont {C.~M.}\ \bibnamefont
  {Caves}},\ }\bibfield  {title} {\bibinfo {title} {One from many: estimating a
  function of many parameters},\ }\href
  {https://doi.org/10.1088/1751-8121/abb9ed} {\bibfield  {journal} {\bibinfo
  {journal} {J. Phys. A: Math. Theor.}\ }\textbf {\bibinfo {volume} {54}},\
  \bibinfo {pages} {014001} (\bibinfo {year} {2020})}\BibitemShut {NoStop}%
\bibitem [{\citenamefont {Barnett}(2009)}]{2009barnett}%
  \BibitemOpen
  \bibfield  {author} {\bibinfo {author} {\bibfnamefont {S.~M. S.~M.}\
  \bibnamefont {Barnett}},\ }\href@noop {} {\emph {\bibinfo {title} {Quantum
  information}}},\ Oxford master series in physics. Atomic, optical and laser
  physics; 16\ (\bibinfo  {publisher} {Oxford University Press},\ \bibinfo
  {address} {Oxford},\ \bibinfo {year} {2009})\BibitemShut {NoStop}%
\bibitem [{\citenamefont {Demkowicz-Dobrza{\'{n}}ski}\ \emph
  {et~al.}(2020)\citenamefont {Demkowicz-Dobrza{\'{n}}ski}, \citenamefont
  {G{\'{o}}recki},\ and\ \citenamefont
  {Gu{\c{t}}{\u{a}}}}]{demkowicz2020multiparameter}%
  \BibitemOpen
  \bibfield  {author} {\bibinfo {author} {\bibfnamefont {R.}~\bibnamefont
  {Demkowicz-Dobrza{\'{n}}ski}}, \bibinfo {author} {\bibfnamefont
  {W.}~\bibnamefont {G{\'{o}}recki}},\ and\ \bibinfo {author} {\bibfnamefont
  {M.}~\bibnamefont {Gu{\c{t}}{\u{a}}}},\ }\bibfield  {title} {\bibinfo {title}
  {{Multi-parameter estimation beyond quantum Fisher information}},\ }\href
  {https://doi.org/10.1088/1751-8121/ab8ef3} {\bibfield  {journal} {\bibinfo
  {journal} {J. Phys. A: Math. Theor.}\ }\textbf {\bibinfo {volume} {53}},\
  \bibinfo {pages} {363001} (\bibinfo {year} {2020})}\BibitemShut {NoStop}%
\bibitem [{\citenamefont {Goldberg}\ \emph
  {et~al.}(2021{\natexlab{b}})\citenamefont {Goldberg}, \citenamefont {Klimov},
  \citenamefont {Leuchs},\ and\ \citenamefont
  {Sánchez-Soto}}]{goldberg2021rotation}%
  \BibitemOpen
  \bibfield  {author} {\bibinfo {author} {\bibfnamefont {A.~Z.}\ \bibnamefont
  {Goldberg}}, \bibinfo {author} {\bibfnamefont {A.~B.}\ \bibnamefont
  {Klimov}}, \bibinfo {author} {\bibfnamefont {G.}~\bibnamefont {Leuchs}},\
  and\ \bibinfo {author} {\bibfnamefont {L.~L.}\ \bibnamefont
  {Sánchez-Soto}},\ }\bibfield  {title} {\bibinfo {title} {Rotation sensing at
  the ultimate limit},\ }\href {https://doi.org/10.1088/2515-7647/abeb54}
  {\bibfield  {journal} {\bibinfo  {journal} {J. Phys.: Photonics.}\ }\textbf
  {\bibinfo {volume} {3}},\ \bibinfo {pages} {022008} (\bibinfo {year}
  {2021}{\natexlab{b}})}\BibitemShut {NoStop}%
\bibitem [{\citenamefont {Larocca}\ \emph {et~al.}(2022)\citenamefont
  {Larocca}, \citenamefont {Sauvage}, \citenamefont {Sbahi}, \citenamefont
  {Verdon}, \citenamefont {Coles},\ and\ \citenamefont
  {Cerezo}}]{larocca2022group}%
  \BibitemOpen
  \bibfield  {author} {\bibinfo {author} {\bibfnamefont {M.}~\bibnamefont
  {Larocca}}, \bibinfo {author} {\bibfnamefont {F.}~\bibnamefont {Sauvage}},
  \bibinfo {author} {\bibfnamefont {F.~M.}\ \bibnamefont {Sbahi}}, \bibinfo
  {author} {\bibfnamefont {G.}~\bibnamefont {Verdon}}, \bibinfo {author}
  {\bibfnamefont {P.~J.}\ \bibnamefont {Coles}},\ and\ \bibinfo {author}
  {\bibfnamefont {M.}~\bibnamefont {Cerezo}},\ }\bibfield  {title} {\bibinfo
  {title} {Group-invariant quantum machine learning},\ }\href
  {https://doi.org/10.1103/PRXQuantum.3.030341} {\bibfield  {journal} {\bibinfo
   {journal} {PRX Quantum}\ }\textbf {\bibinfo {volume} {3}},\ \bibinfo {pages}
  {030341} (\bibinfo {year} {2022})}\BibitemShut {NoStop}%
\bibitem [{\citenamefont {Perrier}(2024)}]{perrier2024quantum}%
  \BibitemOpen
  \bibfield  {author} {\bibinfo {author} {\bibfnamefont {E.}~\bibnamefont
  {Perrier}},\ }\emph {\bibinfo {title} {{Quantum Geometric Machine
  Learning}}},\ \href {https://opus.lib.uts.edu.au/handle/10453/184991} {Ph.D.
  thesis},\ \bibinfo  {school} {Sydney, U. Technol.} (\bibinfo {year}
  {2024})\BibitemShut {NoStop}%
\bibitem [{\citenamefont {Wakeham}\ and\ \citenamefont
  {Schuld}(2024)}]{wakeham2024inference}%
  \BibitemOpen
  \bibfield  {author} {\bibinfo {author} {\bibfnamefont {D.}~\bibnamefont
  {Wakeham}}\ and\ \bibinfo {author} {\bibfnamefont {M.}~\bibnamefont
  {Schuld}},\ }\href {https://arxiv.org/abs/2409.00172} {\bibinfo {title}
  {{Inference, interference and invariance: How the Quantum Fourier Transform
  can help to learn from data}}},\ \bibinfo {howpublished} {arXiv:2409.00172}
  (\bibinfo {year} {2024})\BibitemShut {NoStop}%
\end{thebibliography}%

\end{document}